\documentclass[a4paper,onecolumn,11pt,accepted=2023-03-14]{quantumarticle}
\pdfoutput=1
\usepackage[utf8]{inputenc}
\usepackage[english]{babel}
\usepackage[T1]{fontenc}
\usepackage{amsmath,amssymb,amsfonts,stmaryrd,wasysym,graphicx,multirow,color,textcomp}
\usepackage{url}
\usepackage{tikz-cd}
\usepackage{lipsum}
\usepackage{multirow}
\usepackage{rotating}
\usepackage[numbers]{natbib}
\usepackage{hyperref}

\newcommand{\CFT}{\hyperlink{CFT}{CFT} }
\newcommand{\OPE}{\hyperlink{OPE}{OPE} }
\newcommand{\WZW}{\hyperlink{WZW}{WZW} }
\newcommand{\KM}{\hyperlink{KM}{KM} }
\newcommand{\RG}{\hyperlink{RG}{RG} }
\newcommand{\UMTC}{\hyperlink{UMTC}{UMTC} }
\newcommand{\PH}{\hyperlink{PH}{PH} }
\newcommand{\SPT}{\hyperlink{SPT}{SPT} }
\newcommand{\SET}{\hyperlink{SET}{SET} }

\begin{document}

\title{Dihedral twist liquid models from emergent Majorana fermions}
\author{Jeffrey C. Y. Teo}\affiliation{Department of Physics, University of Virginia, Charlottesville, VA22904, USA}
\author{Yichen Hu}\affiliation{The Rudolf Peierls Centre for Theoretical Physics, University of Oxford, Oxford OX1 3PU, UK}

\begin{abstract}
We present a family of electron-based coupled-wire models of bosonic orbifold topological phases, referred to as twist liquids, in two spatial dimensions. All local fermion degrees of freedom are gapped and removed from the topological order by many-body interactions. Bosonic chiral spin liquids and anyonic superconductors are constructed on an array of interacting wires, each supports emergent massless Majorana fermions that are non-local (fractional) and constitute the $SO(N)$ Kac-Moody Wess-Zumino-Witten algebra at level 1. We focus on the dihedral $D_k$ symmetry of $SO(2n)_1$, and its promotion to a gauge symmetry by manipulating the locality of fermion pairs. Gauging the symmetry (sub)group generates the $\mathcal{C}/G$ twist liquids, where $G=\mathbb{Z}_2$ for $\mathcal{C}=U(1)_l$, $SU(n)_1$, and $G=\mathbb{Z}_2$, $\mathbb{Z}_k$, $D_k$ for $\mathcal{C}=SO(2n)_1$. We construct exactly solvable models for all of these topological states. We prove the presence of a bulk excitation energy gap and demonstrate the appearance of edge orbifold conformal field theories corresponding to the twist liquid topological orders. We analyze the statistical properties of the anyon excitations, including the non-Abelian metaplectic anyons and a new class of quasiparticles referred to as Ising-fluxons. We show an eight-fold periodic gauging pattern in $SO(2n)_1/G$ by identifying the non-chiral components of the twist liquids with discrete gauge theories.
\end{abstract}

\maketitle

\tableofcontents

\section{Introduction}
Topological phases in two spatial dimensions are long-range entangled states of quantum matter that support fractional quasiparticle excitations, known as anyons (a review of these ideas can be found in ref.~\cite{Wilczekbook,Wenbook,Fradkinbook,WenRMP2017}). The topological order of a topological phase is the characterization of the fusion, exchange, and braiding properties of its anyonic excitations. When symmetry is present, these phases are further distinguished by their symmetry-protected or symmetry-enriched topological ({\color{blue}\hypertarget{SPT}{SPT}}/{\color{blue}\hypertarget{SET}{SET}}) orders.~\cite{ChenGuLiuWen12, LuVishwanathE8, MesarosRan12, EssinHermele13, Kapustin14, BiRasmussenSlagleXu14, ElseNayak14, WangGuWen15, LuVishwanath16, ZaletelLuVishwanath17, Chen17} Symmetries in topological phases can be analyzed theoretically by studying the statistical properties of fluxes, referred to as twist defects~\cite{Kitaev06, EtingofNikshychOstrik10, Barkeshli2010, Bombin, Bombin11, KitaevKong12, kong2012A, YouWen, YouJianWen, PetrovaMelladoTchernyshyov14, BarkeshliQi, BarkeshliQi13, BarkeshliJianQi, TeoRoyXiao13long, teo2013braiding, khan2014, TeoHughesFradkin15, BaisHaaker15, TarantinoLindnerFidkowski15, KhanTeoVishveshwara15, Teotwistdefectreview, BarkeshliBondersonChengWang14, BridgemanHahnOsborneWolf20}. These topological point defects ``rotate'' the local winding low-energy degrees of freedom according to the symmetry actions. For example, a twist defect associated to an anyonic symmetry~\cite{khan2014} (or outer automorphism) permutes the anyon type of an orbiting quasiparticle excitation. Such defects must exhibit non-Abelian statistical behaviors even when the underlying topological order is Abelian, and may be useful in the construction of a topological quantum computer~\cite{Preskill97, Freedman98, Kitaev97, OgburnPreskill99, Preskilllecturenotes, FreedmanLarsenWang2002a, FreedmanKitaevLarsenWang01, ChetanSimonSternFreedmanDasSarma, Wangbook, SternLindner13}. 

Twist liquids~\cite{TeoHughesFradkin15} are topological phases where (1) the symmetries are ``elevate'' into local gauge symmetries, and (2) the gauge fluxes are themselves deconfined quantum excitations. The promotion of a (parent) \hyperlink{SPT}{SPT}/\SET phase to a (descendant) twist liquid phase is known as gauging.~\cite{TeoHughesFradkin15, BarkeshliBondersonChengWang14} The fundamental examples are discrete gauge theories~\cite{BaisDrielPropitius92,Propitius-1995,PropitiusBais96,Preskilllecturenotes} $D^{[\omega]}(G)$ in 2+1D. These are twist liquids promoted from a short-range entangled (non-fractional) \SPT phase equipped with the symmetry group $G$. The cohomological classification~\cite{ChenLiuWen11,ChenGuLiuWen11} $[\omega]$ of the parent \SPT phase determines the Dijkgraaf-Witten deformation~\cite{DijkgraafWitten90,DijkgraafPasquierRoche91,AltschulerCoste92,BaisvanDrielPropitius93,Propitius-1995} of the discrete gauge theory that dictates the anyon braiding statistics~\cite{LevinGu12} after gauging. Despite their structural simplicity (excitations have integral quantum dimensions and their braid groups have finite image~\cite{EtingofRowellWitherspoon08}), the exact evaluation of link invariants in some gauge groups, such as the alternating group $G=A_m$ for $m\geq 5$, are in the \#P-complete computational complexity class~\cite{KroviRussell15}. Anyons are capable of universal quantum computing when the finite gauge group $G$ is non-solvable~\cite{Mochon03} (e.g.~$S_3$) or non-nilpotent~\cite{Mochon04} (e.g.~$A_5$) if the gate-set involves measurements~\cite{Preskill97, OgburnPreskill99, BondersonFreedmanNayak08}.

In general, a twist liquid $\mathcal{C}/G$ is the resulting phase when a symmetry group $G$ of a long-range entangled (fractional) \SET phase $\mathcal{C}$ is gauged. In this paper, we focus on a family of chiral twist liquids. In a chiral topological phase, the 1+1D conformal field theory~\cite{GinspargLectureNotes, bigyellowbook, Blumenhagenbook} ({\color{blue}\hypertarget{CFT}{CFT}}) that describes the gapless edge excitations is identified with the unitary modular tensor category~\cite{Walkernotes91, Turaev92, BakalovKirillovlecturenotes, FuchsRunkelSchweigert02, Rowell05, Kitaev06, Bondersonthesis, RowellStongWang09, Turaevbook, Delaneylecturenotes} ({\color{blue}\hypertarget{UMTC}{UMTC}}) that describes the 2+1D gapped topological order by the bulk-edge correspondence~\cite{FrohlichGabbiani90, MooreRead, Wenedgereview, ReadRezayi, Kitaev06}. In a chiral twist liquid, symmetry gauging acts as an orbifolding procedure~\cite{DixonHarveyVafaWitten85I,DixonHarveyVafaWitten85II,Ginsparg88,DijkgraafVerlindeVerlinde88,MooreSeiberg89zoo} on the edge \CFT so that the bulk-edge correspondence is preserved~\cite{ChenRoyTeoRyu17}. The deconfined gauge fluxes in the bulk associate to twist fields that change the boundary conditions on the edge. Examples include orbifold fractional quantum Hall states~\cite{Barkeshli2010,BarkeshliWen10PRL,BarkeshliWen11,BarkeshliWen12,MollerHormoziSlingerlandSimon14,KaneStern18,TamHuKane20} and quantum spin liquids~\cite{LevinGu12, LuVishwanath16}. In this paper, with the interest on the topological and modular properties, we do not distinguish between the twist liquid bulk \UMTC and its edge orbifold \hyperlink{CFT}{CFT}, and we will refer to both as $\mathcal{C}/G$.

While non-chiral twist liquids can be constructed microscopically from exactly-solvable string-net models~\cite{TeoHughesFradkin15, BarkeshliBondersonChengWang14, LevinWen05} and this provides a proof of principle that the orbifold topological order can be supported by a gapped local Hamiltonian, the method is not designed to analyze chiral electronic twist liquid phases. Moreover, the fundamental relationship between symmetry gauging and the change of locality is obscure. On one hand, the gauge charges in a twist liquid are non-local and fractional because they carry non-trivial mutual braiding statistics with the gauge fluxes. On the other hand, they are derived from field operators in the parent globally-symmetric \SET phase that are local integral combinations of electrons and ``condense'' in the anyon condensation~\cite{BaisSlingerlandCondensation,Kong14,NeupertHeKeyserlingkSierraBernevig16,Burnell18} picture. In other words, gauging is the reversal of gauge charge condensation. In this paper, we seek a microscopic model description based on electrons that explicitly demonstrates such reconfiguration of gauge charge locality. At the same time, we explore models that display the bulk-edge correspondence between chiral twist liquid bulk topological order and edge orbifold \hyperlink{CFT}{CFT}.

The coupled-wire construction~\cite{KaneMukhopadhyayLubensky02, TeoKaneCouplewires} provides a strategy to obtain exactly-solvable model Hamiltonians for these purposes. The theoretical technique was inspired by sliding Lutthinger liquids~\cite{OHernLubenskyToner99,EmeryFradkinKivelsonLubensky00,VishwanathCarpentier01,SondhiYang01,MukhopadhyayKaneLubensky01} and was first implemented by Kane, Mukhopadhyay, and Lubensky~\cite{KaneMukhopadhyayLubensky02} to describe the Abelian Laughlin~\cite{Laughlin83} and Haldane-Halperin hierarchy~\cite{Haldane83,Halperin84} fractional quantum Hall states. The method was later generalized to other (Abelian or non-Abelian, integer or fractional) quantum Hall states~\cite{TeoKaneCouplewires,KlinovajaLoss14,MengStanoKlinovajaLoss14,SagiOregSternHalperin15,FujiHeBhattacharjeePollmann16,KaneSternHalperin17,FujiLecheminant17,KaneStern18,FujiFurusaki18,SirotaSahooChoTeo18,FontanaGomesHernaski19,LopesQuitoHanTeo19,ImamuraTotsukaHansson19,TamHuKane20,TamKane21}, anyon models~\cite{OregSelaStern14,StoudenmireClarkeMongAlicea15,IadecolaNeupertChamonMudry19,TamVenderbosKane21}, spin liquids~\cite{MengNeupertGreiterThomale15,GorohovskyPereiraSela15,HuangChenGomesNeupertChamonMudry16,PatelChowdhury16}, (fractional) topological insulators~\cite{NeupertChamonMudryThomale14,KlinovajaTserkovnyak14,SagiOreg14,MrossEssinAlicea15,SantosHuangGefenGutman15,SirotaRazaTeo17,HanTeo18} and superconductors~\cite{mongg2,SeroussiBergOreg14,SahooZhangTeo15,HuKane18,ParkRazaGilbertTeo18,Meng18,YangPerrinPetrescuGarateLeHur20}, fracton models~\cite{SullivanIadecolaWilliamson21,SullivanDuaCheng21} and higher-dimensional fractional phases~\cite{IadecolaNeupertChamonMudry16,FujiFurusaki19,SagiOreg15,Meng15,MengGrushinShtengelBardarson16}, as well as the exploration of dualities~\cite{MrossAliceaMotrunich16,MrossAliceaMotrunich17} and quantum entanglement~\cite{CanoHughesMulligan15,SohalHanSantosTeo20,PakHamedTeoMulligan21}. The coupled-wire construction is a highly anisotropic microscopic description where the low-energy electronic degrees of freedom are confined on a 2D array of one-dimensional wires. Local many-body intra- and inter-wire electron backscattering interactions are responsible for the finite excitation energy gap in the bulk but leave behind gapless modes on the edge.

A universal procedure of constructing a coupled-wire model for a general \UMTC has not been established, and it is out of the scope of this paper to construct electronic models for general twist liquids. Among orbifold phases $\mathcal{C}/G$, there is a sub-collection where the parent $G$-symmetric \SET phase $\mathcal{C}$ corresponds to an edge \CFT described by an affine Kac-Moody~\cite{Kac68,Moody68} ({\color{blue}\hypertarget{KM}{KM}}) Wess-Zumino-Witten~\cite{WessZumino71,WittenWZW,witten1984} ({\color{blue}\hypertarget{WZW}{WZW}}) current Lie algebra. These \WZW phases are of significant interest in the coupled-wire construction because (1) some of the many-body electron interactions can be introduced as backscattering of the local \KM currents (such as the Gross-Neveu interactions~\cite{GrossNeveu,ZamolodchikovZamolodchikov78,Witten78,ShankarWitten78}), and (2) some non-local \KM current operators can be associated with the gauge charges in the twist liquid. In particular, when $\mathcal{C}$ is an affine simply-laced \KM \WZW Lie algebra at level 1 (i.e.~the $A_r=SU(r+1)$, $D_r=SO(2r)$ series and the exceptional $E_6$, $E_7$, $E_8$), it has an Abelian topological order. $N$-dimensional real (complex) point groups are discrete finite subgroups $G$ in $SO(N)$ ($SU(N)$), and they correspond to the orbifold phases $SO(N)_k/G$ ($SU(N)_k/G$), where $k$ is the level of the \WZW algebra. 3-dimensional point group orbifold \hyperlink{CFT}{CFT}s $SU(2)_1/G$ are known~\cite{Ginsparg88,DijkgraafVerlindeVerlinde88}. In this paper, we focus on gauging a dihedral symmetry group $D_k$ in $SO(2n)_1$. In doing so, the construction also encompasses the anyon relabeling $\mathbb{Z}_2$ conjugation symmetry in $U(1)_l$ and $SU(n)_1$.

We summarize the main results of this paper below. We discuss the implications and possible future directions in section~\ref{sec:conclusion}.

\subsection{Summary of results}\label{sec:overview}
In this paper, we investigate various aspects and consequences of gauging a non-Abelian dihedral symmetry group $D_k$ of topological phases of matter in the orthogonal family. The results we achieved are threefold. 

First, we present two different ways of constructing a one-dimensional quantum wire of gapless bosons with an effective $SO(N)$ symmetry in low energy. We begin with a one-dimensional quantum wire of interacting spinful electrons. By taking advantage of either an umklapp scattering term or a superconducting pairing term, we lift all charge modes and local fermion excitations above a finite energy gap. The residual low-energy degrees of freedom below the energy gap are gapless bosons, which can be effectively described by the $SO(N)_1$ \WZW \hyperlink{CFT}{CFT}. These boson wires serve as the basic building blocks for the subsequent coupled wire constructions of the globally symmetric ``parent'' topological order phases ($\mathcal{C}$) and their gauged and locally symmetric ``descendant'' counterparts (twist liquid / orbifold phases $\mathcal{C}/G$) in the orthogonal and unitary family. They include (1) $\mathcal{C}=SO(2n)_1$ and its $G=\mathbb{Z}_2$, $\mathbb{Z}_k$, or the non-Abelian $D_k=\mathbb{Z}_2\ltimes\mathbb{Z}_k$ twist liquid phases, (2) $\mathcal{C}=U(1)_l$, $SU(n)_1$, or $U(1)_l\times SU(n)_1$ (conformally embedded as \WZW subalgebras of $SO(2n)_1$) and their $G=\mathbb{Z}_2$ twist liquid phases. The degree, rank, and level are related by $k=n$ and $l=4n$ (or $k=n/2$ and $l=n$) when $n$ is odd (resp.~even). The level $l$ of $U(1)_l$ corresponds to the radius $R=\sqrt{l}/2$ of the compactified free boson \CFT at central charge $c=1$. The coupled-wire constructions based on the umklapp (pairing) approach lead to a topologically ordered spin liquid (superconductor). 

\begin{figure}[htbp]
\centering\includegraphics[width=0.49\textwidth]{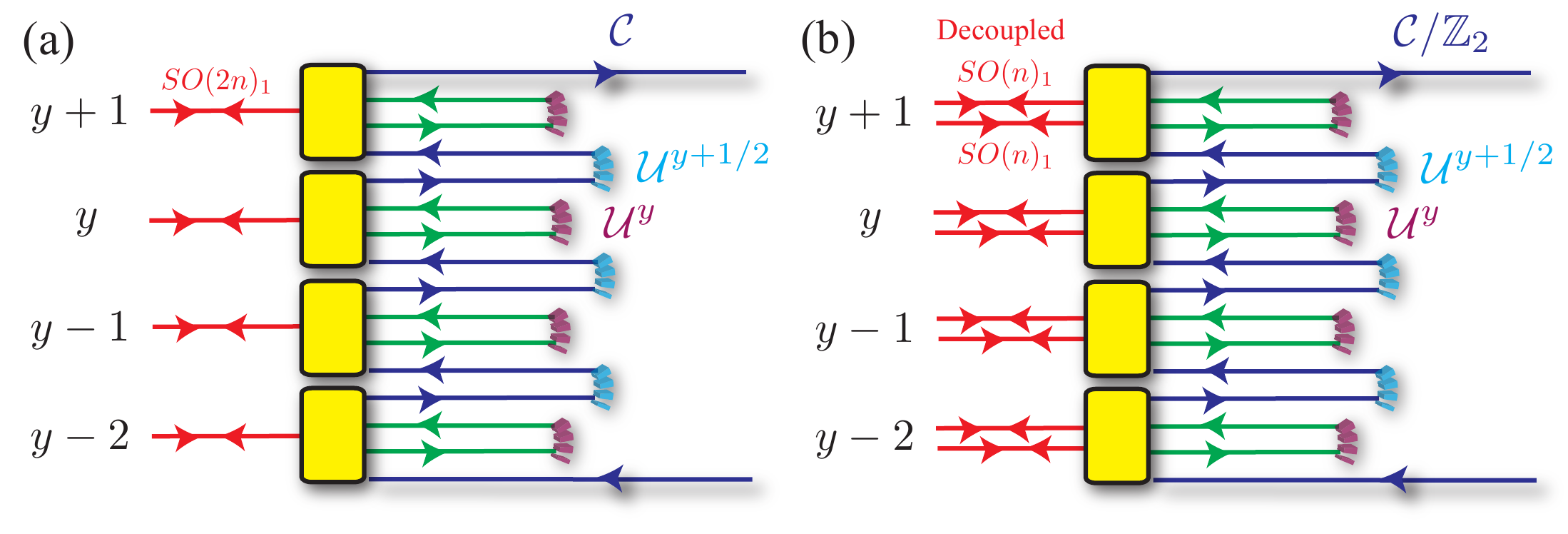}
\caption{Coupled-wire models for (a) $\mathcal{C}=U(1)_l$ or $SU(n)_1$ Abelian topological states, and (b) $\mathcal{C}/\mathbb{Z}_2=U(1)_l/\mathbb{Z}_2$ or $SU(n)_1/\mathbb{Z}_2$ twist liquid topological states from bosonic $SO(N)_1$ wires. Intra-/inter-wire interactions $\mathcal{U}^y$, $\mathcal{U}^{y+1/2}$ introduce a finite bulk excitation energy gap and leave behind chiral gapless boundaries described by CFTs that correspond to the bulk topological orders. Yellow boxes represent splitting of channels by (a) the conformal embedding $SO(2n)_1=U(1)_l\times SU(n)_1$, and (b) the coset decomposition $SO(n)_1^2=SO(n)_2\times[SO(n)_1^2/SO(n)_2]$.}\label{fig:CWMschematic}
\end{figure}

\begin{figure}[htbp]
\centering\includegraphics[width=0.45\textwidth]{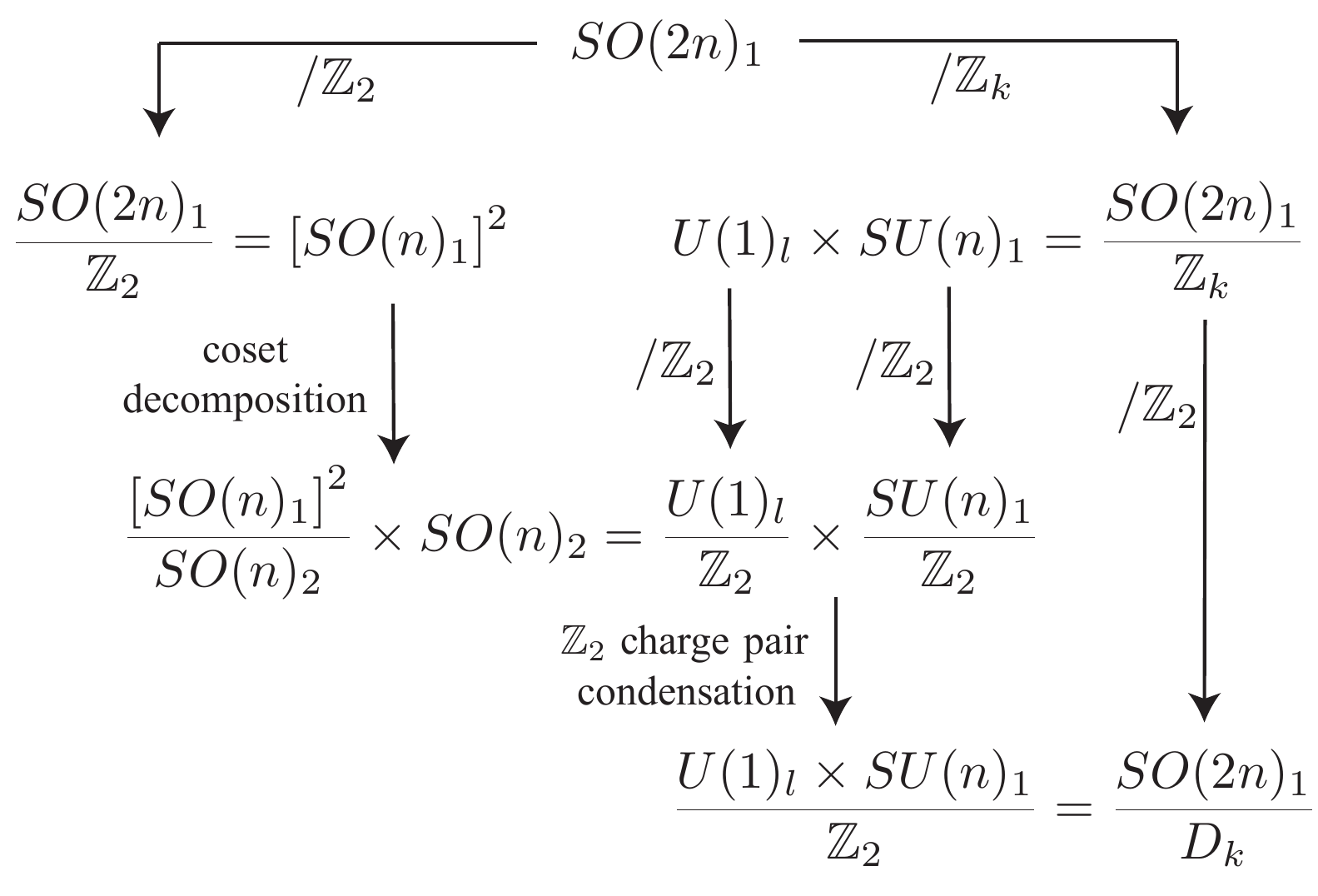}
\caption{Gauging flowchart of $SO(2n)_1$ and twist liquid orbifold phases constructed in this paper.}\label{fig:familytree}
\end{figure}

Since any intra- and inter-wire coupling term in a coupled-wire Hamiltonian has to be made out of integral combinations of local electrons, it is crucial to keep track of the locality of field operators and their physical origin. This is our second key result. Starting from our wires of bosons, upon re-fermionization, each wire carries $N$ pairs of counter-propagating Majorana fermions. Different from the local electronic Majorana fermions based on the Bogoliubov--de Gennes (BdG) formalism in a conventional Bardeen--Cooper--Schrieffer (BCS) superconductor, our Majorana fermions here are emergent and fractional. Only even fermion products are integral combinations of local electrons. More importantly, we explicitly connect symmetry gauging and the modification of the locality of field operators. For a general twist liquid phase $\mathcal{C}/G$, local fields in $\mathcal{C}/G$ are exclusively $G$-invariant local fields in $\mathcal{C}$. Taking $\mathcal{C}=SO(2n)_1$ as an example, for its Majorana fermions $\psi^{i=1,\cdots,2n}$, the $\mathbb{Z}_2$ symmetry sends $\psi_A^{i=1,\cdots,n} \equiv \psi^i \rightarrow \psi_A^i$ and $\psi_B^{i=1,\cdots,n}\equiv \psi^{i+n} \rightarrow -\psi_B^i$. At the same time, the $\mathbb{Z}_k$ symmetry adds a phase to the Dirac fermion $d^i\sim\psi_A^i+i\psi_B^i \rightarrow e^{i2\pi/k}d^i$. While $\psi_A^i\psi_B^i$ are integral in $SO(2n)_1$, they become fractional in $SO(2n)_1/\mathbb{Z}_2$ because they are not invariant under the internal $\mathbb{Z}_2$ symmetry. Similarly, the Dirac pairs $d^id^j$ are not local fields in $SO(2n)_1/\mathbb{Z}_k$ because their phases are rotated by the internal $\mathbb{Z}_k$ symmetry. Moving on to the non-Abelian $D_k$ symmetry, we find that the \begin{align}\frac{SU(n)_1}{\mathbb{Z}_2}=SO(n)_2\quad\mbox{and}\quad\frac{U(1)_l}{\mathbb{Z}_2}=\frac{SO(n)_1\times SO(n)_1}{SO(n)_2}\label{SUnU1Z2}\end{align} topological orders are based on local fields of $SO(2n)_1$ that are invariant under the $D_k$ symmetry. Inside these topological phases, we highlight the existence of metaplectic anyons in the form of $\mathbb{Z}_2$ fluxes, when $n$ is odd, together with a microscopic electronic Hamiltonian from the coupled-wire construction. (See figure~\ref{fig:CWMschematic} for a diagrammatic outline of the coupled-wire models and figure~\ref{fig:familytree} for a summary of the gauging relations between different twist liquids constructed in this paper.) Moreover, we discover a novel type of anyon with quantum dimension $\sqrt{2}n$ in \begin{align}\frac{SO(2n)_1}{D_k}=\frac{U(1)_l\times SU(n)_1}{\mathbb{Z}_2}\end{align} when $n$ is odd. It is a composite quasiparticle that combines twist fields from both the $SO(n)_2$ and $U(1)_l/\mathbb{Z}_2$ theories. We dubbed it an ``Ising-fluxon''. 

Third, we provide a systematic characterization of the $G$-twist liquids ($G=\mathbb{Z}_2$, $\mathbb{Z}_k$, or $D_k)$ at the level of unitary modular tensor categories (\hyperlink{UMTC}{UMTC}). Loosely speaking~\cite{TeoHughesFradkin15}, anyons in $\mathcal{C}/G$ are super-selection combinations of anyons in $\mathcal{C}$ attached with deconfined $G$-gauge fluxes and gauge charges. If $G$ contains only inner automorphisms that do not permute anyon types, such as the case when $\mathcal{C}=SO(2n)_1$ and (i) $G=\mathbb{Z}_k$ for all $n$, and (ii) $G=\mathbb{Z}_2$, or $D_k$ for even $n$, we demonstrate the \UMTC equivalence \begin{align}SO(2n)_1/G=SO(2n)_1\boxtimes D^{[\omega]}(\widehat{G}).\label{innerautgauging}\end{align} Here, the quantum symmetry group $\widehat{G}$ is a subgroup of an extension of $G$ by the Abelian fusion group $\mathcal{A}$ of anyons in $\mathcal{C}$. In general, the group extension can be projective~\cite{Wenspinliquid02,EtingofNikshychOstrik10,BarkeshliBondersonChengWang14} (i.e.~non-symmorphic~\cite{TeoHughesFradkin15}) and is classified by the group cohomology~\cite{Cohomologybook} $H^2(G,\mathcal{A})$. $D^{[\omega]}(\widehat{G})$ denotes the discrete $\widehat{G}$ gauge theory~\cite{BaisDrielPropitius92,Propitius-1995,PropitiusBais96,Preskilllecturenotes} deformed by the Dijkgraaf-Witten invariant~\cite{DijkgraafWitten90,DijkgraafPasquierRoche91,AltschulerCoste92,BaisvanDrielPropitius93,Propitius-1995} $[\omega]$ in $H^3(\widehat{G},U(1))$. The invariant $[\omega]$ specifies the quantum basis transformations (i.e.~associations) between flux combinations with different orders, $\hat{g}_1(\hat{g}_2\hat{g}_3)\cong(\hat{g}_1\hat{g}_2)\hat{g}_3$, and in turn, dictates the spin-exchange statistics of gauge fluxes. The tensor product $\boxtimes$ is {\em relative} to a set of bosonic pairs of anyons from both sides that are in fact local and are ``condensed'' in the anyon condensation picture~\cite{BaisSlingerlandCondensation,Kong14,NeupertHeKeyserlingkSierraBernevig16,Burnell18}. (The decoupled product $\times$ is a special case when there is no anyon pair other than the vacuum is ``condensed''.) 
Projective quantum symmetry groups $\widehat{G}$ relevant to this paper include (i) $\widehat{\mathbb{Z}_2}=\mathbb{Z}_4$, $\widehat{D_k}=Q_{4k}$ when $n=2k\equiv2$ modulo 4, and (ii) $\widehat{\mathbb{Z}_k}=\mathbb{Z}_{2k}$, $\widehat{D_k}=D_{2k}$ when $n=2k\equiv0$ modulo 4. In particular, the dicyclic group $Q_{4k}$ is the double cover of the dihedral group $D_k$ that contains the perpendicular axes of $k$-fold and 2-fold rotations in a half-integral spin representation. 

On the other hand, the $\mathbb{Z}_2$ symmetry is an outer automorphism for $U(1)_l$ and $SU(n)_1$. The twofold symmetry non-trivially transposes anyon classes by conjugation. Gauging these symmetries then takes us beyond conventional discrete gauge theories. This, in general, enables non-Abelian quasiparticles with non-integral quantum dimensions, such as the $\mathbb{Z}_2$ twist fields in the two prototypical $\mathbb{Z}_2$ twist liquid phases in \eqref{SUnU1Z2}. For $SO(2n)_1$ with odd rank $n=k$, the $\mathbb{Z}_2$, and subsequently, the $D_k$ symmetries contain outer automorphisms (also known as anyon relabeling symmetries~\cite{khan2014,Teotwistdefectreview}). Upon gauging, we show \begin{align}SO(2n)_1/G=SO(2n)_1\boxtimes\mathcal{Z}_n(G),\label{outerautgauging}\end{align} where $\mathcal{Z}_n(G)$ is a non-chiral quantum double~\cite{Kasselbook,BakalovKirillovlecturenotes,LevinWen05} carrying a $G$-gauge symmetry. $\mathcal{Z}_n(\mathbb{Z}_2)$ has a non-chiral Ising topological order and is the twist liquid phase from gauging the twofold electric-magnetic symmetry of the $\mathbb{Z}_2$ discrete gauge theory.~\cite{BarkeshliWen12,TeoHughesFradkin15,ChenRoyTeoRyu17} $\mathcal{Z}_n(D_k)$ is the twist liquid phase of gauging a {\em mixed} $\mathbb{Z}_2$ symmetry of the $\mathbb{Z}_k$ discrete gauge theory. It supports the Ising-fluxon excitations and is referred to as the Ising-fluxon topological phase in this article. Combining both even and odd $n$ cases in \eqref{innerautgauging} and \eqref{outerautgauging}, we observe an eight-fold periodic pattern of the topological orders of the twist liquids $SO(2n)_1/G$.

The paper is organized as the following. In section~\ref{sec:bosonwires}, we present the superconductor and spin liquid constructions of 1+1D gapless boson wires from interacting spinful electrons. The $SO(N)_1$ \WZW symmetry, locality and internal $\mathbb{Z}_2$ symmetry of field operators within a single wire are also discussed. In section~\ref{sec:SONSUN1}, we proceed to describe the coupled-wire constructions for 2+1D topological phases of the orthogonal and unitary families. The $SO(N)_1$ models are presented in section~\ref{sec:SON} while the $U(1)_l$ and $SU(n)_1$ models are presented in section~\ref{sec:unitaryfamily}. The global $D_k$ symmetry of the ``parent'' $SO(2n)_1$ phase is described in section~\ref{sec:gaugingZkZ2}. Its $\mathbb{Z}_2$ and $\mathbb{Z}_k$ twist liquid ``descendants'' $SO(2n)_1/\mathbb{Z}_2=SO(n)_1^2$ and $SO(2n)_1/\mathbb{Z}_k=U(1)_l\times SU(n)_1$ are introduced in section~\ref{sec:SO2nZ2TL} and \ref{sec:SO2nZkTL} respectively. In section~\ref{sec:dihedralTL}, we develop the dihedral twist liquid phase $SO(2n)_1/D_k=\left[U(1)_l\times SU(n)_1\right]/\mathbb{Z}_2$. Its construction relies on the simpler $SU(n)_1/\mathbb{Z}_2=SO(n)_2$ and $U(1)_l/\mathbb{Z}_2=SO(n)_1^2/SO(n)_2$ orbifold phases, which are presented in section~\ref{sec:SUnZ2TL} and \ref{sec:U1Z2orbifold}. For each case, the relation between locality and internal symmetries of field operators is carefully examined at the beginning. Then, we explicitly present the exactly solvable intra- and inter-wire interaction terms from local electrons. The interactions introduce a finite 2D bulk excitation energy gap but leave behind gapless edge \CFT that corresponds to the bulk topological order. The novel eight-fold periodic topological order of $SO(2n)_1/D_k$ is fully narrated in section~\ref{SO2nDkTL}. 

While the new and essential content is presented in the main sections, we include comprehensive revisions on the relevant background in four appendices. The \WZW \KM algebras and their primary field modular contents of $SO(N)_1$, $SO(n)_2$ and the $U(1)_l/\mathbb{Z}_2$ orbifold \CFT are reviewed in appendix~\ref{app:SO(N)algebra} and \ref{app:orbifoldanyons}. Discrete gauge theories with cyclic, dihedral, and dicyclic gauge groups, $\mathbb{Z}_k$, $D_k$ and $Q_{4k}$, appear as non-chiral components of the various $SO(2n)_1/G$ twist liquids introduced in the main sections. They are reviewed in appendix~\ref{app:DGT}. The group extensions and Dijkgraaf-Witten invariants relevant to the discrete gauge theories are classified by the group cohomologies $H^2(G,\mathcal{A})$ and $H^3(G,U(1))$. We present these cohomology classifications and the explicit cycle representations in appendix~\ref{app:cohomology}.

\section{Boson wires from interacting electrons}\label{sec:bosonwires}
The building blocks of the coupled-wire models are gapless bosons in one spatial dimension that carry an effective $SO(N)$ symmetry in low energy. These wires can be realized by 1D electrons under strong interactions that lift all charge modes and local fermion excitations above a finite energy gap. The remaining electrically neutral gapless modes below the gap are described by a $SO(N)$ Wess-Zumino-Witten~\cite{WessZumino71,WittenWZW,witten1984} (\hyperlink{WZW}{WZW}) Kac-Moody~\cite{Kac68,Moody68} (\hyperlink{KM}{KM}) conformal field theory~\cite{GinspargLectureNotes,bigyellowbook,Blumenhagenbook} (\hyperlink{CFT}{CFT}) at level 1. We illustrate this using $\mathcal{N}$ copies of Rashba spin-orbit coupled electrons. The electron operators $c_{sj}(x)\approx c_{sj}^L(x)e^{i(k_f+sk_{SO})x}+c_{sj}^R(x)e^{i(-k_f+sk_{SO})x}$ near the Fermi level can be represented by vertex operators of bosonized variables $c_{sj}^\sigma(x)\sim e^{i\Phi_{sj}^\sigma(x)}$, where $s=\uparrow(\downarrow)=+(-)$ labels spins, $\sigma=L(R)=+(-)$ specifies the left/right propagation directions, $j=1,\ldots,\mathcal{N}$ designates the electron channels, and the Fermi momenta of the channels are $k_s^\sigma=\sigma k_f+sk_{SO}$. The low-energy kinetic behavior of the electrons can be described by the Luttinger liquid~\cite{Tomonaga50,Luttinger63} Lagrangian density \begin{align}\begin{split}\mathcal{L}_0&=\frac{1}{4\pi}\sigma\delta_{\sigma\sigma'}\delta^{ss'}\delta^{jj'}\partial_x\Phi_{sj}^\sigma\partial_t\Phi_{s'j'}^{\sigma'}-\mathcal{H}_0,\\\mathcal{H}_0&=v^{sj,s'j'}_{\sigma\sigma'}\partial_x\Phi_{sj}^\sigma\partial_x\Phi_{s'j'}^{\sigma'}.\end{split}\label{freeL0}\end{align} The Hamiltonian density $\mathcal{H}_0=\mathcal{H}_{\mathrm{Dirac}}+\mathcal{H}_{\mathrm{int}}$ includes the single-body massless Dirac theory $\mathcal{H}_{\mathrm{Dirac}}=iv\sigma {c_{sj}^\sigma}^\dagger\partial_x c_{sj}^\sigma=\frac{v}{4\pi}(\partial_x\Phi^\sigma_{sj})^2$ and density-density interactions $\mathcal{H}_{\mathrm{int}}=u^{sj,s'j'}_{\sigma\sigma'}n_{sj}^\sigma n_{s'j'}^{\sigma'}$, where $n_{sj}^\sigma={c_{sj}^\sigma}^\dagger c_{sj}^\sigma=\sigma\partial_x\Phi_{sj}^\sigma/(2\pi)$. The ``$p\dot{q}$'' term of the Lagrangian sets the equal-time commutation relation \begin{align}\left[\Phi^\sigma_{sj}(x),\partial_{x'}\Phi^{\sigma'}_{s'j'}(x')\right]=2\pi i\sigma\delta^{\sigma\sigma'}\delta_{ss'}\delta_{jj'}\delta(x-x').\label{ETCR0}\end{align} 

The charged $U(1)$ sector can be gapped by either a $\mathcal{N}$-body umklapp scattering or a superconducting pairing \begin{gather}\mathcal{U}_{\mathrm{umklapp}}=u\cos\theta_\rho\sim u\prod_{j=1}^\mathcal{N}\prod_{s=\uparrow,\downarrow}{c_{sj}^L}^\dagger c_{sj}^R+h.c.,\nonumber\\\mathcal{U}_{\mathrm{SC}}=\Delta\cos\varphi_\rho\sim\Delta\prod_{j=1}^\mathcal{N}\prod_{s=\uparrow,\downarrow}c_{sj}^Lc_{sj}^R+h.c.,\label{umpklappSC}\\\varphi_\rho=\sum_{sj}\left(\Phi^L_{sj}+\Phi^R_{sj}\right),\quad\theta_\rho=\sum_{sj}\left(\Phi^L_{sj}-\Phi^R_{sj}\right).\nonumber\end{gather} This can be shown by re-expressing the kinetic Lagrangian density \eqref{freeL0} in terms of new variables \begin{align}\mathcal{L}_0&=\frac{1}{4\pi}\sum_{\sigma=+,-}\sigma\left[2\partial_x\phi_\rho^\sigma\partial_t\phi_\rho^\sigma+\sum_{a=1}^{2\mathcal{N}-1}\partial_x\phi_a^\sigma\partial_t\phi_a^\sigma\right]-\mathcal{H}_0.\label{freeL}\end{align} This can be achieved by using the basis transformation \begin{align}2\phi_\rho^\sigma=\varphi_0+\sigma\theta_\rho,\quad\phi^\sigma_a=\varphi_0-\Phi^\sigma_{s_aj_a},\label{basistransformationumklapp}\end{align} for the umklapp scattering case, or \begin{align}2\phi_\rho^\sigma=\varphi_\rho+\sigma\theta_0,\quad\phi^\sigma_a=\theta_0-\sigma\Phi^\sigma_{s_aj_a},\label{basistransformationSC}\end{align} for the superconducting case, where \begin{align}\varphi_0=\frac{\Phi^L_{\uparrow1}+\Phi^R_{\downarrow1}}{2},\quad\theta_0=\frac{\Phi^L_{\uparrow1}-\Phi^R_{\downarrow1}}{2}.\end{align} Here, in \eqref{basistransformationumklapp} and \eqref{basistransformationSC}, the $a$ index ranges in $1,\ldots,2\mathcal{N}-1$. It enumerates the spins $s=\uparrow,\downarrow$ and channels $j=1,\ldots,\mathcal{N}$ but exclude $(s,j)=(\uparrow,1)$ for the $\sigma=L$ sector and $(\downarrow,1)$ for the $\sigma=R$ sector. The enumeration can be explicitly chosen by assigning in the $L$ sector, \[\begin{split}&s_a=\uparrow,\quad j_a=a/2+1,\quad\mbox{if $a$ is even}\\&s_a=\downarrow,\quad j_a=(a+1)/2,\quad\mbox{if $a$ is odd},\end{split}\] or in the $R$ sector, \[\begin{split}&s_a=\downarrow,\quad j_a=a/2+1,\quad\mbox{if $a$ is even}\\&s_a=\uparrow,\quad j_a=(a+1)/2,\quad\mbox{if $a$ is odd}.\end{split}\] The set of bosonized variables in \eqref{basistransformationumklapp} or in \eqref{basistransformationSC} obey the equal-time commutation relations \begin{align}\begin{split}\left[\phi^\sigma_\rho(x),\partial_{x'}\phi^{\sigma'}_\rho(x')\right]&=\pi i\sigma\delta^{\sigma\sigma'}\delta(x-x'),\\\left[\phi^\sigma_a(x),\partial_{x'}\phi^{\sigma'}_{a'}(x')\right]&=2\pi i\sigma\delta^{\sigma\sigma'}\delta(x-x'),\\\left[\phi^\sigma_\rho(x),\partial_{x'}\phi^{\sigma'}_a(x')\right]&=0.\end{split}\end{align} With a particular set of electron density interactions, the Hamiltonian density takes a diagonal form \begin{align}\begin{split}\mathcal{H}_0&=\frac{v}{4\pi}\sum_{\sigma=L,R}\sum_{a=1}^{2\mathcal{N}-1}\left(\partial_x\phi_a^\sigma\right)^2+v_\rho\left[g_\rho\left(\partial_x\varphi\right)^2+\frac{1}{g_\rho}\left(\partial_x\theta\right)^2\right]\end{split}\label{freeH0}\end{align} where $(\varphi,\theta)=(\varphi_0,\theta_\rho)$ for the umklapp case or $(\varphi_\rho,\theta_0)$ for the superconducting case.

The umklapp scattering \eqref{umpklappSC} requires translation symmetry breaking and can arise in a half-filled lattice when the Fermi momentum is commensurate, $4k_f=2\pi/l$, where $l$ is a microscopic lattice translation. The process is relevant in the renormalization group~\cite{Giamarchibook,Senechal2004,Tsvelikbook,Fradkinbook} ({\color{blue}\hypertarget{RG}{RG}}) sense when $g_\rho<4$ and leads to a spin liquid. The charge-violating pairing \eqref{umpklappSC} can be induced by proximity with a bulk $s$-wave superconductor and is relevant when $g_\rho>1/4$. The unaffected low-energy theory is generated by $2\mathcal{N}-1$ counter-propagating pairs of Dirac fermions $d_a^\sigma\sim e^{i\phi_a^\sigma}$, which are {\em non-local} and must come in pairs. Their fractional nature is revealed by their charge and spin numbers $Q(e^{ib^{sj}_\sigma\Phi^\sigma_{sj}})=e\sum_{sj\sigma}b^{sj}_\sigma$, $S(e^{ib^{sj}_\sigma\Phi^\sigma_{sj}})=\sum_{sj\sigma}sb^{sj}_\sigma/2$, where the coefficients $b^{sj}_\sigma$ can be any numbers. Contrary to an electron/hole which carries charge $Q=\pm e$ and angular spin $S=\pm1/2$, the Dirac fermion $d_a$ has $Q=0$ and $S=\pm1/2$ ($Q=e$ modulo $2e$ and $S=0,\pm1$) in the spin liquid (resp.~superconductor). In the spin liquid, the momentum of these Dirac fermions becomes 0 or $\pi/a$ when the electron Fermi level is tuned near the band crossing of the two spins so that $k_f=k_{SO}$. In the superconductor, the momentum is zero when spin-orbit coupling is negligible $k_{SO}=0$. The Dirac fermions split into Majorana components $\psi=\psi^\dagger$ by the real and imaginary decomposition \begin{align}d_a^\sigma=\frac{1}{\sqrt{2}}\left(\psi_{2a-1}^\sigma+i\psi_{2a}^\sigma\right)\sim e^{i\phi^\sigma_a}.\end{align} A subset of counter-propagating pairs can turn massive and be removed from low-energy by introducing the fermion backscattering \begin{align}\mathcal{U}_{\mathrm{mass}}&=im\sum_{j=N+1}^{4\mathcal{N}-2}\psi_j^L\psi_j^R,\label{Majoranamass}\end{align} where $\psi^L\psi^R$ are a local integral combination of electrons operators. In energy much smaller than $m$ and $u$ or $\Delta$, the Hamiltonian \eqref{freeH0} is effectively truncated to a theory with $N$ counter-propagating pairs of massless Majorana fermions \begin{align}\mathcal{H}_0=\frac{iv}{2}\sum_{\sigma=+,-}\sum_{j=1}^N\sigma\psi_j^\sigma\partial_x\psi_j^\sigma.\label{freeH}\end{align} We recall $\mathcal{N}$ is the number of spinful non-chiral electron channels in each wire, and $N$ is the number of Majorana fermions that remain massless and unaffected by \eqref{Majoranamass}. The number of Majorana's $N$ can take any positive integer value as long as there are enough number of electron channels $\mathcal{N}>(N+2)/4$.

The low-energy theory \eqref{freeH} has an emergent $SO(N)$ symmetry and is invariant under an orthogonal transformation $\psi_j\to\mathcal{O}_j^i\psi_i$ of the fermion vector. Each chiral sector $\sigma=L,R$ carries a $SO(N)$ \WZW \KM algebra at level 1 generated by the current operators $J^\sigma_{jk}=i\psi_j^\sigma\psi_k^\sigma$, for $1\leq j<k\leq N$ (see appendix~\ref{app:SO(N)algebra} for a review). The conformal primary fields are vector and spinor representations of $SO(N)$.
From \eqref{basistransformationumklapp} or \eqref{basistransformationSC}, the vertex operator representation of the Dirac fermion $d_a^\sigma\sim e^{i\phi^\sigma_a}=e^{ia^{sj}_{\sigma\sigma'}\Phi^{\sigma'}_{sj}}$ in terms of the electronic bosonized variables $\Phi^\sigma_{sj}$ contains the half-integral coefficients $a^{\uparrow1}_L,a^{\downarrow1}_R=\pm1/2$. The electronic origin dictates that the Dirac fermions $d_a^\sigma$ and Majorana fermions $\psi_j^\sigma$ are non-local fractional operators and cannot exist individually. On the other hand, all even products of fermions, such as the $SO(N)_1$ current operators and the fermion backscattering $\psi_j^L\psi_k^R$, are linear combination of integral products of electron operators, and are therefore local. 
A spinor field can be represented by a vertex operator \begin{align}s_{\boldsymbol\varepsilon}^\sigma=e^{i\varepsilon^a\phi^\sigma_a/2},\label{spinorvertex}\end{align} where $\boldsymbol\varepsilon=(\varepsilon^1,\ldots,\varepsilon^{2\mathcal{N}-1})$ has unit entries $\varepsilon^a=\pm1$. The spinor fields separate into even and odd primary sectors $[s^\sigma_\pm]$ distinguished by the parity $\prod_a\varepsilon^a=\pm1$. A spinor pair of the same chirality $s^\sigma_{\boldsymbol\epsilon}s^\sigma_{-\boldsymbol\epsilon'}$ is a local integral combination of electrons when $\prod_a\varepsilon_a\varepsilon'_a=1$, or is equivalent to a non-local odd fermion combination when $\prod_a\varepsilon_a\varepsilon'_a=-1$. Moreover, when projected onto the ground state Hilbert space of the umklapp (or pairing) potential in \eqref{umpklappSC}, the corresponding sine-Gordon variable $\langle\theta_\rho\rangle$ (resp.~$\langle\varphi_\rho\rangle$) admits a ground state expectation value. Consequently, all pair of spinor fields with the same parity but opposite chirality $s^L_{\boldsymbol\epsilon}s^R_{\boldsymbol\epsilon'}$ is effectively local in low-energy because it is equivalent to an integral combination of electrons up to the vertex operator $e^{i\theta_\rho/2}$ ($e^{i\varphi_\rho/2}$), which becomes a $U(1)$ phase when projected onto the ground state. In addition, when projecting to the ground states of \eqref{Majoranamass}, the local spinor pairs $s^\sigma_{\boldsymbol\varepsilon}s^\sigma_{-\boldsymbol\varepsilon'}$ and $s^L_{\boldsymbol\varepsilon}s^R_{-\boldsymbol\varepsilon'}$ operate as pairs of spinor twist fields in the low-energy $SO(N)_1$ sector if the ground state projection is non-vanishing. In general, an operator is effectively integral and local if and only if it is invariant under the internal $\mathbb{Z}_2$ gauge symmetry at {\em any} given wire $y$ \begin{align}\mathbb{Z}_2(y):\quad\psi_{y'j}^\sigma\to(-1)^{\delta_{yy'}}\psi_{y'j}^\sigma,\quad\phi^\sigma_{y'a}\to\phi^\sigma_{y'a}+\sigma\pi\delta_{yy'}.\label{Z2gauge0}\end{align}

The bosonic $SO(N)_1$ wire in \eqref{freeH} will be the building blocks of the 2D coupled-wire models described in the following sections. After an energy gap has been established by inter-wire interactions, the topological phase will be stable against perturbations smaller than the energy gap. In particular, the fine-tuning of the electron density interactions $u^{sjs'j'}_{\sigma\sigma'}$ and the Fermi level that enable the bosonic $SO(N)_1$ theory can be relaxed in the upcoming 2D topological model.

\section{Topological phases in the orthogonal and unitary families}\label{sec:SONSUN1}
We construct exactly-solvable coupled-wire models of topological phases in two spatial dimensions with $SO(N)_1$, $U(1)_l$, and $SU(n)_1$ topological orders. The models are built on a 2D array of boson wires, each being effectively described by a $SO(N)_1$ \WZW \KM \CFT \eqref{freeH} in low-energy and originated from strongly correlated electrons. In a closed torus geometry, a finite excitation energy gap is created by backscattering interactions between neighboring wires. In an open geometry, the 1D boundary hosts chiral gapless degrees of freedom that propagate in a single forward direction and are effectively described by a \CFT corresponding to the topological order~\cite{FrohlichGabbiani90,MooreRead,Wenedgereview,ReadRezayi,Kitaev06}. We show the topological phases in the orthogonal and unitary families are related by gauging~\cite{TeoHughesFradkin15,BarkeshliBondersonChengWang14} (also referred to as orbifolding~\cite{DixonHarveyVafaWitten85I,DixonHarveyVafaWitten85II,Ginsparg88,DijkgraafVerlindeVerlinde88,MooreSeiberg89zoo}) the discrete $\mathbb{Z}_2$ and $\mathbb{Z}_k$ symmetries \begin{align}\begin{split}\frac{SO(2n)_1}{\mathbb{Z}_2}&=SO(n)_1\times SO(n)_1\\\frac{SO(2n)_1}{\mathbb{Z}_k}&=U(1)_l\times SU(n)_1\end{split}\end{align} where $k=n$ ($k=n/2$) and $l=4n$ ($l=n$) for odd (even) $n$.

\subsection{The \texorpdfstring{$SO(N)_1$}{SO(N)} family}\label{sec:SON}

The model begin with a 2D array of bosonic $SO(N)_1$ wires. All the wires are parallel to the horizontal $x$ direction and each one is labeled by its vertical $y$ position, for $y=1,2,\ldots,\mathsf{L}$. The array can occupy an open plane with top (bottom) edges at $y=\mathsf{L}$ ($y=1$), or it can wrap around a periodic cylinder geometry when $y\equiv y+\mathsf{L}$. The boson wires can emerge from strongly interacting 1D electrons in a spin liquid or a superconducting setting presented in the previous section. In both cases, all fermionic local excitations are gapped and the remaining gapless boson modes are represented by even combinations of non-local emergent Majorana fermions $\psi_{yj}^\sigma(x)=\psi_j^\sigma(x,y)$, where $j=1,\ldots,N$ and $\sigma=L,R=+,-$ labels the left and right propagating directions. Unlike BdG fermions which are integral combinations of electrons and holes in a conventional BCS superconductor, these Majorana fermions are fractional and cannot appear individually. Only pairs of them on the {\em same} wire are combinations of integral products of electrons and holes. This includes the current generators $J^\sigma_{y,jk}=i\psi^\sigma_{yj}\psi^\sigma_{yk}$ of the $SO(N)_1$ \WZW \KM algebra and the intra-wire backscattering $\psi^L_{yj}\psi^R_{yk}$. The effective low-energy kinetic Hamiltonian density is $\mathsf{L}$ copies of \eqref{freeH} \begin{align}\mathcal{H}_0=\frac{iv}{2}\sum_{y=1}^{\mathsf{L}}\sum_{j=1}^N\sum_{\sigma=+,-}\sigma\psi_{yj}^\sigma\partial_x\psi_{yj}^\sigma.\label{kineticH}\end{align} 

Interwire single-fermion tunnelings $\psi^\sigma_{yj}\psi^{\sigma'}_{y'j'}$, for $y\neq y'$, are forbidden by locality, but two-fermion processes are allowed and can be represented by local integral combinations of electrons. In particular, the interwire Gross-Neveu interaction~\cite{GrossNeveu,ZamolodchikovZamolodchikov78,Witten78,ShankarWitten78} that backscatters the $SO(N)_1$ currents ${\bf J}_y^\sigma=(J_{y,jk}^\sigma)$ \begin{align}\mathcal{U}_{SO(N)_1}&=u\sum_y{\bf J}_{y+1}^L\cdot{\bf J}_y^R\label{SONpotential}\\&=-u\sum_y\sum_{1\leq j<k\leq N}\psi_{y+1,j}^L\psi_{y+1,k}^L\psi_{y,j}^R\psi_{y,k}^R\nonumber\end{align} gaps all bulk degrees of freedom and leads to the $SO(N)_1$ topological phase~\cite{Kitaev06}. The interaction involves interwire backscattering of the $SO(N)_1$ \WZW \KM currents $J_{y,jk}^\sigma=i\psi_{y,j}^\sigma\psi_{y,k}^\sigma$. The model $\mathcal{H}_0+\mathcal{U}_{SO(N)_1}$ admits a mean-field approximation with real order parameters $\hat{\mathcal{O}}_{y+1/2,j}=\int dxi\psi^L_{y+1,j}\psi^R_{y,j}$, whose ground state expectation values $\mathcal{O}_{y+1/2}=\langle\hat{\mathcal{O}}_{y+1/2,j}\rangle$ are non-vanishing and independent from $j$ due to the $SO(N)$ symmetry. The mean-field approximation of the the Gross-Neveu potential is quadratic in $\psi$. It introduces a finite excitation energy gap and turns the Majorana fermions massive.

Bosonizing~\cite{bigyellowbook,Giamarchibook,bosonizationbook,Senechal2004,Fradkinbook} according to \begin{align}d_{y,a}^\sigma=\frac{1}{\sqrt{2}}\left(\psi_{y,a}^\sigma+i\psi_{y,n+a}^\sigma\right)\sim e^{i\phi^\sigma_{y,a}},\label{bosonization}\end{align} the $SO(N)_1$ Gross-Neveu interaction is equivalent to the sine-Gordon potential \begin{align}\mathcal{U}_{SO(2n)_1}&=-\tilde{u}\sum_y\sum_{a=1}^n\partial_x\phi^L_{y+1,a}\partial_x\phi^R_{y,a}\nonumber\\&-\Delta\sum_y\sum_{1\leq a<b\leq n}\sum_{\epsilon=\pm}\cos\left(\Theta_{y+1/2}^a+\epsilon\Theta_{y+1/2}^b\right),\nonumber\\\mathcal{U}_{SO(2n+1)_1}&=\mathcal{U}_{SO(2n)_1}\label{SONsineGordon}\\&-\tilde\Delta\sum_y\left(\sum_{a=1}^n\cos\Theta_{y+1/2}^a\right)i\psi^L_{y+1,2n+1}\psi^R_{y,2n+1},\nonumber\end{align} where the angle variables $\Theta^a_{y+1/2}=\phi^L_{y+1,a}-\phi^R_{y,a}$ are non-chiral combinations of the bosonized variables. The potentials $\Delta$ and $\tilde\Delta$ are relevant in the \RG sense under repulsive density interactions when $\tilde{u}>0$. They introduces a finite energy gap above the ground state, where the order parameter quantum fluctuate about the expectation values \begin{align}\left\langle\Theta^a_{y+1/2}(x)\right\rangle&=m^a_{y+1/2}\pi,\label{GEVSON}\\\mathrm{sgn}\mathcal{O}_{y+1/2}&=\mathrm{sgn}\left\langle\int dxi\psi^L_{y+1,j}\psi^R_{y,j}\right\rangle=(-1)^{m^a_{y+1/2}},\nonumber\end{align} where ${\bf m}_{y+1/2}=(m^1_{y+1/2},\ldots,m^n_{y+1/2})$ either have all even or all odd integral entries. 

In a periodic geometry where $y\equiv y+\mathsf{L}$, all excitations are separated away from the ground state by a finite energy gap. In an open geometry where the system terminates at wire $y=1$ and $\mathsf{L}$, the Gross-Neveu interaction \eqref{SONpotential} leaves behind the gapless left (right) moving $SO(N)_1$ \WZW \CFT on the boundary at $y=1$ ($y=\mathsf{L}$), where the emergent Majorana fermions $\psi^L_{y=1,j}$ and $\psi^R_{y=\mathsf{L},j}$ remain massless. The primary field excitations on the boundaries and bulk anyon quasiparticle excitations are correlated due to the bulk topology and boson locality. They are created and non-locally connected by Wilson string operators, which are integral combination of electrons. The edge primary fields and bulk anyons are classified by the super-selection sectors $X=1,\psi,s_+,s_-$ when $N=2n$ is even, or $X=1,\psi,\sigma$ when $N=2n+1$ is odd. Fields and anyons in the same class differ from each other by local electronic combinations. The vacuum sector 1 contains all integral fields. $\psi$ contains all odd combinations of the non-local Majorana fermions. $s_\pm$ and $\sigma$ are non-local spinor fields with spin (or conformal scaling dimension) $h=N/16$. A full braid between a spinor and a fermion associates a $\pi$ monodromy phase. Therefore a spinor field carries a $\mathbb{Z}_2$ flux component with respect to the internal fermion parity symmetry \eqref{Z2gauge0}. The $SO(2n)_1$ topological order is Abelian and the anyons obey the single-channel fusion rules \begin{align}&\psi\times\psi=1,\quad\psi\times s_\pm=s_\mp,\nonumber\\&s_\pm\times s_\pm=\left\{\begin{array}{*{20}l}1,&\mbox{for even $n$}\\\psi,&\mbox{for odd $n$}\end{array}\right..\label{SO2nfusion}\end{align} The $SO(2n+1)_1$ topological order is non-Abelian. The Ising twist field $\sigma$ obeys the fusion rules \begin{align}\psi\times\sigma=\sigma,\quad\sigma\times\sigma=1+\psi.\label{SO2n+1fusion}\end{align} Since all excitations are created by strings of electron operators, the edge primary fields and bulk anyons must have balancing topological charges $X_{\mathrm{top-edge}}\times X_{\mathrm{bottom-edge}}\times X_{\mathrm{bulk}}=1$.

The $SO(N)_1$ topological orders are sixteen-fold periodic~\cite{Kitaev06}. The anyon fusion and braiding structures of $SO(N)_1$ and $SO(N+16)_1$ are identical. Therefore, as \hyperlink{UMTC}{UMTC}s, $SO(N)_1=SO(N+16)_1$. As \hyperlink{CFT}{CFT}s, their chiral charge charges differ by 8. This can be compensated by subtracting from $SO(N+16)_1$ the $E_8$ \WZW \CFT at level 1.~\cite{SahooZhangTeo15} The subtraction does not alter the topological order because $(E_8)_1$ does not support fractional excitations.

\subsection{The \texorpdfstring{$U(1)_l$}{U(1)} and \texorpdfstring{$SU(n)_1$}{SU(n)} family}\label{sec:unitaryfamily}
The coupled-wire models in the unitary family $U(1)_l$ and $SU(n)_1$ are summarized in figure~\ref{fig:CWMschematic}(a). The level of $U(1)$ is $l=4n$ when $n$ is odd or $l=n$ when $n$ is even. It corresponds to the compactification radius $R=\sqrt{l}/2$ of the free boson on a circle. The models are constructed based on the conformal embedding $U(1)_l\times SU(n)_1\subseteq SO(2n)_1$. Among the $SO(2n)$ rotations of the Majorana fermions $\psi_j$, unitary rotations $d_a\to U_a^bd_b$ of the Dirac fermions $d_a=(\psi_{a}+i\psi_{n+a})/\sqrt{2}\sim e^{i\phi_a}$ form a unitary subgroup $U(n)$ that commutes with the artificial $U(1)$ symmetry $\phi_a\to\phi_a+\vartheta$. The corresponding $U(1)$ invariant \WZW \KM subalgebra is generated by the current operators ${\bf J}_{U(1)_l\times SU(n)_1}$ consisting of the Cartan-Weyl generators $\partial_x\phi_a\sim d_a^\dagger d_a$ and roots $e^{i\left(\phi_a-\phi_b\right)}\sim d_b^\dagger d_a$ for $a\neq b$. The subalgebra decomposes into a diagonal $U(1)_l$ sector generated by $\partial_x\phi_\perp$, where \begin{align}\phi_\perp=\phi_1+\ldots+\phi_n,\label{phiperp}\end{align} and an off-diagonal $SU(n)_1$ sector. The Cartan-Weyl generators of $SU(n)_1$ can be chosen to be $\partial_x\tilde\phi_p$, for $p=1,\ldots,n-1$, using an orthogonal transformation \begin{align}\tilde\phi_p=\frac{1}{\sqrt{p(p+1)}}\left(-p\phi_{p+1}+\sum_{a=1}^p\phi_a\right),\end{align} all of which being orthogonal to \eqref{phiperp}. Together with the roots $e^{i\left(\phi_a-\phi_b\right)}$, they generate the $SU(n)$ \WZW \KM algebra at level 1. 

The kinetic Hamiltonian density \eqref{kineticH} of the array of bosonic $SO(2n)_1$ wires can be re-expressed in the bosonized form \begin{align}\mathcal{H}_0&=\frac{v}{4\pi}\sum_{y=1}^{\mathsf{L}}\sum_{\sigma=L,R}\sum_{a=1}^n\left(\partial_x\phi_{ya}^\sigma\right)^2=\mathcal{H}_0^{U(1)}+\mathcal{H}_0^{SU(n)},\nonumber\\\mathcal{H}_0^{U(1)}&=\frac{v/n}{4\pi}\sum_{y=1}^{\mathsf{L}}\sum_{\sigma=L,R}\left(\partial_x\phi_{y\perp}^\sigma\right)^2,\label{kineticH2}\\\mathcal{H}_0^{SU(n)}&=\frac{v}{4\pi}\sum_{y=1}^{\mathsf{L}}\sum_{\sigma=L,R}\sum_{p=1}^{n-1}\left(\partial_x\tilde\phi_{yp}^\sigma\right)^2.\nonumber\end{align} We consider backscattering interactions that are composed of inter-wire or intra-wire sine-Gordon potentials. The interactions among the $U(1)_l$ sectors are either the intra-wire or inter-wire backscattering \begin{align}\mathcal{U}^y_{U(1)_l}&=-u_\perp^{\mathrm{intra}}\partial_x\phi_{y\perp}^L\partial_x\phi_{y\perp}^R-\Delta_\perp^{\mathrm{intra}}\cos\Theta_y^\perp,\label{U1sineGordon}\\\mathcal{U}^{y+1/2}_{U(1)_l}&=-u_\perp^{\mathrm{inter}}\partial_x\phi_{y+1,\perp}^L\partial_x\phi_{y\perp}^R-\Delta_\perp^{\mathrm{inter}}\cos\Theta_{y+1/2}^\perp,\nonumber\end{align} where $\Theta_y^\perp=\phi_{y\perp}^L-\phi_{y\perp}^R$ and \begin{align}\Theta_{y+1/2}^\perp=q\left(\phi_{y+1,\perp}^L-\phi_{y,\perp}^R\right),\quad q=\left\{\begin{array}{*{20}l}1,&\mbox{for even $n$}\\2,&\mbox{for odd $n$}\end{array}\right..\end{align} The extra factor of 2 for the odd $n$ case is to ensure an even number of Dirac fermions are backscattered from one wire to the next so that the interaction can be constructed by an integral combination of electrons. No such distinction is necessary for the intra-wire interaction because $\cos\Theta_y^\perp$ is already an even product of Dirac fermions on the {\em same} wire. The interactions within the $SU(n)_1$ sector consist of backscattering processes of the $SU(n)_1$ currents ${\bf J}_{SU(n)_1}$ inside a wire or in-between wires \begin{align}\mathcal{U}_{SU(n)_1}^y&=u^{\mathrm{intra}}{\bf J}_{y,SU(n)_1}^L\cdot{\bf J}_{y,SU(n)_1}^R\nonumber\\&=-\tilde{u}^{\mathrm{intra}}\sum_{p=1}^{n-1}\partial_x\tilde\phi_{yp}^L\partial_x\tilde\phi_{yp}^R\nonumber\\&\;\;\;-\Delta^{\mathrm{intra}}\sum_{1\leq a<b\leq n}\cos\left(\Theta_y^a-\Theta_y^b\right),\label{SUnsineGordon}\\\mathcal{U}_{SU(n)_1}^{y+1/2}&=u^{\mathrm{inter}}{\bf J}_{y+1,SU(n)_1}^L\cdot{\bf J}_{y,SU(n)_1}^R\nonumber\\&=-\tilde{u}^{\mathrm{inter}}\sum_{p=1}^{n-1}\partial_x\tilde\phi_{y+1,p}^L\partial_x\tilde\phi_{yp}^R\nonumber\\&\;\;\;-\Delta^{\mathrm{inter}}\sum_{1\leq a<b\leq n}\cos\left(\Theta_{y+1/2}^a-\Theta_{y+1/2}^b\right),\nonumber\end{align} where $\Theta_y^a=\phi_{ya}^L-\phi_{ya}^R$ and $\Theta_{y+1/2}^a=\phi_{y+1,a}^L-\phi_{y,a}^R$.

The $U(1)_l$ topological phase is constructed by combining inter-wire interactions \eqref{U1sineGordon} in the $U(1)_l$ sector and intra-wire interactions \eqref{SUnsineGordon} in the $SU(n)_1$ sector \begin{align}\mathcal{H}_{U(1)_l}=\mathcal{H}_0+\sum_{y=1}^{\mathsf{L}}\left(\mathcal{U}^{y+1/2}_{U(1)_l}+\mathcal{U}^y_{SU(n)_1}\right).\label{HU1}\end{align} This gaps all excitations in the bulk. For $\Delta^{\mathrm{inter}}_\perp,\Delta^{\mathrm{intra}}>0$, the sine-Gordon potentials pin the finite ground state expectation values of the angle variables $\left\langle\Theta^\perp_{y+1/2}\right\rangle$ and $\left\langle\Theta^a_y-\Theta^b_y\right\rangle$ to integer multiples of $2\pi$. The $SU(n)_1$ potential $\Delta^{\mathrm{intra}}$ is relevant in the \RG sense when the density interaction is ``repulsive'', i.e.~$\tilde{u}^{\mathrm{intra}}>0$. The $U(1)_l$ sine-Gordon potential $\Delta^{\mathrm{inter}}_\perp$ is relevant when $u^{\mathrm{inter}}_\perp>v(q^4n^2-4)/[2\pi n(q^4n^2+4)]$. In an open geometry, the model \eqref{HU1} leaves behind the gapless chiral $U(1)_l$ \KM \CFT on the edges generated by $\phi^R_{y=\mathsf{L},\perp}$ and $\phi^L_{y=1,\perp}$. 

The $U(1)_l$ topological phase supports $l$ Abelian anyon types $1,[e],\left[e^2\right],\ldots,\left[e^{l-1}\right]$. They carry spins $h_{e^m}=m^2/(2l)$ and obey the fusion rules $e^m\times e^{m'}=e^{m+m'}$ and $\left[e^l\right]=\left[e^0\right]=1$. The anyon $e^m$ corresponds to the primary field \begin{align}e^m=\left\{\begin{array}{*{20}l}e^{im\phi_\perp/n}&\mbox{if $n$ even}\\e^{im\phi_\perp/(2n)}&\mbox{if $n$ odd}\end{array}\right.\label{U1anyons}\end{align} on the edge \hyperlink{CFT}{CFT}. The level $l$ is determined by the smallest non-trivial local boson $e^l$. When $n$ is even, $e^{i\phi_\perp}\sim d_1\ldots d_n$ is an even product of Dirac fermions and hence it is a local integral combination of electrons. This sets $l=n$ by equating $e^l=e^{i\phi_\perp}$. When $n$ is odd, $e^{i\phi_\perp}$ is not local because it is an odd product of Dirac fermions and has $\pi$ monodromy with $e^m$ for odd $m$. Instead, the primitive local boson is $e^l=e^{i2\phi_\perp}\sim(d_1\partial d_1)\ldots(d_n\partial d_n)$. This sets $l=4n$. The $U(1)_l$ topological order can be described by the $2+1$D Chern-Simons field theory~\cite{WittenJonespolynomials,FrohlichZee,LopezFradkin91,WenZee92} \begin{align}\mathcal{S}_{\mathrm{CS}}\left[U(1)_l\right]=\frac{l}{4\pi}\int_{2+1}a\wedge da.\end{align}

On the other hand, the $SU(n)_1$ topological phase -- which is the ``particle-hole'' conjugate of the $U(1)_l$ -- is constructed by combining inter-wire interactions in the $SU(n)_1$ sector and intra-wire interactions in the $U(1)_l$ sector \begin{align}\mathcal{H}_{SU(n)_1}=\mathcal{H}_0+\sum_{y=1}^{\mathsf{L}}\left(\mathcal{U}^y_{U(1)_l}+\mathcal{U}^{y+1/2}_{SU(n)_1}\right).\label{HSUn}\end{align} Similar to the $U(1)_l$ case, it creates a finite excitation energy gap in the bulk, but leaves behind the $SU(n)_1$ \WZW \CFT on the edges. The $SU(n)_1$ topological phase supports $n$ Abelian anyon types $1,[\Psi],\left[\Psi^2\right],\ldots,\left[\Psi^{n-1}\right]$. They carry spins $h_{\Psi^m}=m(n-m)/(2n)$ and obey the fusion rules $[\Psi^m]\times[\Psi^{m'}]=\left[\Psi^{m+m'}\right]$ and $\left[\Psi^n\right]=\left[\Psi^0\right]=1$. $\left[\Psi^m\right]$ corresponds not to a single field, but a super-selection sector of fields that rotate irreducibly under the $SU(n)_1$ \WZW \KM algebra. On the edge $SU(n)_1$ \hyperlink{CFT}{CFT}, $\left[\Psi^m\right]$ is spanned by $C^n_m=n!/[m!(n-m)!]$ primary fields \begin{align}\left[\Psi^m\right]=\mathrm{span}\left\{e^{i\left(\phi_{a_1}+\ldots+\phi_{a_m}-m\phi_\perp/n\right)}\right\}_{1\leq a_1<\ldots<a_m\leq n}.\label{SUnanyons}\end{align} The $SU(n)_1$ topological order can be described by the $2+1$D Chern-Simons field theory \begin{align}\mathcal{S}_{\mathrm{CS}}\left[SU(n)_1\right]=\frac{K_{IJ}}{4\pi}\int_{2+1}a^I\wedge da^J\end{align} with $n-1$ components $a^1,\ldots,a^{n-1}$, where $K_{IJ}=2\delta_{IJ}-\delta_{I,J+1}-\delta_{I,J-1}$ is the Cartan matrix of $SU(n)$.

The trivial topological phase with a finite excitation energy gap can be constructed using solely intra-wire backscattering interactions \begin{align}\mathcal{H}_{\mathrm{trivial}}=\mathcal{H}_0+\sum_{y=1}^{\mathsf{L}}\left(\mathcal{U}^y_{U(1)_l}+\mathcal{U}^y_{SU(n)_1}\right).\label{trivialH1}\end{align} The model is a stack of decoupled 1D gapped boson wires. The edge of an open system does not carry gapless modes, and the ground state of any system, regardless of being open or closed, is non-degenerate. On the contrary, a non-trivial topological phase can be constructed using solely inter-wire backscattering interactions \begin{align}\mathcal{H}_{SO(2n)_1}=\mathcal{H}_0+\sum_{y=1}^{\mathsf{L}}\left(\mathcal{U}^{y+1/2}_{U(1)_l}+\mathcal{U}^{y+1/2}_{SU(n)_1}\right).\label{HU1SUncond}\end{align} It is worth observing that the topological order is {\em not} $U(1)_l\times SU(n)_1$, but instead $SO(2n)_1$, which was presented in the previous subsection. This is because the bosonic anyon pair \begin{align}z=\left\{\begin{array}{*{20}l}\left[e^2\right]\times\left[\Psi^2\right]&\mbox{if $n$ even}\\\left[e^4\right]\times\left[\Psi^2\right]&\mbox{if $n$ odd}\end{array}\right.\label{Zkcharges0}\end{align} along with its higher powers $z^p$ are even products of Dirac fermions ($z\sim d_ad_b$) and are therefore actually integral combinations of local electrons. They should belong in the vacuum sector, and are {\em condensed} in the context of anyon condensation~\cite{BaisSlingerlandCondensation,Kong14,NeupertHeKeyserlingkSierraBernevig16,Burnell18}. The topological phase only carries 4 distinct anyon types -- the trivial class 1 of local fields, the fermion $\psi=[e]\times[\Psi]$ if $n$ even or $\psi=[e^2]\times[\Psi]$ if $n$ odd, the even spinor $s_+=[e^{n/2}]$ if $n$ even or $s_+=[e^n]$ if $n$ odd, and the odd spinor $s_-=s_+\times\psi$ -- matching the $SO(2n)_1$ topological order.

The topological phase with $U(1)_l\times SU(n)_1$ order can be constructed by stacking the $U(1)_l$ and $SU(n)_1$ models \begin{align}\mathcal{H}_{U(1)_l\times SU(n)_1}=\mathcal{H}_{U(1)_l}\oplus\mathcal{H}_{SU(n)_1}\label{HU1xSUn}\end{align} from \eqref{HU1} and \eqref{HSUn}. The two Hamiltonian components act independently on decoupled Hilbert spaces. Here, the boson wires that constitute the $U(1)_l$ layer are completely distinct from those that constitute the $SU(n)_1$ layer. Consequently, the anyon pair $z$ in \eqref{Zkcharges0} is {\em not} a local integral combination of electrons in this construction. The $U(1)_l\times SU(n)_1$ topological phase is a promotion from $SO(2n)_1$ by gauging a $\mathbb{Z}_k$ symmetry as explained in the subsection below.

\subsection{Gauging the \texorpdfstring{$\mathbb{Z}_k$}{Zk} or \texorpdfstring{$\mathbb{Z}_2$}{Z2} symmetries in \texorpdfstring{$SO(2n)_1$}{SO(2n)}}\label{sec:gaugingZkZ2}
The $SO(2n)_1$ topological phase admits global $\mathbb{Z}_k$ and $\mathbb{Z}_2$ symmetries, where $k=n$ when $n$ is odd or $k=n/2$ when $n$ is even. They originate from the rotation and conjugation of the Dirac fermions $d_{ya}^\sigma=(\psi_{y,a}^\sigma+i\psi_{y,n+a}^\sigma)/\sqrt{2}\sim e^{i\phi^\sigma_{ya}}$ \begin{align}\begin{split}&\mathbb{Z}_k:\quad d^\sigma_{ya}\to e^{2\pi i/k}d^\sigma_{ya},\quad\phi^\sigma_{ya}\to\phi^\sigma_{ya}+2\pi/k\\&\mathbb{Z}_2:\quad d^\sigma_{ya}\to {d^\sigma_{ya}}^\dagger,\quad\phi^\sigma_{ya}\to-\phi^\sigma_{ya}\end{split}\label{Dksymm}\end{align} where the symmetries apply to all species $a=1,\ldots,n$, propagation directions $\sigma=L,R$ on all wires $y=1,\ldots,\mathsf{L}$. Together, they form the dihedral group $D_k=\mathbb{Z}_2\ltimes\mathbb{Z}_k$. The symmetry group is non-Abelian since the $\mathbb{Z}_k$ and $\mathbb{Z}_2$ actions do not mutually commute. Unlike the internal $\mathbb{Z}_2$ symmetry in \eqref{Z2gauge0}, which is a local gauge symmetry of $SO(2n)_1$ dictated by locality, the $\mathbb{Z}_k$ and $\mathbb{Z}_2$ symmetries in \eqref{Dksymm} are global symmetries of the $SO(2n)_1$ Hamiltonian $\mathcal{H}_0+\mathcal{U}_{SO(2n)_1}$ in \eqref{kineticH} and \eqref{SONsineGordon}. While all local operators are necessarily invariant under the internal $\mathbb{Z}_2$ symmetry \eqref{Z2gauge0}, they may change under the global symmetries. For example, the fermion pair $\psi^\sigma_{yj}\psi^{\sigma'}_{yj'}$ on a given wire is an integral combination of electrons and is invariant under \eqref{Z2gauge0}, but in general is not fixed by the global $\mathbb{Z}_k$ or $\mathbb{Z}_2$ symmetries. In the following, we discuss topological phases that promote the $\mathbb{Z}_k$ or $\mathbb{Z}_2$ symmetries into local gauge symmetries.

\subsubsection{The \texorpdfstring{$SO(2n)_1/\mathbb{Z}_2$}{SO(2n)/Z2} twist liquid}\label{sec:SO2nZ2TL}
The $SO(2n)_1$ \WZW \KM algebra is generated by current operators $J_{jk}=i\psi_j\psi_k$, for $1\leq j<k\leq 2n$. The $\mathbb{Z}_2$ symmetry from \eqref{Dksymm} that sends $\psi_a\to\psi_a$ and $\psi_{n+a}\to-\psi_{n+a}$, for $a=1,\ldots,n$, fixes the subalgebra $SO(n)_1^A\times SO(n)_1^B$, where the first and second components are generated by the $\mathbb{Z}_2$ invariant current operators $J_{ab}^A=i\psi_a\psi_b$ and $J_{ab}^B=i\psi_{n+a}\psi_{n+b}$, for $1\leq a<b\leq n$. The corresponding $2+1$D topological phase can be constructed using the coupled-wire model consisting of two identical but decoupled copies of $SO(n)_1$ \begin{align}\mathcal{H}_{SO(n)_1^A\times SO(n)_1^B}=\mathcal{H}_{SO(n)_1^A}\oplus\mathcal{H}_{SO(n)_1^B}\end{align} where the Hamiltonian of each copy was presented in section~\ref{sec:SON}.

The $\mathbb{Z}_2$ symmetry fixes all fermions in the $A$ layer but flips the sign of all fermions in the $B$ layer. Restricting to the $SO(n)_1^B$, the symmetry is identical to \eqref{Z2gauge0}, which is a local gauge symmetry. Overall, the $SO(n)_1^A\times SO(n)_1^B$ double has two sets of local $\mathbb{Z}_2$ gauge symmetries at any given wire $y$ \begin{align}\begin{split}\mathbb{Z}_2^A(y):\quad\psi_{y'a}^\sigma\to(-1)^{\delta_{yy'}}\psi_{y'a}^\sigma,\quad\psi_{y',n+a}^\sigma\to\psi_{y',n+a}^\sigma\\\mathbb{Z}_2^B(y):\quad\psi_{y'a}^\sigma\to\psi_{y'a}^\sigma,\quad\psi_{y',n+a}^\sigma\to(-1)^{\delta_{yy'}}\psi_{y',n+a}^\sigma\end{split}\label{Z2gauge}\end{align} for $\sigma=L,R$. An operator is a local integral combination of electrons (up to intra-wire ground state expectation values) if and only if it is symmetric under the local $\mathbb{Z}_2^A\times\mathbb{Z}_2^B$ symmetry.

The $SO(n)_1^A\times SO(n)_1^B$ topological phase is the $\mathbb{Z}_2$ twist liquid of $SO(2n)_1$, where the symmetry is gauged. On the $1+1$D edge, the $SO(n)_1^A\times SO(n)_1^B$ \WZW edge \CFT is the $\mathbb{Z}_2$ orbifold of $SO(2n)_1$. 
\begin{equation}\begin{tikzcd}[column sep=large]SO(2n)_1\arrow[r,rightharpoonup,yshift=0.5ex,"\mathrm{gauging}"]&SO(2n)_1/\mathbb{Z}_2=SO(n)_1^A\times SO(n)_1^B\arrow[l,rightharpoonup,yshift=-0.5ex,"\mathrm{condensation}"]\end{tikzcd}\label{Z2TL}\end{equation}
The bosonic $\mathbb{Z}_2$ charge is the fermion pair $\zeta=\psi^A\psi^B$, which consists of non-local products of fermions $\psi_a\psi_{n+b}$ from both layers, and is odd under the $\mathbb{Z}_2$ symmetry. The spinor fields -- $s^{A/B}_\pm$ for $n$ even and $\sigma^{A/B}$ for $n$ odd -- from either the $A$ or the $B$ layer exhibit a $\pi$ monodromy with $\zeta$ and therefore carry a $\mathbb{Z}_2$ flux component. During anyon condensation, the $\mathbb{Z}_2$ charge $\zeta$ belongs to the vacuum sector as it is part of the local currents in $SO(2n)_1$. The fermions in the two layers are identified $\psi^A\equiv\psi^B$. The spinor fields $s^{A/B}_\pm$ or $\sigma^{A/B}$ in individual layers are confined. The spinor field pairs become the spinor fields $s_\pm$ in $SO(2n)_1$. When $n$ is even, $s_+=s^A_\pm s^B_\pm$ and $s_-=s^A_\pm s^B_\mp$. When $n$ is odd, $\sigma^A\sigma^B=s_++s_-$ decomposes. 

When $n$ is even, the global $\mathbb{Z}_2$ symmetry \eqref{Dksymm} on $SO(2n)_1$ is an inner automorphism and do not alter the anyon types. When $n\equiv0$ modulo 8, the twist liquid $SO(2n)_1/\mathbb{Z}_2=[SO(n)_1]^2$, as an unitary modular tensor category (\hyperlink{UMTC}{UMTC}), is equivalent to the product $SO(2n)_1\times D^{[0]}(\mathbb{Z}_2)$, where $D^{[0]}(\mathbb{Z}_2)$ is the $\mathbb{Z}_2$ discrete gauge theory~\cite{BaisDrielPropitius92,Propitius-1995,PropitiusBais96,Preskilllecturenotes} in its deconfined phase~\cite{JalabertSachdev91,SenthilMatthew00,MoessnerSondhiFradkin01,ArdonneFendleyFradkin04} or equivalently the Kitaev toric code~\cite{Kitaev97}. When $n\equiv4$ modulo 8, the twist liquid \eqref{Z2TL} is equivalent to $SO(2n)_1\times D^{[1]}(\mathbb{Z}_2)$, where $D^{[1]}(\mathbb{Z}_2)$ is the double semion theory~\cite{LevinWen05}. $D^{[0]}(\mathbb{Z}_2)$ and $D^{[1]}(\mathbb{Z}_2)$ can be described by a 2-component Chern-Simons field theory $\mathcal{S}=\int_{2+1}K_{IJ}a^I\wedge da^J/(4\pi)$, where $K=2\sigma_x$ ($2\sigma_z$) for the toric code (double semion) and $\sigma_{x,z}$ are $2\times2$ Pauli matrices. They both carry 4 Abelian anyon types $1,\zeta,\mu,\zeta\times\mu$, each being self-conjugate, and the $\mathbb{Z}_2$ flux $\mu$ -- can be identified as one of $s^{A/B}_\pm$ -- is bosonic with spin 0 (semionic with spin $1/4$) for the toric code (double semion). The distinction between the two $\mathbb{Z}_2$ gauge theories $D^{[v]}(\mathbb{Z}_2)$, for $v=0,1$, stems from the $F$-symbol~\cite{LevinWen05,Kitaev06} of the $\mathbb{Z}_2$ flux that governs the fusion associativity $\mu\times(\mu\times\mu)=(\mu\times\mu)\times\mu$. $F^{\mu\mu\mu}_\mu=(-1)^v$ takes opposite signs for the toric code and double semion theories. The two gauge inequivalent classes of $F$-symbols represent distinct Dijkgraaf-Witten invariants $v=0,1$ modulo 2 in the group cohomology~\cite{Cohomologybook} $H^3(\mathbb{Z}_2,U(1))=\mathbb{Z}_2$. The details of Dijkgraaf-Witten deformation~\cite{DijkgraafWitten90,DijkgraafPasquierRoche91,AltschulerCoste92,BaisvanDrielPropitius93,Propitius-1995} of discrete gauge theories can be found in appendix~\ref{app:DGT}.

When $n\equiv2$ modulo 4, the twist liquid $SO(2n)_1/\mathbb{Z}_2=SO(n)_1^A\times SO(n)_1^B$ cannot be factorized by a $\mathbb{Z}_2$ gauge theory. This is because the $\mathbb{Z}_2$ fluxes are no longer self-conjugate, but follow the fusion rule $s^{A/B}_\pm\times s^{A/B}_\pm=\psi^{A/B}$ instead. The $G=\mathbb{Z}_2$ symmetry group is extended by the anyon fusion group $\mathcal{A}=\{1,\psi,s_+,s_-\}=\mathbb{Z}_2\times\mathbb{Z}_2$ of $SO(2n)_1$ into a quantum symmetry group~\cite{Wenspinliquid02,EtingofNikshychOstrik10,TeoHughesFradkin15,BarkeshliBondersonChengWang14} $\widehat{G}$, which is analogous to a space-group consisting of ``rotations'' $G$ and ``translations'' $\mathcal{A}$. The central extension, captured by the exact sequence \begin{align}1\to\mathcal{A}\hookrightarrow\widehat{G}\to G\to1,\label{groupextension}\end{align} is classified by the group cohomology~\cite{Cohomologybook} $H^2(G,\mathcal{A})$ (see appendix~\ref{app:DGT}), which in our current situation is $H^2(\mathbb{Z}_2,\mathcal{A})=\mathcal{A}$. The trivial cohomology class corresponds to a {\em symmorphic} quantum symmetry group, which is a direct product $\widehat{G}=G\times\mathcal{A}$, and applies in the previous cases when $n\equiv0$ modulo 4. Here, when $n\equiv2$ modulo 4, the cohomology class is non-trivial and corresponds to the {\em non-symmorphic} quantum symmetry group $\widehat{\mathbb{Z}_2}=\mathbb{Z}_4\times\mathbb{Z}_2$. This is because the $\mathbb{Z}_2$ fluxes are of order 4, $(s^{A/B}_\pm)^4=1$. The $G=\mathbb{Z}_2$ ``rotation'' becomes a twofold ``screw rotation'' and squares to the $SO(2n)_1$ fermion $s^2_\pm=\psi$, which can be viewed as a ``translation'' in the fusion group $\mathcal{A}$.

As an \hyperlink{UMTC}{UMTC}, the $SO(n)_1$ topological order is equivalent to the coset $SO(2n)_1/SO(n)_1=SO(2n)_1\boxtimes\overline{SO(n)_1}$, where the denominator group $SO(n)_1$ sits inside the numerator group $SO(2n)_1$ under the matrix embedding $O\to O\oplus\mathbb{I}_n$, $\overline{SO(n)_1}$ is the time-reversal conjugate of $SO(n)_1$, and the fermion pair $\psi\bar\psi$ is condensed in the relative tensor product $\boxtimes$. Hence, the twist liquid $SO(2n)_1/\mathbb{Z}_2=[SO(n)_1]^2$ is equivalent to \begin{align}SO(2n)_1/\mathbb{Z}_2=SO(2n)_1\boxtimes\overline{SO(n)_1}\times SO(n)_1.\label{Z2quotient}\end{align} When $n\equiv2$ modulo 4, the non-chiral product $\overline{SO(n)_1}\times SO(n)_1$ is equivalent, as an \hyperlink{UMTC}{UMTC}, to the discrete gauge theory $D^{[2]}(\mathbb{Z}_4)$ deformed by the Dijkgraaf-Witten invariant $[2]$ in the cohomology group $H^3(\mathbb{Z}_4,U(1))=\mathbb{Z}_4$ (see appendix~\ref{app:DGT}). $D^{[2]}(\mathbb{Z}_4)$ can be described by a 2-component Chern-Simons field theory $\mathcal{S}=\int_{2+1}K_{IJ}a^I\wedge da^J/(4\pi)$ with the $K$-matrix $K=4\sigma_z$. This identifies the twist liquid $SO(2n)_1/\mathbb{Z}_2$ with the relative tensor product $SO(2n)_1\boxtimes D^{[2]}(\mathbb{Z}_4)$, for $n\equiv2$ modulo 4. Here, the fermion $\psi$ in $SO(2n)_1$ is pair condensed with $\bar\psi=\bar{s}^2_\pm$ (the square of the $\mathbb{Z}_4$ flux) in $D^{[2]}(\mathbb{Z}_4)$ in the product.

Overall, we summarize the $\mathbb{Z}_2$ orbifold phases $SO(2n)_1/\mathbb{Z}_2$ for a general even integer $n$ by the \UMTC equivalence \begin{align}&\frac{SO(2n)_1}{\mathbb{Z}_2}=SO(n)_1\times SO(n)_1\label{Z2orbifoldseven}\\&=\left\{\begin{array}{*{20}l}SO(2n)_1\times D^{[0]}(\mathbb{Z}_2),&\mbox{for $n\equiv0$ mod 8}\\SO(2n)_1\times D^{[1]}(\mathbb{Z}_2),&\mbox{for $n\equiv4$ mod 8}\\SO(2n)_1\boxtimes D^{[2]}(\mathbb{Z}_4),&\mbox{for $n\equiv2$ mod 4}\end{array}\right..\nonumber\end{align} 

When $n$ is odd, the global $\mathbb{Z}_2$ symmetry \eqref{Dksymm} on $SO(2n)_1$ is an outer automorphism. It corresponds to the mirror symmetry of the Dynkin diagram of the simply-laced $D_n=SO(2n)$ Lie algebras~\cite{bigyellowbook}. It exchanges the even and odd spinor fields, $s_+\leftrightarrow s_-$, in $SO(2n)_1$. Consequently, the $\mathbb{Z}_2$ fluxes $\sigma^{A/B}$ in the twist liquid $SO(2n)_1/\mathbb{Z}_2=SO(n)_1^A\times SO(n)_1^B$ are non-Abelian and obey the Ising fusion rule $\sigma\times\sigma=1+\psi$. When applying \eqref{Z2quotient} for odd $n$, the non-chiral product $\mathcal{Z}_n(\mathbb{Z}_2)=\overline{SO(n)_1}\times SO(n)_1$ is equivalent as \UMTC to \begin{align}\begin{split}&Z(\mathrm{Ising})=\mathrm{Ising}\times\overline{\mathrm{Ising}},\quad\mbox{for $n\equiv\pm1$ modulo 8}\\&Z(SU(2)_2)=SU(2)_2\times\overline{SU(2)_2},\quad\mbox{for $n\equiv\pm3$ modulo 8}.\end{split}\label{Isings}\end{align} (Here $Z(\mathcal{C})$ stands for the Drinfeld center~\cite{Kasselbook,BakalovKirillovlecturenotes,LevinWen05} of the fusion category $\mathcal{C}$.) The $\mathrm{Ising}$ and $SU(2)_2$ topological orders have identical anyon content $1,\psi,\sigma$ and fusion rules. They differ by the Ising spin statistics $h_\sigma=1/16$ and Frobenius-Schur indicator~\cite{Kitaev06} $\varkappa_\sigma=d_\sigma\left[F^{\sigma\sigma\sigma}_\sigma\right]^1_1=1$ for the former, and $h_\sigma=3/16$ and $\varkappa_\sigma=-1$ for the latter. Both non-chiral products in \eqref{Isings} are $\mathbb{Z}_2$ twist liquids of the Kitaev toric code after gauging an electric-magnetic symmetry~\cite{BarkeshliWen12,TeoHughesFradkin15,ChenRoyTeoRyu17}. They are differentiated from each other by the non-trivial Dijkgraff-Witten invariant in $H^3(\mathbb{Z}_2,U(1))$. The $\mathbb{Z}_2$ orbifold phases $SO(2n)_1/\mathbb{Z}_2$ for odd $n$ can be summarized by the \UMTC equivalence \begin{align}&\frac{SO(2n)_1}{\mathbb{Z}_2}=SO(n)_1\times SO(n)_1=SO(2n)_1\boxtimes\mathcal{Z}_n(\mathbb{Z}_2)\nonumber\\&\mathcal{Z}_n(\mathbb{Z}_2)=\left\{\begin{array}{*{20}l}Z(\mathrm{Ising}),&\mbox{for $n\equiv\pm1$ mod 8}\\Z(SU(2)_2),&\mbox{for $n\equiv\pm3$ mod 8}\end{array}\right.,\label{Z2orbifoldsodd}\end{align} where under the relative tensor product $\boxtimes$, the fermion in $SO(2n)_1$ is pair condensed with the fermion in $\overline{\mathrm{Ising}}$ when $n\equiv1$, in $\overline{SU(2)_2}$ when $n\equiv3$, in $SU(2)_2$ when $n\equiv-3$, or in $\mathrm{Ising}$ when $n\equiv-1$ modulo 8.

\subsubsection{The \texorpdfstring{$SO(2n)_1/\mathbb{Z}_k$}{SO(2n)/Zk} twist liquid}\label{sec:SO2nZkTL}
The (complexified) $SO(2n)_1$ \WZW \KM algebra is spanned by the Cartan-Weyl generators $d_a^\dagger d_a\sim\partial\phi_a$, the positive roots $d_ad_b\sim e^{i(\phi_a+\phi_b)}$ and $d_ad_b^\dagger\sim e^{i(\phi_a-\phi_b)}$, and the negative roots which are the hermitian conjugate of the positive ones, where $1\leq a<b\leq n$. The $\mathbb{Z}_k$ symmetry in \eqref{Dksymm} (or the diagonal $U(1)$ symmetry) fixes $\partial\phi_a$ and $e^{\pm i(\phi_a-\phi_b)}$, which generate the $\mathbb{Z}_k$-invariant $U(1)_l\times SU(n)_1$ subalgebra. The corresponding $2+1$D topological phase can be constructed using the coupled-wire model consisting of the decoupled $U(1)_l$ and $SU(n)_1$ layers (see \eqref{HU1xSUn} in section~\ref{sec:unitaryfamily}).

The $U(1)_l\times SU(n)_1$ topological phase is the $\mathbb{Z}_k$ twist liquid of $SO(2n)_1$, where the symmetry is gauged. On the $1+1$D edge, the $U(1)_l\times SU(n)_1$ \WZW edge \CFT is the $\mathbb{Z}_k$ orbifold of $SO(2n)_1$. 
\begin{equation}\begin{tikzcd}[column sep=large]SO(2n)_1\arrow[r,rightharpoonup,yshift=0.5ex,"\mathrm{gauging}"]&SO(2n)_1/\mathbb{Z}_k=U(1)_l\times SU(n)_1\arrow[l,rightharpoonup,yshift=-0.5ex,"\mathrm{condensation}"]\end{tikzcd}\label{ZkTL}\end{equation} 
The $\mathbb{Z}_k$ symmetry is an inner autormophism of $SO(2n)$ and it does not alter the anyon types of $SO(2n)_1$. After gauging, the $\mathbb{Z}_k$ gauge charge can be identified by the anyon pair \begin{align}z=\left\{\begin{array}{*{20}l}\left[e^2\right]\times\left[\Psi^2\right]&\mbox{if $n$ even}\\\left[e^4\right]\times\left[\Psi^2\right]&\mbox{if $n$ odd}\end{array}\right.,\label{Zkcharge}\end{align} where $e^m$ and $\Psi^m$ are the primary fields of $U(1)_l$ and $SU(n)_1$ defined previously in \eqref{U1anyons} and \eqref{SUnanyons}. Unlike in model \eqref{HU1SUncond} where the $\mathbb{Z}_k$ charge is local, here the $U(1)_l$ and $SU(n)_1$ states are constructed independently and therefore the $\mathbb{Z}_k$ charge $z$ does not belong to the vacuum sector. The order of the gauge group $\mathbb{Z}_k$ is set by the fusion rule $z^k=1$ for $k=n$ if $n$ is odd or $n/2$ when $n$ is even.

When $k$ is odd (i.e.~$n$ is not an integer multiple of 4), the $U(1)_l\times SU(n)_1$ topological order factorizes into $SO(2n)_1\times D^{[0]}(\mathbb{Z}_k)$ as an \hyperlink{UMTC}{UMTC}, where $D^{[0]}(\mathbb{Z}_k)$ is the $\mathbb{Z}_k$ discrete gauge theory~\cite{BaisDrielPropitius92,Propitius-1995,PropitiusBais96,Preskilllecturenotes} in its deconfined phase or equivalently the $\mathbb{Z}_k$ Wen plaquette model~\cite{Wenplaquettemodel}. The $D^{[0]}(\mathbb{Z}_k)$ quantum double can be described by a two-component Chern-Simons field theory $\mathcal{S}=\int_{2+1}K_{IJ}a^I\wedge da^J/(4\pi)$, where $K=k\sigma_x$. Its anyons are dyons $m^az^b$ composed of fluxes $m$ and charges $z$, where $a,b=0,1,\ldots, k-1$. The primitive gauge flux can be identified with the bosonic anyon pair \begin{align}m=\left[e^{lr/k}\right]\times\left[\Psi^{-2r}\right].\end{align} Here, $8r\equiv1$ modulo $k=n$ when $n$ is odd, or $4r\equiv1$ modulo $k=n/2$ when $n$ is $2$ modulo 4, so that the monodromy braiding phase between $z$ and $m$ is $e^{2\pi i/k}$. The $SO(2n)_1$ topological order can be embedded in $U(1)_l\times SU(n)_1$ and decouples from $D^{[0]}(\mathbb{Z}_k)$. When $n$ is odd, $\left[e^{2n}\right]$ and $\left[e^{\pm n}\right]$ -- all of which being local with respect with the $\mathbb{Z}_k$ flux and charge -- take the roles of the fermion and the two spinors of $SO(2n)_1$ respectively. When $n$ is 2 modulo 4, $\left[e^{n/2}\right]$ and $\left[\Psi^{n/2}\right]$ can be treated as the two spinor fields of $SO(2n)_1$, and the fermion is the product $\left[e^{n/2}\right]\times\left[\Psi^{n/2}\right]$.

When $k$ is even (i.e.~$n$ is an integer multiple of 4), the $\mathbb{Z}_k$ gauge part of $SO(2n)_1/\mathbb{Z}_k=U(1)_l\times SU(n)_1$ does not decompose. This is because all anyon $m$ that carries a primitive gauge flux component and exhibits a $e^{2\pi i/k}$ monodromy braiding phase with the primitive gauge charge $z$ defined in \eqref{Zkcharge} must be of order $2k$, i.e.~$m^{2k}=1$ but $m^k\neq1$. For example, the generator $\left[\Psi^{k-1}\right]$ of $SU(n)_1$ carries a unit of gauge flux and its $k^{\mathrm{th}}$ power $\left[\Psi^{k-1}\right]^k\equiv\left[\Psi^k\right]$ does not belong to the vacuum sector but the even spinor sector $s_+$ in $SO(2n)_1$. The quantum symmetry group $\widehat{\mathbb{Z}_k}=\mathbb{Z}_{2k}\times\mathbb{Z}_2$ (see \eqref{groupextension}) that extends $G=\mathbb{Z}_k$ by $\mathcal{A}=\mathbb{Z}_2\times\mathbb{Z}_2$ -- the fusion group of Abelian anyons in $SO(2n)_1$ -- is non-symmorphic and is not the direct product $\mathbb{Z}_k\times\mathcal{A}$. The twist liquid is represented by a non-trivial class in the group cohomology $H^2(\mathbb{Z}_k,\mathcal{A})=\mathcal{A}$ when $k$ is even (see appendix~\ref{app:DGT}).

When $k=n/2\equiv0$ modulo 4, $SO(2n)_1$ is equivalent as an \UMTC to the discrete $\mathbb{Z}_2$ gauge theory $D^{[0]}(\mathbb{Z}_2)$ and carries the same topological order as the Kitaev's toric code~\cite{Kitaev97}. The chiral central charge $c=n$ is a multiple of 8 and can be offset by copies of the topological trivial $E_8$ at level 1.~\cite{Kitaev06,SahooZhangTeo15,LopesQuitoHanTeo19} The even spinor $s_+$ in $SO(2n)_1$ can be identified as the $\mathbb{Z}_2$ gauge flux. Gauging the $\mathbb{Z}_k$ symmetry extends the gauge group to $\mathbb{Z}_{2k}=\mathbb{Z}_n$. The $\mathbb{Z}_k$ orbifold phase $SO(2n)_1/\mathbb{Z}_k=U(1)_n\times SU(n)_1$ is equivalent as an \UMTC to the discrete $\mathbb{Z}_n$ gauge theory $D^{[k]}(\mathbb{Z}_n)$ deformed by the Dijkgraaf-Witten invariant $[k]$ in $H^3(\mathbb{Z}_n,U(1))=\mathbb{Z}_n$ (see appendix~\ref{app:DGT}). To see this, we first notice when $n$ is divisible by 8, $SU(n)_1$ has the same topological order as $\overline{U(1)_n}$ -- the time-reversal conjugate of $U(1)_n$. This is because $\left[\Psi^{k-1}\right]$ shares the same fusion and statistical properties of $\bar{e}$, the primitive anyon in $\overline{U(1)_n}$, and all anyons in $SU(n)_1$ can be generated by powers of $\left[\Psi^{k-1}\right]$ since $n=2k$ and $k-1$ are relatively prime. Consequently, the $SO(2n)_1/\mathbb{Z}_k$ twist liquid has the identical topological order of the non-chiral product $D^{[k]}(\mathbb{Z}_n)=U(1)_n\times\overline{U(1)_n}$, which can be described by a two-component Chern-Simons field theory $\mathcal{S}=\int_{2+1}K_{IJ}a^I\wedge da^J/(4\pi)$, where $K=n\sigma_z$. This $K$-matrix is inequivalent to $n\sigma_x$, which instead corresponds to the un-deformed discrete gauge theory $D^{[0]}(\mathbb{Z}_n)$. 

For $k=n/2\equiv2$ modulo 4, the chiral central charge $c=n$ is equivalent to 4 modulo 8, and therefore the orbifold phase $SO(2n)_1/\mathbb{Z}_n=U(1)_n\times SU(n)_1$ cannot be identified to a non-chiral gauge theory. Instead, it is equivalent as an \UMTC to the relative tensor product $SO(8)_1\boxtimes D^{[k]}(\mathbb{Z}_n)$, where the product $\boxtimes$ involves a fermion pair condensation. To see this, we observe the \UMTC equivalence $SU(n)_1\equiv SO(8)_1\boxtimes\overline{U(1)_n}$, where the vector fermion $\psi$ in $SO(8)_1$ and the self-conjugate fermion $\bar\psi=\bar{e}^{n/2}$ in $\overline{U(1)_n}$ are pair condensed in the product, i.e.~$\psi\bar{e}^{n/2}=1$. The $SU(n)_1$ anyon $\left[\Psi^{k-1}\right]$ can be identified with the product $s_+\bar{e}$, where $s_+$ is the even spinor of $SO(8)_1$. This is because they both carry spin $1/2-1/(2n)$ modulo 1, have order $n$, and generate all anyons that are not confined by the condensate. The non-chiral product $U(1)_n\times\overline{U(1)_n}$ is identical to the deformed discrete gauge theory $D^{[k]}(\mathbb{Z}_n)$ for the same reasons in the $k\equiv0$ mod 4 case presented above. 

Overall, we summarize the $\mathbb{Z}_k$ orbifold phases for general integer $k$ by the \UMTC equivalence \begin{align}&\frac{SO(2n)_1}{\mathbb{Z}_k}=U(1)_l\times SU(n)_1\label{Zkorbifolds}\\&=\left\{\begin{array}{*{20}l}SO(2n)_1\times D^{[0]}(\mathbb{Z}_k),&\mbox{for odd $k$}\\D^{[k]}(\mathbb{Z}_{2k}),&\mbox{for $k\equiv0$ mod 4}\\SO(8)_1\boxtimes D^{[k]}(\mathbb{Z}_{2k}),&\mbox{for $k\equiv2$ mod 4}\end{array}\right.\nonumber\end{align} where $k=n$ and $l=4n$ when $n$ is odd, or $k=n/2$ and $l=n$ when $n$ is even. The $\mathbb{Z}_{2k}$ group is the ``non-symmorphic'' $\mathbb{Z}_2$ central extensions of $\mathbb{Z}_k$ and associates to non-trivial cohomology elements in $H^2(\mathbb{Z}_k,\mathcal{A})$. The super-script $[k]$ in $D^{[k]}(\mathbb{Z}_{2k})$ labels the Dijkgraaf-Witten invariant in the cohomology group $H^3(\mathbb{Z}_{2k},U(1))=\mathbb{Z}_{2k}$ that deforms the gauge theory. The two even $k$ cases in \eqref{Zkorbifolds} agree with \eqref{innerautgauging} and admit the decomposition $SO(2n)_1\boxtimes D^{[k]}(\mathbb{Z}_{2k})$. This is because (i) when $k=n/2\equiv2$ modulo 4, $SO(2n)_1$ is equivalent to $SO(8)_1$ as an \hyperlink{UMTC}{UMTC}. (ii) When $k=n/2\equiv0$ modulo 4, $SO(2n)_1$ is equivalent to $D^{[0]}(\mathbb{Z}_2)$ as an \hyperlink{UMTC}{UMTC}, and $D^{[k]}(\mathbb{Z}_{2k})=D^{[0]}(\mathbb{Z}_2)\boxtimes D^{[k]}(\mathbb{Z}_{2k})$ under the $\mathbb{Z}_2$ charge pair condensation in the relative tensor product.

\section{Dihedral twist liquids}\label{sec:dihedralTL}
We construct coupled-wire models of the following non-Abelian twist liquid topological phases (see figure~\ref{fig:CWMschematic}(b)) \begin{align}\frac{SU(n)_1}{\mathbb{Z}_2}&=SO(n)_2,\quad\frac{U(1)_l}{\mathbb{Z}_2}=\frac{SO(n)_1\times SO(n)_1}{SO(n)_2}\nonumber\\\frac{SO(2n)_1}{D_k}&=\frac{SU(n)_1}{\mathbb{Z}_2}\boxtimes\frac{U(1)_l}{\mathbb{Z}_2},\label{NBTL}\end{align} where $n\geq3$. Here, $k=n/2=l/2$ when $n$ is even or $k=n=l/4$ when $n$ is odd. The $\mathbb{Z}_2$ charge pair is condensed in the relative tensor product $\boxtimes$. The $SU(n)_1/\mathbb{Z}_2$ and $U(1)_l/\mathbb{Z}_2$ orbifolds are components of the coset decomposition~\cite{GinspargLectureNotes,bigyellowbook,Blumenhagenbook} \begin{align}\left[SO(n)_1\right]^2=SO(n)_2\times\frac{\left[SO(n)_1\right]^2}{SO(n)_2}\label{coset}\end{align} where the $SO(n)_2$ \WZW \KM theory is the diagonal sub-algebra of $\left[SO(n)_1\right]^2$. From section~\ref{sec:gaugingZkZ2}, we identified $\left[SO(n)_1\right]^2$ with the $\mathbb{Z}_2$ twist liquid $SO(2n)_1/\mathbb{Z}_2$. Therefore the coset decomposition \eqref{coset} is the $\mathbb{Z}_2$ orbifold version of the conformal embedding $SO(2n)_1\supseteq U(1)_l\times SU(n)_1$. In section~\ref{sec:gaugingZkZ2}, we discussed the global $D_k=\mathbb{Z}_2\ltimes\mathbb{Z}_k$ symmetry of the $SO(2n)_1$ \WZW \CFT, where the $\mathbb{Z}_k$ and $\mathbb{Z}_2$ actions were presented in \eqref{Dksymm}. We also showed that the decoupled pair $U(1)_l\times SU(n)_1$ carried a local $\mathbb{Z}_k$ gauge symmetry and could be identified as the $SO(2n)_1/\mathbb{Z}_k$ orbifold phase. In addition, they carry a global $\mathbb{Z}_2$ conjugation symmetry \begin{align}\mathbb{Z}_2:\quad\phi_{y\perp}^\sigma\to-\phi_{y\perp}^\sigma,\quad\phi^\sigma_{ya}-\phi^\sigma_{yb}\to\phi^\sigma_{yb}-\phi^\sigma_{ya}\label{globalZ2}\end{align} where $\phi_a$, for $a=1,\ldots,n$, are the bosonized variables of the Dirac fermions $d_a\sim e^{i\phi_a}$ in $SO(2n)_1$, $\phi_\perp=\phi_1+\ldots+\phi_n$ is the diagonal part that generates $U(1)_l$, and $e^{i(\phi_a-\phi_b)}$ are the roots of the $SU(n)_1$ \WZW \KM algebra. Here the $\mathbb{Z}_2$ symmetry applies globally to all species $a,b=1,\ldots,n$, propagation directions $\sigma=L,R$ on all wires $y=1,\ldots,\mathsf{L}$. 

In this section, we construct a new $D_k$ twist liquid phase by further promoting the $\mathbb{Z}_2$ symmetry into a local gauge symmetry. 

\subsection{The \texorpdfstring{$SU(n)_1/\mathbb{Z}_2$}{SU(n)/Z2} twist liquid}\label{sec:SUnZ2TL}
The $SU(n)_1/\mathbb{Z}_2$ orbifold phase is constructed based on the identification $SU(n)_1/\mathbb{Z}_2=SO(n)_2$.~\cite{Schellekens99} The coupled-wire model is built on a 2D array of wires, each consists of the $SO(n)_1\times SO(n)_1$ pair. 
The kinetic Hamiltonian density is the decoupled direct sum \begin{align}\mathcal{H}_0&=\mathcal{H}_0^A\oplus\mathcal{H}_0^B\label{kineticHSOnxSOn}\\&=\frac{iv}{2}\sum_{y=1}^{\mathsf{L}}\sum_{a=1}^n\sum_{\sigma=+,-}\sigma\left(\psi^{A\sigma}_{ya}\partial_x\psi^{A\sigma}_{ya}+\psi^{B\sigma}_{ya}\partial_x\psi^{B\sigma}_{ya}\right)\nonumber\end{align} where $\psi^{A\sigma}_{ya}=\psi^\sigma_{ya}$ and $\psi^{B\sigma}_{ya}=\psi^\sigma_{y,n+a}$, for $a=1,\ldots,n$, are the two sets of Majorana fermions of the $SO(n)_1^A\times SO(n)_1^B$ pair at the $y^{\mathrm{th}}$ wire, and $\sigma=L,R=+,-$ labels the two propagation directions. Here, the two $SO(n)_1$ \hyperlink{CFT}{CFT}s originate independently from two distinct sets of interacting 1D electrons. (See section~\ref{sec:bosonwires} for the electronic spin liquid and superconductor origins.) Consequently, only fermion pairs $\psi_{ya}^\sigma\psi_{ya'}^{\sigma'}$ and $\psi_{y,n+a}^\sigma\psi_{y,n+a'}^{\sigma'}$ from the same $A/B$ sector on the same wire are integral combinations of local electrons. Fermion pairs from distinct sectors $\psi_{y,a}^\sigma\psi_{y,n+a'}^{\sigma'}$ are all non-local. In general, an operator is an integral combination of local electrons (up to intra-wire ground state expectation values) if and only if it is even under the $\mathbb{Z}_2^A\times\mathbb{Z}_2^B$ gauge symmetry defined in \eqref{Z2gauge}. Grouping Majorana fermions from opposite sectors into Dirac fermions $d^\sigma_{ya}=(\psi^{A\sigma}_{ya}+i\psi^{B\sigma}_{ya})/\sqrt{2}\sim e^{i\phi^\sigma_{ya}}$, the $\mathbb{Z}_2^B$ symmetry is the conjugation \begin{align}\mathbb{Z}_2^B(y):\quad\begin{split}&\phi^\sigma_{y'a}\to(-1)^{\delta_{yy'}}\phi^\sigma_{y'a}\\&\psi^{A\sigma}_{y'a}\to\psi^{A\sigma}_{y'a},\quad\psi^{B\sigma}_{y'a}\to(-1)^{\delta_{yy'}}\psi^{B\sigma}_{y'a}\end{split}\label{Z2gauge1}\end{align} that flips $d^\sigma_{ya}\to{d^\sigma_{ya}}^\dagger$.  In the following, we refer to the $\mathbb{Z}_2^B$ symmetry simply as the $\mathbb{Z}_2$ symmetry.

The coset decomposition \eqref{coset} separates the $[SO(n)_1]^2$ \CFT into the decoupled $U(1)_l/\mathbb{Z}_2=[SO(n)_1]^2/SO(n)_2$ and $SU(n)_1/\mathbb{Z}_2=SO(n)_2$ sectors. The $U(1)_l$ is generated by the diagonal bosonized variable $\phi_\perp=\phi_1+\ldots+\phi_n$. Unlike previously in section~\ref{sec:unitaryfamily} when the $U(1)_l$ coupled-wire model was constructed based on $SO(2n)_1$ wires, now the new notion of locality from the $SO(n)_1\times SO(n)_1$ pair and its local $\mathbb{Z}_2$ gauge symmetry \eqref{Z2gauge1} enforce the $U(1)_l/\mathbb{Z}_2$ orbifold \CFT on each wire. This is because the physical Hilbert space is even under $\mathbb{Z}_2:\phi_\perp\to-\phi_\perp$. The $U(1)_l$ current \begin{align}\partial\phi_\perp\sim\sum_{a=1}^n\psi_a^A\psi_a^B\label{U1Z2charge}\end{align} is no longer local as it is odd under $\mathbb{Z}_2$. It takes the role of the $\mathbb{Z}_2$ gauge charge in $U(1)_l/\mathbb{Z}_2$ and must appear in pairs. The $U(1)_l/\mathbb{Z}_2$ sector on a given wire is gapped under the intra-wire backscattering interaction borrowed from \eqref{U1sineGordon} \begin{align}\mathcal{U}^y_{U(1)_l/\mathbb{Z}_2}=-u_\perp^{\mathrm{intra}}\partial_x\phi_{y\perp}^L\partial_x\phi_{y\perp}^R-\Delta_\perp^{\mathrm{intra}}\cos\Theta_y^\perp,\label{U1sineGordon2}\end{align} where $\Theta_y^\perp=\phi^L_{y\perp}-\phi^R_{y\perp}$. Here, both the density interaction and the sine-Gordon potential are even under \eqref{Z2gauge1} and are integral combinations of local electrons. When $u_\perp^{\mathrm{intra}}>v(n^2-4)/[2\pi n(n^2+4)]$, the sine-Gordon potential is relevant in the \RG sense and introduces a finite excitation energy gap for the $U(1)_l/\mathbb{Z}_2$ sector.

The remaining $SO(n)_2$ sector of the decomposition \eqref{coset} is the diagonal \WZW \KM sub-algebra of the $SO(n)_1^A\times SO(n)_1^B$ pair. It generates the simultaneous orthogonal rotation of both the $A$ and $B$ Majorana fermions $\psi^{A/B}_a\to O_a^b\psi^{A/B}_b$. The $SO(n)_2$ currents are the sums $J_{ab}=J_{ab}^A+J_{ab}^B=i\psi_a^A\psi_b^A+i\psi_a^B\psi_b^B$, for $1\leq a<b\leq n$. Focusing on the $\sigma=L$ chiral sector, the currents obey the operator product expansion~\cite{GinspargLectureNotes,bigyellowbook,Blumenhagenbook} \begin{align}J_{ab}(z)J_{cd}(w)=\frac{2\delta_{ac}\delta_{bd}}{(z-w)^2}+\frac{if_{(ab)(cd)(ef)}}{z-w}J_{ef}(w)+\ldots.\label{SOn2OPE}\end{align} Here, $z$ and $w$ are complex Euclidean space-time parameters $e^{2\pi(v\tau+ix)/l}$, $\tau=it$ is the Wick rotated time, and $l$ is the $x$-circumference of the closed circular wires. The structure constant of the $SO(n)$ Lie algebra is \begin{align}f_{(ab)(cd)(ef)}&=\delta_{ea}\delta_{fd}\delta_{bc}-\delta_{ea}\delta_{fc}\delta_{bd}\nonumber\\&\;\;\;+\delta_{eb}\delta_{ad}\delta_{fc}-\delta_{eb}\delta_{fd}\delta_{ac}.\label{SONstructureconstant}\end{align} The factor of 2 in \eqref{SOn2OPE} sets the level of the $SO(n)_2$ \WZW \KM algebra. At the same time, $SO(n)_2$ is also the \WZW \KM sub-algebra of $SU(n)_1$ that is even under the $\mathbb{Z}_2$ gauge symmetry. Among the unitary rotations $d_a\to U_a^bd_b$ of Dirac fermions, only the real orthogonal ones are compatible and commutes with the $\mathbb{Z}_2$ conjugation $d\to d^\dagger$. The $SO(n)_2$ currents are self-conjugate combinations of $SU(n)_1$ root operators \begin{align}\begin{split}J_{ab}&=J_{ab}^A+J_{ab}^B=i\psi_a^A\psi_b^A+i\psi_a^B\psi_b^B\\&\sim2\cos\left(\phi_a-\phi_b\right),\end{split}\label{SOn2currents}\end{align} which are integral combinations of local electrons. The $SO(n)_2$ complement of $SU(n)_1$ consists of the Cartan-Weyl generators and the anti-conjugate root combinations \begin{align}\begin{split}&\partial\tilde\phi_p\sim\frac{i}{\sqrt{p(p+1)}}\left(-p\psi^A_{p+1}\psi^B_{p+1}+\sum_{a=1}^p\psi_a^A\psi_a^B\right)\\&S_{ab}=i\psi_a^A\psi_b^B-i\psi_a^B\psi_b^A\sim2\sin\left(\phi_a-\phi_b\right)\end{split}\label{SUnZ2charge}\end{align} both of which are odd under $\mathbb{Z}_2$ and are therefore fractional. Similar to \eqref{U1Z2charge}, operators in \eqref{SUnZ2charge} take the role of the $\mathbb{Z}_2$ gauge charge in $SU(n)_1/\mathbb{Z}_2$.

The $SO(n)_2$ sector is gapped using the inter-wire current backscattering interaction \begin{align}&\mathcal{U}^{y+1/2}_{SO(n)_2}=u^{\mathrm{inter}}{\bf J}_{y+1,SO(n)_2}^L\cdot{\bf J}_{y,SO(n)_2}^R\label{SOn2int}\\&=-\Delta^{\mathrm{inter}}\sum_{1\leq a<b\leq n}\cos\left(\phi_{y+1,a}^L-\phi_{y+1,b}^L\right)\cos\left(\phi_{y,a}^R-\phi_{y,b}^R\right)\nonumber\\&=-\frac{1}{2}\Delta^{\mathrm{inter}}\sum_{1\leq a<b\leq n}\left(\cos\theta^{y+1/2}_{ab}+\cos\varphi^{y+1/2}_{ab}\right)\nonumber\end{align} where ${\bf J}_{SO(n)_2}=(J_{ab})_{1\leq a<b\leq n}$ are the $SO(n)_2$ currents defined in \eqref{SOn2currents}, $\theta^{y+1/2}_{ab}=\phi^L_{y+1,a}-\phi^L_{y+1,b}-\phi^R_{y,a}+\phi^R_{y,b}$ and $\varphi^{y+1/2}_{ab}=\phi^L_{y+1,a}-\phi^L_{y+1,b}+\phi^R_{y,a}-\phi^R_{y,b}$. The interaction is marginally relevant~\cite{Cardybook} when $u^{\mathrm{inter}},\Delta^{\mathrm{inter}}>0$. Since the two sets of sine-Gordon variables $\theta$ and $\varphi$ do not all mutually commute, they cannot both take non-vanishing ground state expectation values. The current backscattering interaction \eqref{SOn2int} condenses the bosons $J_{y+1,ab}^LJ_{y,ab}^R$, and when $\Delta^{\mathrm{inter}}>0$, it pins the positive ground state expectation values \begin{align}\left\langle\cos\left(\phi_{y+1,a}^L-\phi_{y+1,b}^L\right)\cos\left(\phi_{y,a}^R-\phi_{y,b}^R\right)\right\rangle>0.\end{align} At the same time, the $SO(n)_2$ currents are quadratic fusion products of the $\mathbb{Z}_2$ gauge charges from \eqref{SUnZ2charge}, $S\times S=J$. Focusing on the $\sigma=L$ chiral sector, the gauge charges obey the operator product expansion \begin{align}&S_{ab}(z)S_{cd}(w)\label{S2J}\\&=\frac{2(\delta_{ac}\delta_{bd}+\delta_{bc}\delta_{ad})}{(z-w)^2}-\frac{ig_{(ab)(cd)(ef)}}{z-w}J_{ef}(w)+\ldots,\nonumber\end{align} where $g_{(ab)(cd)(ef)}=\delta_{ac}\delta_{be}\delta_{df}+\delta_{bd}\delta_{ae}\delta_{cf}+\delta_{bc}\delta_{ae}\delta_{df}+\delta_{ad}\delta_{be}\delta_{cf}$. Since the $SO(n)_2$ current $J$ is generated by squaring the $\mathbb{Z}_2$ gauge boson charges $S$, the interaction \eqref{SOn2int} must also condenses the $\mathbb{Z}_2$ charge pairs $S_{y+1,ab}^LS_{y,ab}^R$ in addition to $J_{y+1,ab}^LJ_{y,ab}^R$. This gives rise to the finite ground state expectation value \begin{align}s_{y+1/2}=\left\langle\sin\left(\phi_{y+1,a}^L-\phi_{y+1,b}^L\right)\sin\left(\phi_{y,a}^R-\phi_{y,b}^R\right)\right\rangle\label{SOn2orderparameter}\end{align} which can be either positive or negative. Its sign determines which set of sine-Gordon variables, $\theta$ or $\varphi$, in \eqref{SOn2int} is pinned. \begin{align}\begin{split}s_{y+1/2}>0:\quad\left\langle\cos\theta^{y+1/2}_{ab}\right\rangle=1,\quad\left\langle\cos\varphi^{y+1/2}_{ab}\right\rangle=0,\\s_{y+1/2}<0:\quad\left\langle\cos\theta^{y+1/2}_{ab}\right\rangle=0,\quad\left\langle\cos\varphi^{y+1/2}_{ab}\right\rangle=1.\end{split}\label{SOn2GS}\end{align} The positive case corresponds to backscattering of $SU(n)_1$ roots $\cos\theta_{ab}\sim e^{i(\phi_a^L-\phi_b^L)}e^{-i(\phi_a^R-\phi_b^R)}+h.c.$ (c.f.~\eqref{SUnsineGordon}), while the negative case corresponds to pairing of $SU(n)_1$ roots $\cos\varphi_{ab}\sim e^{i(\phi_a^L-\phi_b^L)}e^{i(\phi_a^R-\phi_b^R)}+h.c.$. Both lead to a finite excitation energy gap in $SU(n)_1/\mathbb{Z}_2=SO(n)_2$.

The above reasoning of a finite energy gap relies on \eqref{S2J}, which only holds when $n\geq 3$. When $n=2$, the $SO(2)_2$ current $J\sim\cos(\phi_1-\phi_2)$ cannot be generated by squaring the gauge charge $S\sim\sin(\phi_1-\phi_2)$. In fact, the two bosons $S$ and $J$ are decoupled from each other, and the current backscattering interaction $J^LJ^R$ does not condense $S^LS^R$. Our models constructed in this section are therefore restricted to $n\geq3$ where there is a multiplet of $\mathbb{Z}_2$ gauge charges $S_{ab}$ that generate the $SO(n)_2$ currents $J_{ab}$ by squaring.

Overall, the coupled-wire model \begin{align}\mathcal{H}_{SU(n)_1/\mathbb{Z}_2}&=\mathcal{H}_0+\sum_{y=1}^{\mathsf{L}}\left(\mathcal{U}^y_{U(1)_l/\mathbb{Z}_2}+\mathcal{U}^{y+1/2}_{SO(n)_2}\right)\label{HSUn2}\end{align} -- that combines the kinetic Hamiltonian \eqref{kineticHSOnxSOn}, the $U(1)_l/\mathbb{Z}_2$ intra-wire interaction \eqref{U1sineGordon2} and the $SU(n)_1/\mathbb{Z}_2$ inter-wire interaction \eqref{SOn2int} -- describes the $SU(n)_1/\mathbb{Z}_2=SO(n)_2$ twist liquid topological phase. The model has a finite bulk excitation energy gap and carries gapless chiral $SO(n)_2$ \WZW \CFT on the edges.


\begin{figure}[htbp]
\centering\includegraphics[width=0.5\textwidth]{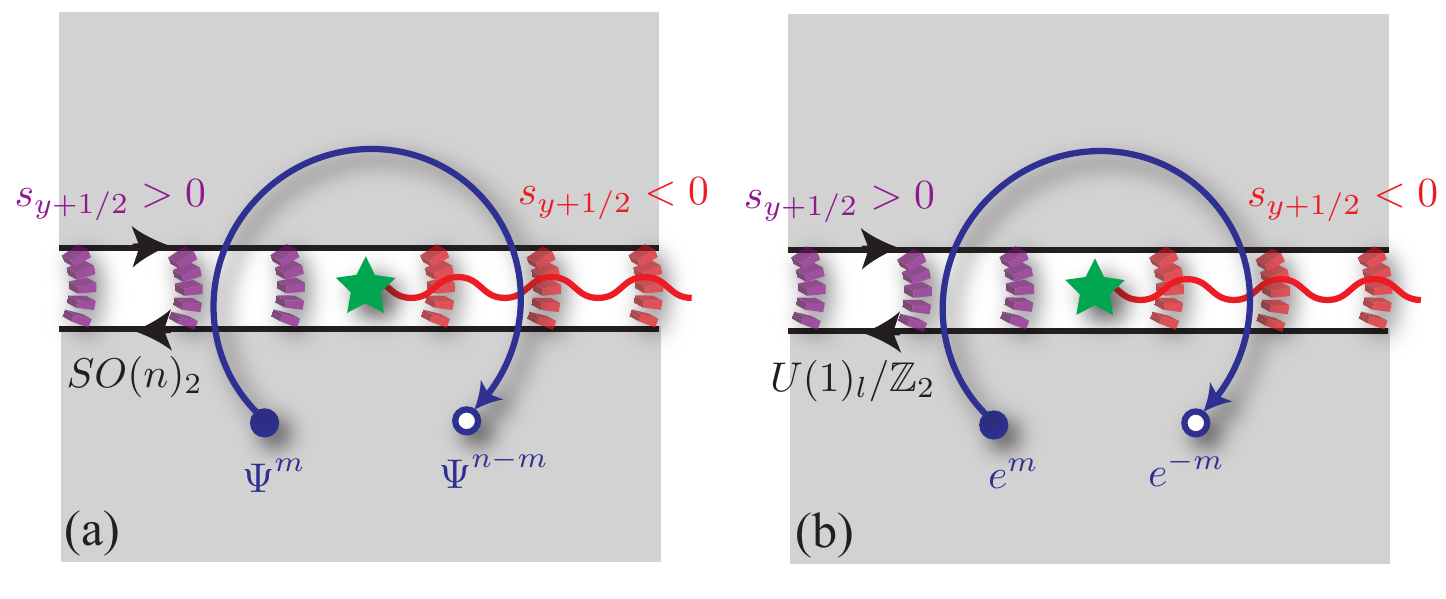}
\caption{A non-Abelian $\mathbb{Z}_2$ gauge flux $\sigma$ or $\tau$ (green star) sits at the domain wall where the order parameter $s_{y+1/2}$ in (a) \eqref{SOn2orderparameter} for $SU(n)_1/\mathbb{Z}_2$ or (b) \eqref{U1Z2orderparameter} for $U(1)_l/\mathbb{Z}_2$ changes sign. An orbiting anyon is conjugated, $\Psi_m\to\Psi_{n-m}$ or $e^m\to e^{-m}$, when passing across the branch cut (red curvy line).}\label{fig:Z2twistdefect}
\end{figure}

Domain wall between the two sets of ground states \eqref{SOn2GS} where the order parameter in \eqref{SOn2orderparameter} changes sign, $s_{y+1/2}(x)\sim\langle S^L_{y+1,ab}(x)S^R_{y,ab}(x)\rangle\sim\pm\mathrm{sgn}(x-x_0)$, traps a $\mathbb{Z}_2$ gauge flux at $x_0$. (See figure~\ref{fig:Z2twistdefect}(a).) In \eqref{SUnZ2charge}, we identified $S^\sigma_{y,ab}$ as one of the non-local fields that represents the $\mathbb{Z}_2$ gauge charge $S$. The sign flip of $s_{y+1/2}(x)$ corresponds to the $\pi$ monodromy braiding phase between the $\mathbb{Z}_2$ gauge flux and gauge charge. At the same time, the $\mathbb{Z}_2$ flux conjugates any orbiting $SU(n)_1$ anyon, $\Psi^{m}\to\Psi^{n-m}$. To see this, we first re-express the $SU(n)_1$ primary fields defined in \eqref{SUnanyons} as $\Psi^m\sim e^{i\sum_{j=1}^m\sum_{b=1}^n(\phi_{a_j}-\phi_b)/n}$. When $\Psi^m$ passes through the side of the domain wall with $s_{y+1/2}(x)>0$, the backscattering term pins $\langle\Psi^m_L(x)\Psi^m_R(x)^\dagger\rangle\sim e^{i\sum_{jb}\langle\theta^{y+1/2}_{a_jb}(x)\rangle/n}\sim1$ to a finite value. Further moving $\Psi^m$ to the other side of the domain wall with $s_{y+1/2}(x)<0$, there is no longer any backscattering term, but instead the pairing term pins $\langle\Psi^m_L(x)\Psi^m_R(x)\rangle\sim e^{i\sum_{jb}\langle\varphi^{y+1/2}_{a_jb}(x)\rangle/n}\sim1$ to a finite value. Therefore, $\Psi^m$ is conjugated to $[(\Psi^m)^\dagger]=[\Psi^{n-m}]$ as we complete this full circle. Consequently, in the $SU(n)_1/\mathbb{Z}_2$ orbifold phase, any pair of conjugated partners are grouped together into a single non-Abelian super-selection sector \begin{align}\Phi^i=\Psi^i+\Psi^{n-i}\label{SUnsupersectors}\end{align} for $1\leq i<n/2$. When $n$ is even, the Abelian anyon $\Psi^{n/2}$ in $SU(n)_1$ is closed under the $\mathbb{Z}_2$ symmetry \eqref{Z2gauge1}. It decomposes into the $\mathbb{Z}_2$ even and odd sectors \begin{align}\begin{split}\left[\Psi^{n/2}_+\right]=\mathrm{span}\left\{\cos\left(\frac{1}{2}\phi_\perp-\sum_{j=1}^{n/2}\phi_{a_j}\right)\right\}_{1\leq a_1<\ldots<a_{n/2}\leq n},\\\left[\Psi^{n/2}_-\right]=\mathrm{span}\left\{\sin\left(\frac{1}{2}\phi_\perp-\sum_{j=1}^{n/2}\phi_{a_j}\right)\right\}_{1\leq a_1<\ldots<a_{n/2}\leq n},\end{split}\label{SUnZ2spinors}\end{align} each forms an irreducible representation of $SO(n)_2$. 

When $n$ is odd, there are two $\mathbb{Z}_2$ fluxes that differ from each other by a $\mathbb{Z}_2$ charge \begin{align}\sigma=\tau\times S,\quad\tau=\sigma\times S.\label{Z2fluxfusion0}\end{align} They obey the fusion rule \begin{gather}\sigma\times\sigma=\tau\times\tau=1+\sum_{i=1}^{(n-1)/2}\Phi^i,\nonumber\\\Phi^i\times\sigma=\Phi^i\times\tau=\sigma+\tau.\end{gather} When $n$ is even, there are four $\mathbb{Z}_2$ fluxes related to each other by \begin{gather}\sigma^\pm=\tau^\pm\times S,\quad\tau^\pm=\sigma^\pm\times S,\nonumber\\\sigma^\pm\times\Phi^i=\tau^\pm\times\Phi^i=\left\{\begin{array}{*{20}l}\sigma^\pm+\tau^\pm,&\mbox{if $i$ even}\\\sigma^\mp+\tau^\mp,&\mbox{if $i$ odd}\end{array}\right..\label{Z2fluxfusion1}\end{gather} They obey the pair fusion rules \begin{gather}\sigma^\pm\times\sigma^\pm=\tau^\pm\times\tau^\pm=1+\Psi^{n/2}_\pm+\sum_{i\mbox{ }\mathrm{even}}\Phi^i\nonumber\\\sigma^\pm\times\sigma^\mp=\tau^\pm\times\tau^\mp=\sum_{i\mbox{ }\mathrm{odd}}\Phi^i\label{Z2fluxfusion2}\end{gather} when $n\equiv0$ modulo 4, or \begin{gather}\sigma^\pm\times\sigma^\pm=\tau^\pm\times\tau^\pm=\Psi^{n/2}_\pm+\sum_{i\mbox{ }\mathrm{odd}}\Phi_i\nonumber\\\sigma^\pm\times\sigma^\mp=\tau^\pm\times\tau^\mp=1+\sum_{i\mbox{ }\mathrm{even}}\Phi_i\label{Z2fluxfusion3}\end{gather} when $n\equiv2$ modulo 4. 

\begin{table}[htbp]
\centering
\begin{tabular}{lll}
anyons&quantum dimension & spin (modulo 1)\\\hline
1&1&0\\
$S$&1&1\\
$\Phi^i$&2&$i(n-i)/(2n)$\\
$\sigma$&$\sqrt{n}$&$(n-1)/16$\\
$\tau$&$\sqrt{n}$&$(n+7)/16$
\end{tabular}
\caption{Anyon content of $SU(n)_1/\mathbb{Z}_2=SO(n)_2$ when $n$ is odd. Here $i=1,\ldots,(n-1)/2$.}\label{SOn2anyonsodd}
\end{table}

\begin{table}[htbp]
\centering
\begin{tabular}{lll}
anyons&quantum dimension & spin (modulo 1)\\\hline
1&1&0\\
$S$&1&1\\
$\Psi_\pm^{n/2}$&1&$n/8$\\
$\Phi^i$&2&$i(n-i)/(2n)$\\
$\sigma^\pm$&$\sqrt{n/2}$&$(n-1)/16$\\
$\tau^\pm$&$\sqrt{n/2}$&$(n+7)/16$
\end{tabular}
\caption{Anyon content of $SU(n)_1/\mathbb{Z}_2=SO(n)_2$ when $n$ is even. Here $i=1,\ldots,n/2-1$.}\label{SOn2anyonseven}
\end{table}

The anyons of $SU(n)_1/\mathbb{Z}_2=SO(n)_2$ together with their quantum dimensions and spins are summarized in table~\ref{SOn2anyonsodd} and \ref{SOn2anyonseven}. The complete modular anyon content of $SO(n)_2$ can be found in appendix~\ref{app:orbifoldanyons}. When $n$ is odd, the $\mathbb{Z}_2$ fluxes $\sigma$ and $\tau$ are referred to as metaplectic anyons by Hastings, Nayak, and Wang in ref.~\cite{HastingsNayakWang12,HastingsNayakWang14}, and the $SO(n)_2$ topological order is known as a metaplectic topological quantum field theory. The braiding operations performed by the metaplectic anyons can be simulated by a classical computer. On the other hand, they showed that the $d=2$ super-sectors $\Phi_i$ in \eqref{SUnsupersectors} computes a $\# P$-hard link invariant, and therefore their braidings cannot be efficiently simulated classically. This suggests the topological order we consider may have (non-universal) quantum computing power beyond a classical computer.

\subsection{The \texorpdfstring{$U(1)_l/\mathbb{Z}_2$}{U(1)/Z2} twist liquid}\label{sec:U1Z2orbifold}
The $U(1)_l/\mathbb{Z}_2$ orbifold~\cite{Ginsparg88,DijkgraafVerlindeVerlinde88,DijkgraafVafaVerlindeVerlinde89} phases were constructed by a coupled-wire model as fractional quantum Hall states in ref.~\cite{KaneStern18,TamHuKane20} based on a 2-fluid model~\cite{KaneSternHalperin17}. Here, we construct the orbifold phases in the electron-based superconducting or spin liquid settings. The coupled-wire model makes use of the coset identification~\footnote{L.~Dixon, cited in ref.~\cite{GinspargLectureNotes}.} $U(1)_l/\mathbb{Z}_2=[SO(n)_1]^2/SO(n)_2$. The level is $l=n$ ($l=4n$) when $n$ is even (odd), and it sets the compactified free boson circle radius $R=\sqrt{l}/2$. The model is the ``particle-hole conjugation'' of the $SU(n)_1/\mathbb{Z}_2=SO(n)_2$ model in \eqref{HSUn2} in the sense that the inter-wire and intra-wire backscattering directions of $SO(n)_2$ and $U(1)_l/\mathbb{Z}_2$ are reversed. \begin{align}\mathcal{H}_{U(1)_l/\mathbb{Z}_2}&=\mathcal{H}_0+\sum_{y=1}^{\mathsf{L}}\left(\mathcal{U}^y_{SO(n)_2}+\mathcal{U}^{y+1/2}_{U(1)_l/\mathbb{Z}_2}\right).\label{HU1l2}\end{align} Like the previous $SU(n)_1/\mathbb{Z}_2$ case, the model here is based on a 2D array of $SO(n)_1^A\times SO(n)_1^B$ wires, each carries two sets of Majorana fermions $\psi^{A\sigma}_{ya}=\psi^\sigma_{ya}$ and $\psi^{B\sigma}_{ya}=\psi^\sigma_{y,n+a}$, for $a=1,\ldots,n$, $\sigma=L,R=+,-$ and $y=1,\ldots,\mathsf{L}$. All fermion pair from the same $C=A,B$ sector on the same wire, $\psi^{C\sigma}_{ya}\psi^{C\sigma'}_{ya'}$, is an integral combination of local bosons. All fermion pairs from opposite sectors, $\psi^{A\sigma}_{ya}\psi^{B\sigma'}_{y'a'}$, are fractional and must come in pairs. The kinetic Hamiltonian $\mathcal{H}_0$ is identical to the previous $SU(n)_1/\mathbb{Z}_2$ model and is given in \eqref{kineticHSOnxSOn}. The intra-wire $SO(n)_2$ current backscattering potential is \begin{align}&\mathcal{U}^y_{SO(n)_2}=u^{\mathrm{intra}}\sum_{1\leq a<b\leq n}J^{ab}_{y,SO(n)_2^L}J^{ab}_{y,SO(n)_2^R}\label{SOn2intra}\end{align} where $J^{ab}_{y,SO(n)_2^\sigma}=i(\psi^{A\sigma}_{ya}\psi^{A\sigma}_{yb}+\psi^{B\sigma}_{ya}\psi^{B\sigma}_{yb})$ are the $SO(n)_2$ currents. It gaps the $SO(n)_2$ degrees of freedom in all wires for the same reasons \eqref{SOn2int} does and has been presented in the previous subsection. 

For the inter-wire interaction $\mathcal{U}^{y+1/2}_{U(1)_l/\mathbb{Z}_2}$, it may be tempting to mimick \eqref{U1sineGordon} and consider the sine-Gordon potentials $\cos\left[q\left(\phi^L_{y+1,\perp}\pm\phi^R_{y,\perp}\right)\right]$, where $e^{i\phi_\perp}=e^{i(\phi_1+\ldots+\phi_n)}$ is a $SO(n)_2$ singlet that generates $U(1)_l/\mathbb{Z}_2$ and is the diagonal product of Dirac fermions $e^{i\phi_a}=(\psi_a^A+i\psi_a^B)/\sqrt{2}$. However, the potentials switch \begin{align}\cos\left[q\left(\phi^L_{y+1,\perp}+\phi^R_{y,\perp}\right)\right]\leftrightarrow\cos\left[q\left(\phi^L_{y+1,\perp}-\phi^R_{y,\perp}\right)\right]\end{align} under the local $\mathbb{Z}_2\times\mathbb{Z}_2$ symmetry (see \eqref{Z2gauge1}) \begin{align}\begin{split}&\mathbb{Z}^A_2(y):\quad\phi^\sigma_{y'\perp}\to(-1)^{\delta_{yy'}}\phi^\sigma_{y'\perp}+n\pi\delta_{yy'}\\&\mathbb{Z}^B_2(y):\quad\phi^\sigma_{y'\perp}\to(-1)^{\delta_{yy'}}\phi^\sigma_{y'\perp}\end{split}\label{Z2gauge2}\end{align} and are therefore {\em not} integral combinations of electrons. Instead, we introduce the interactions \begin{align}\begin{split}\mathcal{U}^{y+1/2}_{U(1)_l/\mathbb{Z}_2}&=-u^{\mathrm{inter}}_\perp\left(\Sigma_{y+1/2}\right)^2\\&\quad-\Delta^{\mathrm{inter}}_\perp\cos\left(q\phi^L_{y+1,\perp}\right)\cos\left(q\phi^R_{y,\perp}\right)\\
\Sigma_{y+1/2}&=g\partial_x\phi^L_{y+1,\perp}\partial_x\phi^R_{y,\perp}\\&\quad+\sin\left(q\phi^L_{y+1,\perp}\right)\sin\left(q\phi^R_{y,\perp}\right)
\end{split}\label{U1sineGordon3}\end{align} where $q=1$ when $n$ is even, or $q=2$ when $n$ is odd. The interactions are local because $\cos\left(q\phi_\perp\right)$, $\left(\partial_x\phi_\perp\right)^2$, $\left[\sin\left(q\phi_\perp\right)\right]^2$, and $\partial_x\phi_\perp\sin\left(q\phi_\perp\right)$ are symmetric under \eqref{Z2gauge2} and are combinations of even products of fermions in both the $A$ and $B$ sectors. 

The $\Delta^{\mathrm{inter}}_\perp$ term can be expressed as \begin{align}\cos\left(q\phi^L_{y+1,\perp}\right)\cos\left(q\phi^R_{y,\perp}\right)=\frac{1}{2}\left(\cos\theta^{y+\frac{1}{2}}_\perp+\cos\varphi^{y+\frac{1}{2}}_\perp\right).\end{align} The angle variables $\theta^{y+1/2}_\perp=q(\phi^L_{y+1,\perp}-\phi^R_{y,\perp})$ and $\varphi^{y+1/2}_\perp=q(\phi^L_{y,\perp}+\phi^R_{y+1,\perp})$ do not commute, and therefore the two sine-Gordon potentials compete. The $u^{\mathrm{inter}}_\perp$ term can be re-expressed -- using a Hubbard-Stratonovich transformation~\cite{Stratonovich57,Hubbard59} by introducing a real bosonic auxiliary field $s_{y+1/2}$ -- into \begin{align}\frac{\tilde{s}_{y+1/2}^2}{u^{\mathrm{inter}}_\perp}-u^{\mathrm{inter}}_\perp\left(\Sigma_{y+1/2}\right)^2=\frac{s_{y+1/2}^2}{u^{\mathrm{inter}}_\perp}-2s_{y+1/2}\Sigma_{y+1/2}.\end{align} $\tilde{s}_{y+1/2}=s_{y+1/2}-u^{\mathrm{inter}}_\perp\Sigma_{y+1/2}$ completes the square and can be integrated out. In low-energy, the auxiliary takes a finite ground state expectation value $\langle s_{y+1/2}\rangle$. The interaction \eqref{U1sineGordon3} admits the mean-field approximation \begin{align}\mathcal{U}^{y+1/2}_{U(1)_l/\mathbb{Z}_2}&=\frac{\langle s_{y+1/2}\rangle^2}{u^{\mathrm{inter}}_\perp}-2g\langle s_{y+1/2}\rangle\partial_x\phi^L_{y+1,\perp}\partial_x\phi^R_{y,\perp}\nonumber\\&-\left(\frac{\Delta^{\mathrm{inter}}_\perp}{2}+\langle s_{y+1/2}\rangle\right)\cos\theta^{y+1/2}_\perp\nonumber\\&\quad\quad-\left(\frac{\Delta^{\mathrm{inter}}_\perp}{2}-\langle s_{y+1/2}\rangle\right)\cos\varphi^{y+1/2}_\perp.\label{HSU1Z2}\end{align} When $\langle s_{y+1/2}\rangle$ is positive (negative) and for large $g$, $g|\langle s_{y+1/2}\rangle|>v(q^4n^2-4)/[4\pi n(q^4n^2+4)]$, $\cos\theta^{y+1/2}_\perp$ is relevant (irrelevant) while $\cos\varphi^{y+1/2}_\perp$ is irrelevant (relevant) in the \RG sense. Therefore, depending on the sign of the order parameter \begin{align}\mathrm{sgn}\langle s_{y+1/2}\rangle=\mathrm{sgn}\left\langle\sin\left(m\phi^L_{y+1,\perp}\right)\sin\left(m\phi^R_{y,\perp}\right)\right\rangle,\label{U1Z2orderparameter}\end{align} either $\cos\theta^{y+1/2}_\perp$ or $\cos\varphi^{y+1/2}_\perp$ dominates and takes non-vanishing ground state expectation value. \begin{align}\begin{split}s_{y+1/2}>0:\quad\left\langle\cos\theta^{y+1/2}_\perp\right\rangle=1,\quad\left\langle\cos\varphi^{y+1/2}_\perp\right\rangle=0,\\s_{y+1/2}<0:\quad\left\langle\cos\theta^{y+1/2}_\perp\right\rangle=0,\quad\left\langle\cos\varphi^{y+1/2}_\perp\right\rangle=1.\end{split}\label{U1Z2GS}\end{align} Both lead to a finite bulk excitation energy gap.

The ground state expectation values pattern \eqref{U1Z2GS} is similar to \eqref{SOn2GS} in the previous $SU(n)_1/\mathbb{Z}_2$ case, and consequently, the anyon structure also resembles. The non-local bosonic fields $\partial_x\phi_\perp$ and  $\sin(q\phi_\perp)$ belong in the $\mathbb{Z}_2$ charge sector $S$ of the $U(1)_l/\mathbb{Z}_2$ orbifold theory. The domain wall separating the two sets of ground states in \eqref{U1Z2GS}, where $\langle s_{y+1/2}(x)\rangle\sim\langle\sin(q\phi^L_{y+1,\perp}(x))\sin(q\phi^R_{y,\perp}(x))\rangle\sim\pm\mathrm{sgn}(x-x_0)$ changes sign, traps a $\mathbb{Z}_2$ gauge flux (twist field) at $x_0$. It associates a $\pi$ monodromy braiding phase for an orbiting $\mathbb{Z}_2$ charge. It also conjugates an orbiting vertex field \eqref{U1anyons} in $U(1)_l$ between $e^m=e^{im\phi_\perp/(qn)}\leftrightarrow e^{-m}=e^{-im\phi_\perp/(qn)}$ (see figure~\ref{fig:Z2twistdefect}(b)) according to the $\mathbb{Z}_2^{A/B}$ action in \eqref{Z2gauge2}. 

When $e^m$ passes through the side of the domain wall with $s_{y+1/2}(x)>0$, the backscattering term pins $\langle e^m_L(x)e^m_R(x)^\dagger\rangle\sim e^{im\langle\theta^{y+1/2}_\perp(x)\rangle/(q^2n)}\sim1$ to a finite value. Further moving $e^m$ to the other side of the domain wall with $s_{y+1/2}(x)<0$, the pairing term pins $\langle e^m_L(x)e^m_R(x)\rangle\sim e^{im\langle\varphi^{y+1/2}_\perp(x)\rangle/(q^2n)}\sim1$. Therefore, $e^m$ is conjugated to $(e^m)^\dagger=e^{-m}$ in a complete cycle. For $m=1,\ldots,l/2-1$ modulo $l$, \begin{align}\mathcal{E}^m=e^m+e^{-m}\label{U1supersectors}\end{align} is now a single super-selection sector. The $\mathbb{Z}_2$ invariant $e^{l/2}=e^{iq\phi_\perp/2}$ decomposes into the even and odd sectors \begin{align}e^{l/2}_+=\cos(q\phi_\perp/2),\quad e^{l/2}_-=\sin(q\phi_\perp/2),\label{el2pm}\end{align} where $q=1$ for $n$ even, or $q=2$ for $n$ odd. There are four $\mathbb{Z}_2$ twist fields $\sigma^\pm$ and $\tau^\pm$. They obey a set of fusion rules identical to those in $SU(n)_1/\mathbb{Z}_2$ and can be obtained by replacing $\Phi^i,\Psi^{n/2}_\pm$ by $\mathcal{E}^i,e^{l/2}_\pm$ in \eqref{Z2fluxfusion0}, \eqref{Z2fluxfusion1}, \eqref{Z2fluxfusion2} and \eqref{Z2fluxfusion3}. The spin statistics and quantum dimensions of the orbifold anyons are listed in table~\ref{Uorbifoldanyons} and the complete modular data is summarized in appendix~\ref{app:orbifoldanyons}.

\begin{table}[htbp]
\centering
\begin{tabular}{lll}
anyons&quantum dimension & spin (modulo 1)\\\hline
1&1&0\\
$S$&1&1\\
$e_\pm^{l/2}$&1&$l/8$\\
$\mathcal{E}^i$&2&$i^2/(2l)$\\
$\sigma^\pm$&$\sqrt{l/2}$&$1/16$\\
$\tau^\pm$&$\sqrt{l/2}$&$9/16$
\end{tabular}
\caption{Anyons in $U(1)_l/\mathbb{Z}_2$. Here $i=1,\ldots,l/2-1$.}\label{Uorbifoldanyons}
\end{table}

The $U(1)_l/\mathbb{Z}_2=[SO(n)_1]^2/SO(n)_2$ topological order can be related to $SU(n)_1/\mathbb{Z}_2=SO(n)_2$ described in the previous subsection~\ref{sec:SUnZ2TL} by the \UMTC equivalence \begin{align}\frac{[SO(n)_1]^2}{SO(n)_2}=[SO(n)_1]^2\boxtimes\overline{SO(n)_2}.\label{cosetPHequivalence}\end{align} Here $\overline{SO(n)_2}$ is the time-reversal conjugate (or spatial reflection) of $SO(n)_2$ that hosts edge modes in the opposite direction and anyons with reversed spins $h\to-h$. The relative tensor product $\boxtimes$ treats the $\mathbb{Z}_2$ charge pair $\zeta\bar{S}$ as an integral combination that ``condenses'' in the vacuum sector in the anyon condensation picture~\cite{BaisSlingerlandCondensation,Kong14,NeupertHeKeyserlingkSierraBernevig16,Burnell18}, where $\zeta=\psi^A\psi^B$ is the fermion pair from $[SO(n)_1]^2$ (see section~\ref{sec:SO2nZ2TL}) and $\bar{S}$ is the $\mathbb{Z}_2$ charge of $\overline{SO(n)_2}$. In addition, when $n$ is even, the bosonic spinor product $s_\pm^As_\pm^B\overline{\Psi^{n/2}_\pm}$ are also ``condensed'' in $\boxtimes$, where $s^As^B$ are spinor pairs in $[SO(2n)_1]^2$ and $\overline{\Psi^{n/2}_\pm}$ are defined in \eqref{SUnZ2spinors}. (A more detailed description of the \UMTC equivalence \eqref{cosetPHequivalence} in the anyon condensation picture can be found in appendix~\ref{app:orbifoldanyons}.) 
 Eq.\eqref{cosetPHequivalence} demonstrates a generalized notion of ``particle-hole'' conjugation. The $U(1)_l/\mathbb{Z}_2$ state can be regarded as the particle-hole conjugate of $SU(n)_1/\mathbb{Z}_2$, where ``holes'' that occupy a $SO(n)_2$ state are subtracted from a ``filled'' $[SO(n)_1]^2$ state. Similar particle-hole relation also applies to the Abelian unitary family described in section~\ref{sec:unitaryfamily} where $SU(n)_1$ and $U(1)_l$ are conjugate dual so that one can be obtained from subtracting the other from $SO(2n)_1$. This is analogous to the anti-Pfaffian fractional quantum Hall state~\cite{LevinHalperinRosenow07,LeeRyuNayakFisher07}, which is the particle-hole conjugate of the Moore-Read Pfaffian state~\cite{MooreRead,GreiterWenWilczekPRL91} and is obtained by subtracting from the filled Landau level, holes that occupy the Moore-Read Pfaffian state. Here, $[SO(n)_1]^2$ for the orbifold theories (or $SO(2n)_1$ for the Abelian unitary family) takes the role of the filled Landau level and is considered to be the base of the ``particle-hole'' conjugation.

Lastly, we conclude this subsection by commenting on alternative coupled-wire models where $SO(n)_2=SU(n)_1/\mathbb{Z}_2$ and $[SO(n)_1]^2/SO(n)_2=U(1)_l/\mathbb{Z}_2$ are both back-scattered within a wire or in-between wires. Similar to \eqref{trivialH1}, a trivial topological phase of decoupled wires is obtained when all degrees of freedom are gapped by complete intra-wire interactions, \begin{align}\mathcal{H}_{\mathrm{trivial}}=\mathcal{H}_0+\sum_{y=1}^{\mathsf{L}}\left(\mathcal{U}^y_{U(1)_l/\mathbb{Z}_2}+\mathcal{U}^y_{SO(n)_2}\right).\label{trivialH2}\end{align} Here, $\mathcal{H}_0$ is the kinectic Hamiltonian \eqref{kineticHSOnxSOn} of the $[SO(2n)_1]^2$ wires, and $\mathcal{U}^y_{U(1)_l/\mathbb{Z}_2}$ and $\mathcal{U}^y_{SO(n)_2}$ are given in \eqref{U1sineGordon2} and \eqref{SOn2intra} respectively. On the other hand, a non-trivial topological phase can be constructed using solely inter-wire interactions (c.f.~\eqref{HU1SUncond}) \begin{align}\mathcal{H}_{[SO(n)_1]^2}=\mathcal{H}_0+\sum_{y=1}^{\mathsf{L}}\left(\mathcal{U}^{y+1/2}_{SO(n)_2}+\mathcal{U}^{y+1/2}_{U(1)_l/\mathbb{Z}_2}\right),\label{HU1Z2SOn2cond}\end{align} where $\mathcal{U}^{y+1/2}_{SO(n)_2}$ and $\mathcal{U}^{y+1/2}_{U(1)_l/\mathbb{Z}_2}$ are presented in \eqref{SOn2int} and \eqref{U1sineGordon3}. It does not carry the topological order of a decoupled product $SO(n)_2\times U(1)_l/\mathbb{Z}_2$. This is because there are product fields, $\Phi^2\times\mathcal{E}^{2q}$ and its higher powers, that are integral electronic combinations and belong to the vacuum sector. (See \eqref{SUnsupersectors} and \eqref{U1supersectors} for the anyon definitions and compare with the anyon pair in \eqref{Zkcharges0}.) Therefore, the anyon pair splits into $\Phi^2\times\mathcal{E}^{2q}=1+$(confined components) and is ``condensed'' in the anyon condensation picture. The resulting \UMTC of deconfined anyons is identical to the $[SO(n)_1]^2$ topological order.

\subsection{The \texorpdfstring{$SO(2n)_1/D_k$}{SO(2n)/Dk} twist liquid}\label{SO2nDkTL}
In section~\ref{sec:gaugingZkZ2}, we defined the global dihedral $D_k=\mathbb{Z}_2\ltimes\mathbb{Z}_k$ symmetry of $SO(2n)_1$ in \eqref{Dksymm}, where $k=n$ ($k=n/2$) for $n$ odd (even). We demonstrated the gauging of the cyclic $\mathbb{Z}_k$ symmetry leads to the twist liquid orbifold phase $SO(2n)_1/\mathbb{Z}_k=U(1)_l\times SU(n)_1$, where $l=4n$ ($l=n$) for $n$ odd (even). The construction was based on altering the notion of locality so that only the current operators in the $\mathbb{Z}_k$-invariant $U(1)_l\times SU(n)_1$ \WZW sub-algebra in $SO(2n)_1$ were integral electronic combinations. In this subsection, we construct the $D_k$ twist liquid orbifold phase \begin{align}\frac{SO(2n)_1}{D_k}=\frac{U(1)_l\times SU(n)_1}{\mathbb{Z}_2}\label{SO2nDk}\end{align} by further gauging the conjugation symmetry, which corresponds to the quotient group $\mathbb{Z}_2=D_k/\mathbb{Z}_k$. \begin{equation}\begin{tikzcd}SO(2n)_1\arrow[r,"\mathbb{Z}_k","\mathrm{gauging}"']&SO(2n)_1/\mathbb{Z}_k\arrow[r,"\mathbb{Z}_2","\mathrm{gauging}"']&SO(2n)_1/D_k\end{tikzcd}\label{DkTL}\end{equation} The local fields in $SO(2n)_1/D_k$ are the $D_k$-invariant local fields in $SO(2n)_1$, or equivalently, the $\mathbb{Z}_2$-invariant local fields in $SO(2n)_1/\mathbb{Z}_k=U(1)_l\times SU(n)_1$ (recall the conjugation action \eqref{globalZ2}). They are generated by (1) the $SO(n)_2$ \WZW currents, which are the self-conjugate $SU(n)_1$ currents, $\cos\left(\phi_a-\phi_b\right)$, for $1\leq a<b\leq n$ (see \eqref{SOn2currents}), (2) the $\mathbb{Z}_2$ even local fields in $U(1)_l/\mathbb{Z}_2$, $\cos\left(q\phi_\perp\right)$, $\left(\partial_x\phi_\perp\right)^2$, $\left[\sin\left(q\phi_\perp\right)\right]^2$, and $\partial_x\phi_\perp\sin\left(q\phi_\perp\right)$ (see below \eqref{U1sineGordon3}), as well as (3) the pairs of $\mathbb{Z}_2$ odd local fields in $U(1)_l$ and $SU(n)_1$,  $\partial\phi_\perp\partial\tilde\phi_p$, $\partial\phi_\perp\sin\left(\phi_a-\phi_b\right)$, $\sin\left(q\phi_\perp\right)\partial\tilde\phi_p$, and $\sin\left(q\phi_\perp\right)\sin\left(\phi_a-\phi_b\right)$. If the local fields were only generated by sets (1) and (2), the $SO(2n)_1/D_k$ theory would be identical to the decoupled product $U(1)/\mathbb{Z}_2\times SU(n)_1/\mathbb{Z}_2$. Local fields in set (3) are pairs of $\mathbb{Z}_2$ charges from both $U(1)_l/\mathbb{Z}_2$ and $SU(n)_1/\mathbb{Z}_2$. Therefore, the $D_k$ twist liquid \eqref{SO2nDk}, as an \hyperlink{UMTC}{UMTC}, is identical to the relative tensor product $U(1)/\mathbb{Z}_2\boxtimes SU(n)_1/\mathbb{Z}_2$, where the $\mathbb{Z}_2$ charge pair $S_{U(1)/\mathbb{Z}_2}S_{SU(n)_1/\mathbb{Z}_2}$ is ``condensed'' in the anyon condensation~\cite{BaisSlingerlandCondensation,Kong14,NeupertHeKeyserlingkSierraBernevig16,Burnell18} context. This notion of locality requires the $\mathbb{Z}_2$ fluxes from $U(1)_l/\mathbb{Z}_2$ and $SU(n)_1/\mathbb{Z}_2$ to appear simultaneously. In other words, the $\mathbb{Z}_2$ local symmetry applies simultaneously on both $U(1)_l$ and $SU(n)_1$ in \eqref{SO2nDk}.

The coupled-wire model, that respects this notion of locality, is constructed by an array of $SO(2n)_1\times SO(2n)_1$ wires. (See section~\ref{sec:bosonwires} for the electronic spin liquid and superconductor origins.) In low-energy, any given wire at $y$ carries the chiral Majorana fermions $\psi^{A\sigma}_{ya}$ and $\psi^{B\sigma}_{ya}$, where $C=A,B$ distinguishes the two $SO(2n)_1$ sectors, $a=1,\ldots,2n$, and $\sigma=L,R=+,-$. Like the previous orbifold phases, only fermion pairs $\psi^{C\sigma}_{ya}\psi^{C\sigma'}_{ya'}$ from the same $C=A,B$ sector on the same wire $y$ are integral combinations of electrons. The $SO(2n)_1/D_k$ twist liquid model is constructed by populating the $SU(n)_1/\mathbb{Z}_2$ Hamiltonian \eqref{HSUn2} with $\psi^{C\sigma}_{y,1},\ldots\psi^{C\sigma}_{y,n}$ and the $U(1)_l/\mathbb{Z}_2$ Hamiltonian \eqref{HU1l2} with $\psi^{C\sigma}_{y,n+1},\ldots\psi^{C\sigma}_{y,2n}$. \begin{align}&\mathcal{H}_{SO(2n)_1/D_k}\label{HSO2nDk}\\&=\mathcal{H}_{SU(n)_1/\mathbb{Z}_2}\left[\psi_1,\ldots\psi_n\right]+\mathcal{H}_{U(1)_l/\mathbb{Z}_2}\left[\psi_{n+1},\ldots\psi_{2n}\right]\nonumber\end{align} The local $\mathbb{Z}_2$ symmetry \begin{align}\mathbb{Z}_2^B(y):\quad\psi^A\to\psi^A,\quad\psi^B_{y'}\to(-1)^{\delta_{yy'}}\psi^B_{y'}\end{align} acts simultaneously on both $SU(n)_1/\mathbb{Z}_2$ and $U(1)_l/\mathbb{Z}_2$. The $\mathbb{Z}_2$ charge $S_{U(1)_l/\mathbb{Z}_2}\sim\psi^A_a\psi^B_b$ and $S_{SU(n)_1/\mathbb{Z}_2}\sim\psi^A_{n+a}\psi^B_{n+b}$, for $1\leq a<b\leq n$, are individually non-local and must come in pairs. On the other hand, the $\mathbb{Z}_2$ charge pair $S_{U(1)/\mathbb{Z}_2}S_{SU(n)_1/\mathbb{Z}_2}$ is integral because $\psi^A_a\psi^A_{n+a}$ and $\psi^B_b\psi^B_{n+b}$ are local bosons by construction. $S_{U(1)/\mathbb{Z}_2}$ and $S_{SU(n)_1/\mathbb{Z}_2}$ belong in the same anyon class $S$ that represents the $\mathbb{Z}_2$ gauge charge in $SO(2n)_1/D_k$. Although $\mathcal{H}_{SU(n)_1/\mathbb{Z}_2}$ and $\mathcal{H}_{U(1)_l/\mathbb{Z}_2}$ in \eqref{HSO2nDk} involve distinct sets of fermions, their combination is {\em not} a direct sum, unlike \eqref{HU1xSUn}. This is because the Hilbert space does not decompose thanks to the intertwined fermion pair locality.

The topological order of $SO(2n)_1/D_k$ is closely related to that of the discrete gauge theory of the dihedral group and its $\mathbb{Z}_2$ extensions. A review of discrete gauge theory can be found in appendix~\ref{app:DGT}. The relation depends on $n$ modulo 8 because of the eight-fold periodicity of the $SO(2n)_1$ \hyperlink{UMTC}{UMTC} and is summarized in \eqref{Dkorbifolds}. The $SO(2n)_1$ anyon classes form an Abelian group $\mathcal{A}=\{1,\psi,s_+,s_-\}$ under fusion product, where $\mathcal{A}=\mathbb{Z}_2\times\mathbb{Z}_2$ when $n$ is even, or $\mathbb{Z}_4$ when $n$ is odd. The fusion group $\mathcal{A}$ extends the global symmetry group $G=D_k$ of $SO(2n)_1$ in to a quantum symmetry group~\cite{Wenspinliquid02,EtingofNikshychOstrik10,TeoHughesFradkin15,BarkeshliBondersonChengWang14} $\widehat{G}$ (see \eqref{groupextension} and \eqref{appgroupextension}). Inequivalent group extensions are classified by the group cohomology~\cite{Cohomologybook} $H^2(G,\mathcal{A})$ (see appendix~\ref{app:DGT} and \ref{app:cohomology}). When $n=k$ is odd, $\widehat{G}=G\ltimes\mathcal{A}=D_k\ltimes\mathbb{Z}_4$ is ``symmorphic'' and corresponds to the trivial element in $H^2(D_k,\mathcal{A})$. The twofold conjugation symmetry in $D_k$ is an outer automorphism of $SO(2n)$. It acts non-trivially on $\mathcal{A}$ in the semi-direct product and flips between the even and odd spinor, $s_+\leftrightarrow s_-$. When $n$ is even, the $D_k$ symmetries are all inner automorphisms and act trivially on $\mathcal{A}$. However, the global quantum symmetry group $\widehat{G}$ is ``non-symmorphic'' and does not decompose in a direct product $G\times\mathcal{A}$. When $k=n/2$ is even, we saw in section~\ref{sec:SO2nZkTL} that the $\mathbb{Z}_k$ flux had order $2k$ and its $k^{\mathrm{th}}$ power was identified as the even spinor $m^k=s_+$ in $SO(2n)_1$. This extends $D_k$ into $D_{2k}$, and the quantum symmetry group is $\widehat{G}=D_{2k}\times\mathbb{Z}_2$. The non-symmorphic central extension corresponds to the non-trivial cohomology element $[s_\pm,1,1]$ (or $[\psi,1,1]$) in $H^2(D_k,\mathcal{A})=\mathcal{A}^3$ when $n\equiv0$ (resp.~$n\equiv4$) modulo 8. When $k=n/2$ is odd, we showed in section~\ref{sec:SO2nZ2TL} that the $\mathbb{Z}_2$ symmetry was extended to $\mathbb{Z}_4$ because the twofold flux squared to the $SO(2n)_1$ fermion $s^2_\pm=\psi$. This extends the global dihedral symmetry group $D_k$ into the dicyclic group $Q_{4k}$ (defined in \eqref{Dic} below), and the quantum symmetry group is $\widehat{G}=Q_{4k}\times\mathbb{Z}_2$. The non-symmorphic central extension corresponds to a non-trivial cohomology element $[\psi]$ in $H^2(D_k,\mathcal{A})=\mathcal{A}$.

We first present the topological order of $SO(2n)_1/D_k$ when $n\equiv0$ modulo 8. From \eqref{Zkorbifolds} in section \ref{sec:gaugingZkZ2}, we saw that $SO(2n)_1/\mathbb{Z}_k=U(1)_l\times SU(n)_1$ is equivalent, as an \hyperlink{UMTC}{UMTC}, to the $D^{[k]}(\mathbb{Z}_{2k})$ discrete gauge theory, where $k=n/2$. Gauging the $\mathbb{Z}_2$ conjugation symmetry leads to the $D^{[k,0,0]}(D_{2k})$  discrete gauge theory, where \begin{align}D_{2k}&=\mathbb{Z}_2\ltimes\mathbb{Z}_{2k}=\left\{\mu^Am^a:A=0,1,a=-k+1,\ldots,k\right\}\nonumber\\&=\left\langle m,\mu|m^{2k}=\mu^2=(\mu m)^2=1\right\rangle\label{D2kgroup}\end{align} is the dihedral group that contains $\mathbb{Z}_{2k}$ as a normal subgroup and $\mathbb{Z}_2=D_{2k}/\mathbb{Z}_{2k}$. Since the twist liquid descends from $SO(2n)_1$, which has the identical topological order as the $D^{[0]}(\mathbb{Z}_2)$ discrete gauge theory, the sequence of gauging \eqref{DkTL} is the quantum double~\cite{Kasselbook,BakalovKirillovlecturenotes,LevinWen05} (or Drinfeld center) of the group filtration \begin{equation}\begin{tikzcd}[column sep=large]\mathbb{Z}_2\arrow[r,hookrightarrow,"\triangleleft"]\arrow[d,Rightarrow]&\mathbb{Z}_{2k}\arrow[r,hookrightarrow,"\triangleleft"]\arrow[d,Rightarrow]&D_{2k}\arrow[d,Rightarrow]\\D^{[0]}(\mathbb{Z}_2)\arrow[r,"\mathbb{Z}_k=\mathbb{Z}_{2k}/\mathbb{Z}_2","\mathrm{gauging}"']&D^{[k]}(\mathbb{Z}_{2k})\arrow[r,"\mathbb{Z}_2=D_{2k}/\mathbb{Z}_{2k}","\mathrm{gauging}"']&D^{[k,0,0]}(D_{2k})\end{tikzcd}.\end{equation} Here, $[k,0,0]$ represents the Dijkgraaf-Witten invariant in $H^3(D_{2k},U(1))=\mathbb{Z}_{2k}\times\mathbb{Z}_2\times\mathbb{Z}_2$ that deforms the gauge theory~\cite{DijkgraafWitten90,DijkgraafPasquierRoche91,AltschulerCoste92,BaisvanDrielPropitius93,Propitius-1995} (see appendix~\ref{app:DGT} for the cohomology classification). The $\mathbb{Z}_{2k}$ component comes from the cohomology classification $H^3(\mathbb{Z}_{2k},U(1))=\mathbb{Z}_{2k}$ of the normal subgroup $\mathbb{Z}_{2k}$ of $D_{2k}$. Each of the two $\mathbb{Z}_2$'s corresponds to the cohomology classification $H^3(\mathbb{Z}_2,U(1))=\mathbb{Z}_2$ for one of the two conjugacy classes $[\mu]$ and $[\mu m]$ of twofold conjugation symmetries. The trivial components in $[k,0,0]$ signifies that the $\mathbb{Z}_2$ fluxes -- which are pair combinations of the $\sigma_\pm$ or $\tau_\pm$ from $U(1)_l/\mathbb{Z}_2$ and $SU(n)_1/\mathbb{Z}_2$ -- are bosonic or fermionic (see table~\ref{SOn2anyonseven} and \ref{Uorbifoldanyons}). 

Next, we consider $n\equiv4$ modulo 8. From \eqref{Zkorbifolds}, we saw that $SO(2n)_1/\mathbb{Z}_k=U(1)_l\times SU(n)_1$ is equivalent, as an \hyperlink{UMTC}{UMTC}, to $SO(8)_1\boxtimes_{\psi\bar\psi}D^{[k]}(\mathbb{Z}_{2k})$, where $k=n/2$ and the relative tensor product $\boxtimes$ involves the anyon condensation of a fermion pair $\psi\bar\psi$ (see section~\ref{sec:SO2nZkTL}). The $\mathbb{Z}_2$ conjugation action applies simultaneously on $SO(8)_1$ and $D^{[k]}(\mathbb{Z}_{2k})$. After gauging, the former becomes $SO(8)_1/\mathbb{Z}_2=SO(8)_1\times D^{[1]}(\mathbb{Z}_2)$ (see \eqref{Z2orbifoldseven}) and the latter becomes $D^{[k,0,0]}(D_{2k})$. The simultaneous $\mathbb{Z}_2$ action requires the $\mathbb{Z}_2$ gauge charges $\zeta$ from $D^{[1]}(\mathbb{Z}_2)$ and $S$ from $D^{[k,0,0]}(D_{2k})$ to pair ``condense'' in the vacuum sector in the anyon condensation picture. This identifies \begin{align}D^{[1]}(\mathbb{Z}_2)\boxtimes_{\zeta S}D^{[k,0,0]}(D_{2k})=D^{[k,1,1]}(D_{2k})\end{align} and modifies the Dijkgraaf-Witten invariant into $[k,1,1]$ (see appendix~\ref{app:DGT}) so that the $\mathbb{Z}_2$ fluxes -- which are pair combinations of the $\sigma_\pm$ or $\tau_\pm$ from $U(1)_l/\mathbb{Z}_2$ and $SU(n)_1/\mathbb{Z}_2$ -- are all semionic with spins $h=\pm1/4$ modulo 1 (see table~\ref{SOn2anyonseven} and \ref{Uorbifoldanyons}). Combining with the remaining $SO(8)_1$, the $SO(2n)_1/D_k$ twist liquid is identical, as an \hyperlink{UMTC}{UMTC}, to the relative tensor product \begin{align}\frac{SO(2n)_1}{D_k}&=\frac{SO(2n)_1/\mathbb{Z}_k}{\mathbb{Z}_2}=\frac{SO(8)_1\boxtimes_{\psi\bar\psi}D^{[k]}(\mathbb{Z}_{2k})}{\mathbb{Z}_2}\nonumber\\&=\frac{SO(8)_1}{\mathbb{Z}_2}\boxtimes_{\psi\bar\psi,\zeta S}\frac{D^{[k]}(\mathbb{Z}_{2k})}{\mathbb{Z}_2}\\&=\left(SO(8)_1\times D^{[1]}(\mathbb{Z}_2)\right)\boxtimes_{\psi\bar\psi,\zeta S}D^{[k,0,0]}(D_{2k})\nonumber\\&=SO(8)_1\boxtimes_{\psi\bar\psi}D^{[k,1,1]}(D_{2k}),\quad\mbox{if $n\equiv4$ mod 8},\nonumber\end{align} where the vector fermion $\psi$ in $SO(8)_1$ is pair condensed with the Abelian fermionic flux $\bar\psi=m^{n/2}$ in $D^{[k,1,1]}(D_{2k})$ (c.f.~the discussion above \eqref{Zkorbifolds}). The anyon content is almost identical to $D^{[k,1,1]}(D_{2k})$. The anyons are represented by a dyon $\chi$ in $D^{[k,1,1]}(D_{2k})$ if $\chi$ and $m^{n/2}$ braid trivially, or the spinor-dyon composite $s_+\chi$ if $\chi$ and $m^{n/2}$ have $\pi$ monodromy, where $s_+$ is the even spinor in $SO(8)_1$. Therefore, the relative tensor product with $SO(8)_1$ only modifies the topological spins of anyons in $D^{[k,1,1]}(D_{2k})$, but leaves the fusion rules and modular $S$ matrix of $D^{[k,1,1]}(D_{2k})$ unaltered. 

Now, we examine the cases when $n\equiv2$ modulo 4. From \eqref{Zkorbifolds}, we saw that $SO(2n)_1/\mathbb{Z}_k=U(1)_l\times SU(n)_1$ is equivalent, as an \hyperlink{UMTC}{UMTC}, to the decoupled product $SO(2n)_1\times D^{[0]}(\mathbb{Z}_k)$, where $k=n/2$ is odd. From \eqref{Z2orbifoldseven}, $SO(2n)_1/\mathbb{Z}_2=SO(2n)_1\boxtimes_{\psi\bar\psi}D^{[2]}(\mathbb{Z}_4)$, where the vector fermion $\psi$ in $SO(2n)_1$ is pair condensed with the fermionic gauge flux $\bar\psi=\bar{s}^2$ in the relative tensor product $\boxtimes_{\psi\bar\psi}$ (see \eqref{Z2quotient}). The $\mathbb{Z}_2$ gauging of $D^{[0]}(\mathbb{Z}_k)$ produces the un-deformed discrete gauge theory $D^{[0]}(D_k)$ with the trivial Dijkgraaf-Witten invariant $[0]$ in $H^3(D_k,U(1))=\mathbb{Z}_{2k}$. Since $\mathbb{Z}_2$ conjugation acts simultaneously on $SO(2n)_1$ and $D^{[0]}(\mathbb{Z}_k)$, the $\mathbb{Z}_2$ charge pair $\zeta S$ from $D^{[2]}(\mathbb{Z}_4)$ and $D^{[0]}(D_k)$ is local and belong in the vacuum sector in the anyon condensation picture. This combines them into the discrete gauge theory \begin{align}D^{[2k]}(Q_{4k})=D^{[2]}(\mathbb{Z}_4)\boxtimes_{\zeta S}D^{[0]}(D_k)\end{align} of the dicyclic group \begin{align}Q_{4k}&=\left\{\tilde\mu^A\tilde{m}^a:A=0,1,a=-k+1,\ldots,k\right\}\label{Dic}\\&=\left\langle\tilde{m},\tilde\mu|\tilde{m}^{2k}=\tilde\mu^4=1,\tilde{m}^k=\tilde\mu^2,\tilde\mu^{-1}\tilde{m}\tilde\mu=\tilde{m}^{-1}\right\rangle.\nonumber\end{align} (Alternatively, the dicyclic group \eqref{Dic} is also the double cover $\tilde{D}_k=Q_{4k}$ of the dihedral group $D_k$, where $\tilde{m}$ and $\tilde\mu$ are represented by perpendicular rotations $e^{i(2\pi/k)S_z}$ and $e^{i\pi S_x}$, for $S_{x,z}=\sigma_{x,z}/2$, in a half-integral spin representation, and $\tilde{m}^k=\tilde\mu^2=-1$ are $2\pi$ rotations. Therefore, $Q_{4k}$ is a subgroup of the unit quaternions $SU(2)=\{a_0+a_1{\bf i}+a_2{\bf j}+a_3{\bf k}:|{\bf a}|=1\}$, where ${\bf i}$, ${\bf j}$, ${\bf k}$ can be represented by $i\sigma_x$, $i\sigma_y$, $i\sigma_z$.) Since we only encounter the dicyclic group with odd degree $k$, \eqref{Dic} has the equivalent presentation \begin{align}Q_{4k}&=\mathbb{Z}_4\ltimes\mathbb{Z}_k\label{Dicodd}\\&=\left\{\tilde\mu^Am^a:\begin{array}{*{20}l}A=-1,0,1,2,\\a=-(k-1)/2,\ldots,(k-1)/2\end{array}\right\}\nonumber\\&=\left\langle m,\tilde\mu|m^k=\tilde\mu^4=\tilde\mu m\tilde\mu^{-1}m=1\right\rangle.\nonumber\end{align} Here, both the $\mathbb{Z}_4$ flux $[\bar{s}]$ in $D^{[2]}(\mathbb{Z}_4)$ and the $\mathbb{Z}_2$ flux $[\mu]$ in $D^{[0]}(D_k)$ individually carries the $\pi$ monodromy braiding phase with the $\mathbb{Z}_2$ charge pair $\zeta S$, and therefore are separately confined. When appear simultaneously, they combines into the fourfold gauge flux $[\tilde\mu]=[\bar{s}][\mu]$. We associate $[\tilde{m}]=[\bar{s}^2][m]$ to the combination of fermion $\bar\psi=\bar{s}^2$ in $D^{[2]}(\mathbb{Z}_4)$ and the $\mathbb{Z}_k$ flux $[m]$ in $D^{[0]}(D_k)$. Since $k$ is odd and $\bar{s}^4=1$, $\tilde{m}^k=\bar{s}^2$, which can be identified as $\tilde\mu^2$ because $\mu^2=1$ as a group element in $D_k$. The discrete gauge theory $D^{[2k]}(Q_{4k})$ is deformed by the Dijkgraaf-Witten cohomology invariant $[2k]$ in $H^3(Q_{4k},U(1))=\mathbb{Z}_{4k}$ (see appendix~\ref{app:DGT}). The deformation requires the spins of the fourfold fluxes to be $\pm1/8$ and $\pm3/8$ modulo 1. Combining with the remaining $SO(2n)_1$, the $SO(2n)_1/D_k$ twist liquid is identical, as an \hyperlink{UMTC}{UMTC}, to the relative tensor product \begin{align}\frac{SO(2n)_1}{D_k}&=\frac{SO(2n)_1/\mathbb{Z}_k}{\mathbb{Z}_2}=\frac{SO(2n)_1\times D^{[0]}(\mathbb{Z}_k)}{\mathbb{Z}_2}\nonumber\\&=\frac{SO(2n)_1}{\mathbb{Z}_2}\boxtimes_{\zeta S}\frac{D^{[0]}(\mathbb{Z}_k)}{\mathbb{Z}_2}\\&=SO(2n)_1\boxtimes_{\psi\bar\psi}D^{[2]}(\mathbb{Z}_4)\boxtimes_{\zeta S}D^{[0]}(D_k)\nonumber\\&=SO(2n)_1\boxtimes_{\psi\bar\psi}D^{[2k]}(Q_{4k}),\quad\mbox{if $n\equiv2$ mod 4},\nonumber\end{align} where the vector fermion $\psi$ in $SO(2n)_1$ is pair ``condensed'' with the Abelian fermionic flux $\bar\psi=\tilde\mu^2$ in $D^{[2k]}(Q_{4k})$. The ``condensate'' $\psi\bar\psi$ forces the semionic spinor $s_\pm$ in $SO(2n)_1$ to appear simultaneously with one of the $\mathbb{Z}_2$ fluxes in $D^{[2k]}(Q_{4k})$. The $\mathbb{Z}_2$ flux-spinor composites have spins $n/16$ or $(n+8)/16$ modulo 1. This matches with the spins of the pair combination of the $\mathbb{Z}_2$ twist fields, $\sigma_\pm$ or $\tau_\pm$, in $SU(n)_1/\mathbb{Z}_2$ and $U(1)_l/\mathbb{Z}_2$ (see table~\ref{SOn2anyonseven} and \ref{Uorbifoldanyons}). 

Lastly, we present the $SO(2n)_1/D_k$ topological order when $n=k$ is odd. Eq.\eqref{Zkorbifolds} identifies $SO(2n)_1/\mathbb{Z}_k=SO(2n)_1\times D^{[0]}(\mathbb{Z}_k)$. Like the previous cases, the $\mathbb{Z}_2$ conjugation symmetry acts simultaneously on $SO(2n)_1$ and $D^{[0]}(\mathbb{Z}_k)$. The $\mathbb{Z}_2$ gauging of latter produces $D^{[0]}(D_k)$. From \eqref{Z2orbifoldsodd}, the $\mathbb{Z}_2$ gauging of the former gives $SO(2n)_1\boxtimes_{\psi\bar\psi}Z(\mathrm{Ising})$ for $n\equiv\pm1$ modulo 8, or $SO(2n)_1\boxtimes_{\psi\bar\psi}Z(SU(2)_2)$ for $n\equiv\pm3$ modulo 8. (See below \eqref{Z2orbifoldsodd} for the details of fermion pair condensation.) The quantum doubles $Z(\mathrm{Ising})$ and $Z(SU(2)_2)$ are not discrete gauge theories because they support Ising anyons with non-integral quantum dimension $\sqrt{2}$. This stems from the outer automorphism nature of the $\mathbb{Z}_2$ conjugation, which switches the even and odd spinors, $s_+\leftrightarrow s_-$, in $SO(2n)_1$, when $n$ is odd. The $SO(2n)_1/D_k$ topological order therefore carries both Ising and discrete gauge theory components \begin{align}\frac{SO(2n)_1}{D_k}&=SO(2n)_1\boxtimes_{\psi\bar\psi}\mathcal{Z}_n(D_k)\label{DkTLoddn}
\end{align} where the quantum double \begin{align}\begin{split}\mathcal{Z}_n(D_k)&=D^{[0]}(\mathbb{Z}_{2k})\sslash\mathbb{Z}_2\\&=\mathcal{Z}_n(\mathbb{Z}_2)\boxtimes_{\zeta S}D^{[0]}(D_k)\end{split}\label{Isingfluxonphase}\end{align} is the non-chiral \UMTC \begin{align}\begin{split}&Z(\mathrm{Ising})\boxtimes_{\zeta S}D^{[0]}(D_k),\quad\mbox{if $n\equiv\pm1$ mod 8},\\&Z(SU(2)_2)\boxtimes_{\zeta S}D^{[0]}(D_k),\quad\mbox{if $n\equiv\pm3$ mod 8}.\end{split}\label{DZkmZ2}\end{align} Like previous cases, the relative tensor product $\boxtimes_{\zeta S}$ anyon ``condenses'' the $\mathbb{Z}_2$ charge pair $\zeta S$, where $\zeta$ is the fermion pair in $Z(\mathrm{Ising})=\mathrm{Ising}\times\overline{\mathrm{Ising}}$ or $Z(SU(2)_2)=SU(2)_2\times\overline{SU(2)_2}$ and $S=([1],A_1)$ is the $\mathbb{Z}_2$ charge in $D^{[0]}(D_k)$. Consequently, any Ising anyon (such as $\sigma$ and $\bar\sigma$) with quantum dimension $\sqrt{2}$ in $Z(\mathrm{Ising})$ or $Z(SU(2)_2)$ must appear simultaneously with a $\mathbb{Z}_2$ flux $([\mu],\zeta^\lambda)$, for $\lambda=0,1$, in $D^{[0]}(D_k)$. (See table~\ref{tab:dyonsDkodd} for the list of dyons in the $D^{[0]}(D_k)$ discrete gauge theory.) These ``Ising-fluxon''  composites \begin{align}\begin{split}&\begin{array}{*{20}l}\Gamma^\lambda=\sigma\boxtimes([\mu],\zeta^\lambda),\\\overline{\Gamma}^\lambda=\bar\sigma\boxtimes([\mu],\zeta^\lambda),\end{array}\;\mbox{if $n\equiv1,3$ mod 16}\\&\begin{array}{*{20}l}\Gamma^\lambda=\bar\sigma\boxtimes([\mu],\zeta^{1-\lambda}),\\\overline{\Gamma}^\lambda=\sigma\boxtimes([\mu],\zeta^{1-\lambda}),\end{array}\;\mbox{if $n\equiv5,7$ mod 16}\\&\begin{array}{*{20}l}\Gamma^\lambda=\bar\sigma\boxtimes([\mu],\zeta^\lambda),\\\overline{\Gamma}^\lambda=\sigma\boxtimes([\mu],\zeta^\lambda),\end{array}\;\mbox{if $n\equiv-1,-3$ mod 16}\\&\begin{array}{*{20}l}\Gamma^\lambda=\sigma\boxtimes([\mu],\zeta^{1-\lambda}),\\\overline{\Gamma}^\lambda=\bar\sigma\boxtimes([\mu],\zeta^{1-\lambda}),\end{array}\;\mbox{if $n\equiv-5,-7$ mod 16}\end{split}\end{align} in $\mathcal{Z}_n(D_k)$ have quantum dimension $\sqrt{2}n$ and spins $h_{\Gamma^\lambda}=k/16+\lambda/2$, $h_{\overline{\Gamma}^\lambda}=-k/16+\lambda/2$ modulo 1. $\Gamma^0$ and $\Gamma^1=S\times\Gamma^0$ have trivial statistics with the condensate $\psi\bar\psi=\psi_{SO(2n)_1}\bar\psi_{\mathcal{Z}_n(D_k)}$ in \eqref{DkTLoddn} and are deconfined. Here, similar to \eqref{Z2orbifoldsodd}, $\bar\psi_{\mathcal{Z}_n(D_k)}$ is the fermion in $\overline{\mathrm{Ising}}$ when $n\equiv1$ mod 8, in $\overline{SU(2)_2}$ when $n\equiv3$ mod 8, in $\mathrm{Ising}$ when $n\equiv-1$ mod 8, or in $SU(2)_2$ when $n\equiv3$ mod 8. On the other hand, $\overline{\Gamma}^0$ and $\overline{\Gamma}^1=S\times\overline{\Gamma}^0$ have semionic mutual statistics with $\psi\bar\psi$. Therefore, they must appear simultaneously with a spinor $s_\pm$ in $SO(2n)_1$ in \eqref{DkTLoddn} to respect locality of $\psi\bar\psi$. This modifies the spins of the Ising-fluxons to $h_{\Gamma^\lambda}=h_{s_\pm\overline{\Gamma}^\lambda}=k/16+\lambda/2$ (mod 1) in $SO(2n)_1/D_k$. The quantum dimensions and spins of these Ising-fluxons match the pair combination of the $\mathbb{Z}_2$ twist fields, $\sigma$ and $\tau$, in $U(1)_l/\mathbb{Z}_2$ and $SU(n)_1/\mathbb{Z}_2$ (see table~\ref{SOn2anyonsodd} and \ref{Uorbifoldanyons}). They can be identified according to \begin{align}\begin{split}\Gamma^0&=\sigma_{SU(n)_1/\mathbb{Z}_2}\sigma^+_{U(1)_l/\mathbb{Z}_2}\\\Gamma^1&=\tau_{SU(n)_1/\mathbb{Z}_2}\sigma^+_{U(1)_l/\mathbb{Z}_2}\\s_+\overline{\Gamma}^0&=\sigma_{SU(n)_1/\mathbb{Z}_2}\sigma^-_{U(1)_l/\mathbb{Z}_2}\\s_+\overline{\Gamma}^1&=\tau_{SU(n)_1/\mathbb{Z}_2}\sigma^-_{U(1)_l/\mathbb{Z}_2}\end{split}\end{align} (up to the $\mathbb{Z}_2$ charge pair condensate $S_{SU(n)_1/\mathbb{Z}_2}S_{U(1)_l/\mathbb{Z}_2}$ and the fermion pair condensate $\psi_{SO(2n)_1}\bar\psi_{\mathcal{Z}_n(D_k)}$). 

We observe that the quantum double $\mathcal{Z}_n(D_k)$ in \eqref{Isingfluxonphase} is the $\mathbb{Z}_2$ twist liquid of the discrete gauge theory $D^{[0]}(\mathbb{Z}_{2k})$ after gauging an unconventional twofold {\em mixed} symmetry (see \eqref{unconventionalZ2}) that is neither the electric-magnetic symmetry nor the conjugation symmetry. It is a known result~\cite{BarkeshliWen12,TeoHughesFradkin15,ChenRoyTeoRyu17} that the quantum doubles $Z(\mathrm{Ising})$ and $Z(SU(2)_2)$ are $\mathbb{Z}_2$ twist liquids when the electric-magnetic (e-m) symmetry in $D^{[0]}(\mathbb{Z}_2)$ is gauged. The $D^{[0]}(\mathbb{Z}_2)$ discrete gauge theory can be described by a two-component Chern-Simons field theory $\mathcal{S}=\int_{2+1}K_{IJ}a^I\wedge da^J/(4\pi)$, where $K=2\sigma_x$. It exhibits the $\mathbb{Z}_2$ anyon relabeling symmetry~\cite{khan2014,Teotwistdefectreview} $M=\sigma_x$, which exchanges the $\mathbb{Z}_2$ charge and flux (represented by the unit vectors ${\bf e}_1$ and ${\bf e}_2$) and leaves the $K$ matrix invariant, $MKM^T=K$. Gauging the symmetry gives the quantum double \begin{align}\begin{split}\mathcal{Z}_n(\mathbb{Z}_2)&=D^{[0]}(\mathbb{Z}_2)\sslash\mathbb{Z}_2\\&=Z(\mathrm{Ising})\quad\mbox{or}\quad Z(SU(2)_2)\end{split}\label{DZ2mZ2}\end{align} depending on the Frobenius-Schur indicator~\cite{Kitaev06} $\varkappa_\sigma=d_\sigma\left[F^{\sigma\sigma\sigma}_\sigma\right]^1_1=(-1)^{(n^2-1)/8}$ of the $\mathbb{Z}_2$ twist field (see discussion above \eqref{Z2orbifoldsodd}). Here, the double quotient ``$\sslash$'' emphasizes the non-trivial flux-charge switching nature of the $\mathbb{Z}_2$ symmetry, and consequently, the non-gauge theory topological order of $D^{[0]}(\mathbb{Z}_2)\sslash\mathbb{Z}_2$. [Had the $\mathbb{Z}_2$ symmetry preserved the anyon types, the gauging would only extend the gauge group and $D^{[0]}(\mathbb{Z}_2)/\mathbb{Z}_2$ would be a discrete gauge theory $D^{[\omega]}(\mathbb{Z}_4)$ or $D^{[\eta]}(\mathbb{Z}_2\times\mathbb{Z}_2)$.] 

With the help of \eqref{DZ2mZ2}, equation \eqref{Isingfluxonphase} can be re-expressed as \begin{align}&\Big(D^{[0]}(\mathbb{Z}_2)\sslash\mathbb{Z}_2\Big)\boxtimes_{\zeta S}D^{[0]}(D_k)\nonumber\\&=\Big(D^{[0]}(\mathbb{Z}_2)\sslash\mathbb{Z}_2\Big)\boxtimes_{\zeta S}\frac{D^{[0]}(\mathbb{Z}_k)}{\mathbb{Z}_2}\nonumber\\&=\Big(D^{[0]}(\mathbb{Z}_2)\times D^{[0]}(\mathbb{Z}_k)\Big)\sslash\mathbb{Z}_2.\end{align} Here, the product theory $D^{[0]}(\mathbb{Z}_2)\times D^{[0]}(\mathbb{Z}_k)$ has a four-component Chern-Simons description $\mathcal{S}=\int_{2+1}K_{IJ}a^I\wedge da^J/(4\pi)$, the $K$-matrix is the direct sum $K=(2\sigma_x)\oplus(k\sigma_x)$. The $\mathbb{Z}_2$ symmetry acts simultaneously on $D^{[0]}(\mathbb{Z}_2)$ and $D^{[0]}(\mathbb{Z}_k)$ and can be represented by the matrix $M=\sigma_x\oplus(-\mathbb{I})$, which combines the e-m symmetry in $D^{[0]}(\mathbb{Z}_2)$ and the conjugation symmetry in $D^{[0]}(\mathbb{Z}_k)$. Since $k$ is odd, the product $D^{[0]}(\mathbb{Z}_2)\times D^{[0]}(\mathbb{Z}_k)$ has identical topological order as the discrete gauge theory $D^{[0]}(\mathbb{Z}_{2k})$ with gauge group $\mathbb{Z}_{2k}=\mathbb{Z}_2\times\mathbb{Z}_k$. Labeling the $D^{[0]}(\mathbb{Z}_{2k})$ dyons according to their flux-charge content, $m^az^b$, where $a,b=0,1,\ldots,2k-1$ modulo $2k$, the twofold action of the {\em mixed} symmetry $\mathbb{Z}_2:m^az^b\to m^{a'}z^{b'}$ can be represented by the matrix transformation \begin{align}\begin{pmatrix}a\\b\end{pmatrix}\to\begin{pmatrix}a'\\b'\end{pmatrix}=U\begin{pmatrix}a\\b\end{pmatrix},\quad U=\begin{pmatrix}k-1&k\\k&k-1\end{pmatrix}.\label{unconventionalZ2}\end{align} The symmetry has order 2 because the flux-charge labels have $2k$ periodicity $a\equiv a+2k$, $b\equiv b+2k$, and $U$ squares to the identity matrix $\mathbb{I}$ modulo $2k$. The symmetry preserves the spin and braiding data of $D^{[0]}(\mathbb{Z}_{2k})$ because $U^TK^{-1}U\equiv K^{-1}$ modulo $2\mathbb{Z}$, where $K=2k\sigma_x$. Gauging the $\mathbb{Z}_2$ symmetry in $D^{[0]}(\mathbb{Z}_{2k})$ gives the $\mathbb{Z}_2$ twist liquids $\mathcal{Z}_n(D_k)=D^{[0]}(\mathbb{Z}_{2k})\sslash\mathbb{Z}_2$ in \eqref{Isingfluxonphase}, which we referred to as the ``Ising-fluxon'' phase. 
We conjecture in passing that this non-chiral Ising-fluxon topological order may be identical to the Drinfeld center of a (non-commutative) fusion category (see discussion in section~\ref{sec:quantumdoubles}).

Overall, we summarize the $SO(2n)_1/D_k$ twist liquid phases for general integer $n$ by the \UMTC equivalence \begin{align}&\frac{SO(2n)_1}{D_k}=\frac{U(1)_l\times SU(n)_1}{\mathbb{Z}_2}\label{Dkorbifolds}\\&=\left\{\begin{array}{*{20}l}D^{[k,0,0]}(D_{2k}),&\mbox{for $n\equiv0$ mod 8}\\SO(8)_1\boxtimes D^{[k,1,1]}(D_{2k}),&\mbox{for $n\equiv4$ mod 8}\\SO(2n)_1\boxtimes D^{[2k]}(Q_{4k}),&\mbox{for $n\equiv2$ mod 4}\\SO(2n)_1\boxtimes\mathcal{Z}_n(D_k),&\mbox{for $n$ odd}\end{array}\right.\nonumber\end{align} where $k=n$ and $l=4n$ when $n$ is odd, and $k=n/2$ and $l=n$ when $n$ is even. The $D_{2k}$ and $Q_{4k}$ groups are ``non-symmorphic'' $\mathbb{Z}_2$ central extensions of $D_k$ and associates to non-trivial cohomology elements in $H^2(D_k,\mathcal{A})$. The super-script $[\omega]$ in $D^{[\omega]}(G)$ labels the Dijkgraaf-Witten invariant in the cohomology group $H^3(G,U(1))$ that deforms the gauge theory. Similar to \eqref{Zkorbifolds}, the topological orders in the $n\equiv0$ or 4 modulo 8 cases in \eqref{Dkorbifolds} agrees with \eqref{innerautgauging} and admit the decomposition $SO(2n)_1\boxtimes D^{\omega}(D_{2k})$, for $\omega=[k,0,0]$ or $[k,1,1]$. This is because, as \hyperlink{UMTC}{UMTC}s, $SO(2n)_1=D^{[0]}(\mathbb{Z}_2)$ or $SO(8)_1$ when $n\equiv0$ or 4 modulo 8 respectively, and $D^{[0]}(\mathbb{Z}_2)\boxtimes D^{[\omega]}(D_{2k})=D^{[\omega]}(D_{2k})$ after condensing the $\mathbb{Z}_2$ charge pair.

\section{Conclusion and discussion}\label{sec:conclusion}

We constructed the 2+1D electronic coupled-wire models that describe the twist liquid topological phases $SO(2n)_1/G$ in the orthogonal family, for $G=\mathbb{Z}_2$, $\mathbb{Z}_k$ and $D_k$, as well as $SU(n)_1/\mathbb{Z}_2$ and $U(1)_l/\mathbb{Z}_2$ in the unitary family. The degree, rank and level are related by $k=n$ and $l=4n$ ($k=n/2$ and $l=n$) when $n$ is odd (even). We presented their bosonic topological orders and described the fusion and braiding properties of anyon excitations. A more detailed summary of results can be found in section~\ref{sec:overview} in the introduction. Here, we discuss the implications, unanswered questions, speculations, and possible future directions.

\paragraph{Generalized particle-hole conjugations}
Some fractional quantum Hall states in a partially-filled Landau level are pairwise related by particle-hole ({\color{blue}\hypertarget{PH}{PH}}) conjugation~\cite{Girvin84,BalramJain17,NguyenGolkarRobertsSon18,PanKangLillyRenoBaldwinWestPfeifferTsui20}. For example, the Laughlin FQH state $U(1)_3$ of particles at $\nu=1/3$ is conjugate to the Laughlin state of holes $U(1)_1\times\overline{U(1)_3}$ at $\nu=2/3$. Here, $U(1)_1$ is the filled Landau level at $\nu=1$, and $\overline{\mathcal{C}}$ is the time-reversal conjugate of $\mathcal{C}$. In particular, \PH conjugation relates FQH states at filling $\nu=1/2$. For example, the Moore-Read Pfaffian state~\cite{MooreRead,GreiterWenWilczekPRL91} $U(1)_8\otimes_f\mathrm{Ising}$ is conjugate to the anti-Pfaffian state~\cite{LevinHalperinRosenow07,LeeRyuNayakFisher07} $U(1)_1\times\overline{U(1)_8\otimes_f\mathrm{Ising}}=U(1)_8\otimes_f\overline{\mathrm{Ising}}^3$. The \PH symmetric Pfaffian state~\cite{Son15} $U(1)_8\otimes_f\overline{\mathrm{Ising}}$ is its own \PH conjugate. The notion of \PH conjugation can be generalized by replacing the Landau level by a prototype topological phase $\mathcal{P}$. Any topological state $\mathcal{C}$ that ``partially occupies'' $\mathcal{P}$ has a \PH conjugate $\mathrm{PH}(\mathcal{C})=\mathcal{P}\boxtimes\overline{\mathcal{C}}$, where the tensor product, in general, is relative to a certain set of condensed bosonic pairs of anyons. Generalized \PH conjugation has been explored in paired parton FQH states~\cite{SirotaSahooChoTeo18} at $\nu=1/6$ that half-fill the parton Landau level $\mathcal{P}=U(3)_1=U(1)_3\times SU(3)_1$ at $\nu=1/3$, and Fibonacci QH states~\cite{LopesQuitoHanTeo19} at $\nu=8$ that half-fill $\mathcal{P}=(E_8)_1$ at $\nu=16$. In section~\ref{sec:SO2nZkTL} and \ref{sec:dihedralTL}, we made use of the conformal embedding $U(1)_l\times SU(n)_1\subseteq SO(2n)_1$ in \eqref{ZkTL} and the coset decomposition $SO(n)_2\times SO(n)_1^2/SO(n)_2=SO(n)_1^2$ in \eqref{coset}. They imply the generalized \PH conjugations based on the prototype topological phases $\mathcal{P}=SO(2n)_1$ or $SO(n)_1^2$. For example, (i) $U(1)_l$ and $SU(n)_1$ is a \PH conjugate pair based on $\mathcal{P}=SO(2n)_1$ because $U(1)_l=SO(2n)_1\boxtimes\overline{SU(n)_1}$ and $SU(n)_1=SO(2n)_1\boxtimes\overline{U(1)_l}$. The tensor product $\boxtimes$ may be the decoupled one $\times$ or a relative one that condenses a fermion pair or a $\mathbb{Z}_2$ charge pair. The \PH conjugation in example (i) extends to the corresponding $\mathbb{Z}_2$ twist liquids. (ii) $U(1)_l/\mathbb{Z}_2=SO(n)_1^2/SO(n)_2$ and $SU(n)_1/\mathbb{Z}_2=SO(n)_2$ is a \PH conjugate pair based on $\mathcal{P}=SO(2n)_1/\mathbb{Z}_2=SO(n)_1^2$. We expect the generalized notion of \PH conjugation can be made precise in the coupled-wire setting~\cite{MrossAliceaMotrunich17,FujiFurusaki18} by a non-local representation of the \PH operator, which may be interpreted as the 2+1D analog of the Kramers-Wannier duality in 1+1D.

\paragraph{Coupled-wire construction of related twist liquid topological phases}
In this paper, we focus on the $D_k$ symmetry of $SO(2n)_1$ and the $\mathbb{Z}_2$ conjugation of $SU(n)_1$ and $U(1)_l$. While the latter is always an outer automorphism that non-trivially switches anyon types, the former only transposes anyon types when $n$ is odd. In other words, the $\mathbb{Z}_2$ symmetry in $SO(2n)_1$ considered in this paper is the mirror symmetry of the $SO(2n)_1$ Dynkin diagram only when $n$ is odd. Mirror outer automorphisms for even rank $n$, and the $S_3$ triality symmetry of $SO(8)$ was not investigated. While the topological orders of the $SO(2n)_1/\mathbb{Z}_2$, $SO(8)_1/\mathbb{Z}_3$ and $SO(8)_1/S_3$ are known~\cite{TeoHughesFradkin15,ChenRoyTeoRyu17,BarkeshliBondersonChengWang14}, microscopic descriptions of these chiral twist liquids states are missing. (There exist non-chiral exactly-solvable string-net models~\cite{LevinWen05,TeoHughesFradkin15}, whose topological orders contain the above states as chiral sub-sectors.) Moreover, the $D_k$ symmetry is only a particular point group symmetry in $2n$ dimensions. In three dimensions, the cyclic, dihedral, tetrahedral, octahedral, and icosahedral groups, $\mathbb{Z}_n$, $D_n$, $T$, $O$ and $I$, are the orientation preserving point groups in $SO(3)$. Their corresponding orbifold phases $SU(2)_1/\mathbb{Z}_n=U(1)_{2n^2}$, $SU(2)_1/D_n=U(1)_{2n^2}/\mathbb{Z}_2$, $SU(2)_1/T$, $SU(2)_1/O$, and $SU(2)_1/I$, were studied in the \CFT context.~\cite{Ginsparg88,DijkgraafVafaVerlindeVerlinde89,GinspargLectureNotes} The topological orders, as well as their microscopic descriptions, of higher-dimensional point group orbifold phases $SO(N)_l/G$ is largely unknown. Coupled-wire models of these topological phases may be constructed by extending the methods used in this paper. For example, the manipulation of fermion pair locality could be generalized to $G$-invariant local fields in the $SO(N)_l$ \WZW \hyperlink{CFT}{CFT}. Further investigation may lead to exotic symmetries or dualities and quantum critical phase transitions similar to those discussed above.


\paragraph{Quantum doubles and non-chiral twist liquids}\label{sec:quantumdoubles}
In \eqref{Z2orbifoldsodd}, we saw that the quantum double $\mathcal{Z}_n(\mathbb{Z}_2)$ appeared as the non-chiral component of the $SO(2n)_1/\mathbb{Z}_2$ twist liquid when $n$ is odd. $\mathcal{Z}_n(\mathbb{Z}_2)$ is the Drinfeld center~\cite{Kasselbook,BakalovKirillovlecturenotes,LevinWen05} of either the Ising or the $SU(2)_2$ fusion category, which are generated by the simple fusion objects $1,\psi,\sigma$ with the inequivalent Frobenius-Schur indicators~\cite{Kitaev06} $\varkappa_\sigma=d_\sigma\left[F^{\sigma\sigma\sigma}_\sigma\right]^1_1=1$ for the former or $\varkappa_\sigma=-1$ for the latter. The quantum double is not a discrete gauge theory because the quantum dimension of the Ising twist field, $d_\sigma=\sqrt{2}$, is non-integral. The distinction from a gauge theory is a consequence of the $\mathbb{Z}_2$ symmetry of $SO(2n)_1$ being outer automorphic, i.e.~exchanging anyon types. Similarly, we saw in \eqref{DkTLoddn} that the $SO(2n)_1/D_k$ twist liquid, when $n$ is odd, carries the non-chiral \UMTC component $\mathcal{Z}_n(D_k)=\mathcal{Z}_n(\mathbb{Z}_2)\boxtimes D^{[0]}(D_k)$. $\mathcal{Z}_n(D_k)$ is not a gauge theory because it supports the Ising-fluxon with non-integral dimension $d=\sqrt{2}k$. In this paper, we also encountered the outer automorphic $\mathbb{Z}_2$ conjugation symmetry in $SU(n)_1$ and $U(1)_l$. The corresponding twist liquids can be decomposed tautologically into $SU(n)_1/\mathbb{Z}_2=SU(n)_1\boxtimes Z(\mathcal{F}_n)$ and $U(1)_l/\mathbb{Z}_2=U(1)_l\boxtimes Z(\mathcal{F}_l)$, where \begin{align}\mathcal{F}_n=\overline{SU(n)_1}\times\frac{SU(n)_1}{\mathbb{Z}_2},\quad\mathcal{F}_l=\overline{U(1)_l}\times\frac{U(1)_l}{\mathbb{Z}_2}\label{SUnU1Dcenters}\end{align} and $\boxtimes$ anyon condenses $\Psi^a\overline{\Psi^a}$ and $e^a\overline{e^a}$ respectively. We conjecture that the non-chiral \UMTC $\mathcal{Z}_n(D_k)$ and $Z(\mathcal{F}_n)$ are Drinfeld centers of the following fusion categories, and therefore admit exactly-solvable string-net model~\cite{LevinWen05} Hamiltonians. 

First, the non-chiral \UMTC $Z(\mathcal{F}_n)$ in \eqref{SUnU1Dcenters} may be the Drinfeld center of the defect fusion category~\cite{TeoHughesFradkin15,Teotwistdefectreview} (or $G$-crossed category~\cite{EtingofNikshychOstrik10,BarkeshliBondersonChengWang14}) \begin{align}\mathcal{F}_n=\left\{\begin{array}{*{20}l}\mathbb{Z}_n\cup\{\sigma\},\quad\mbox{if $n$ is odd}\\\mathbb{Z}_n\cup\{\sigma_+,\sigma_-\},\quad\mbox{if $n$ is even}\end{array}\right.\end{align} Its simple fusion objects are the quasiparticles $\Psi^a$ in $SU(n)_1$ (or $e^a$ in $U(1)_l$) that obey the $\mathbb{Z}_n$ fusion algebra $\Psi^a\times\Psi^{a'}=\Psi^{[a+a']}$, together with the $\mathbb{Z}_2$ twist defect $\sigma$ when $n$ is odd, or $\sigma_\pm$ when $n$ is even. The twist defects obey the (commutative) fusion rules \begin{gather}\sigma\times\Psi^a=\sigma,\quad\sigma_s\times\Psi_a=\sigma_{s(-1)^a}\nonumber\\\sigma\times\sigma=\sum_{a=0}^{n-1}\Psi^a,\quad\mbox{if $n$ is odd}\nonumber\\\sigma_s\times\sigma_s=\sum_{a\mbox{ }\mathrm{even}}\Psi_a,\quad\mbox{if $n\equiv0$ mod 4},\nonumber\\\sigma_s\times\sigma_s=\sum_{a\mbox{ }\mathrm{odd}}\Psi_a,\quad\mbox{if $n\equiv2$ mod 4},\end{gather} where $s=\pm$. When $n$ is odd, $\mathcal{F}_n$ is a Tambara-Yamagami category~\cite{TambaraYamagami98}. 

Second, for the non-chiral Ising-fluxon topological order $\mathcal{Z}_n(D_k)$, it may be equivalent to the Drinfeld center $Z(\mathrm{Ising}\ltimes\mathbb{Z}_k)$, where $\mathrm{Ising}\ltimes\mathbb{Z}_k$ is the fusion category \begin{align}\mathrm{Ising}\ltimes\mathbb{Z}_k=\left\langle\mathbb{I}^r,\psi^r,\sigma^r:r=0,1,\ldots,k-1\right\rangle.\end{align} The (non-commutative) fusion rules~\cite{TeoRoyXiao13long,khan2014,TeoHughesFradkin15} are \begin{gather}
\mathbb{I}^r\times\mathbb{I}^{r'}=\psi^r\times\psi^{r'}=\mathbb{I}^{[r+r']},\quad
\mathbb{I}^r\times\psi^{r'}=\psi^{[r+r']},\nonumber\\
\sigma^r\times\mathbb{I}^{r'}=\sigma^r\times\psi^{r'}=\sigma^{[r+r']},\nonumber\\
\mathbb{I}^r\times\sigma^{r'}=\psi^r\times\sigma^{r'}=\sigma^{[r'-r]}\nonumber\\
\sigma^r\times\sigma^{r'}=\mathbb{I}^{[r'-r]}+\psi^{[r'-r]},\end{gather} where $[r]$ wraps any integer $r$ to the range between 0 and $k-1$ by adding or subtracting $k$ if necessary. The Frobenius-Schur indicator of $\mathrm{Ising}\ltimes\mathbb{Z}_k$ is $\varkappa_{\sigma^0}=d_{\sigma^0}\left[F^{\sigma^0\sigma^0\sigma^0}_{\sigma^0}\right]_1^1=(-1)^{(k^2-1)/8}$. The Ising-fluxon fields group into the super-selection sector $\Gamma=\{\sigma^0,\ldots,\sigma^{k-1}\}$ in the Drinfeld center $Z(\mathrm{Ising}\ltimes\mathbb{Z}_k)$ and has the combined quantum dimension $d=\sqrt{2}k$. 

\section*{Acknowledgments}
We thank Steven H.~Simon and Henry Chan for valuable discussions. JCYT is supported by the National Science Foundation under Grant Number DMR-1653535. YH is in part supported by grant EP/S020527/1 from EPSRC.

\appendix
\section{The \texorpdfstring{$SO(N)_1$}{SO(N)} current algebra and its primary fields}\label{app:SO(N)algebra}
We review the chiral $SO(N)$ \WZW \KM algebra at level 1 and its primary fields.~\cite{GinspargLectureNotes,bigyellowbook,Blumenhagenbook} In section~\ref{sec:bosonwires}, we presented the electronic origins of the $SO(N)_1$ boson wires, which are the fundamental building block of the topological models in this article. The low-energy degrees of freedom of each boson wire is effectively described by $N$ counter-propagating pairs of massless Majorana fermions $\psi^\sigma_r=(\psi^\sigma_r)^\dagger$, where $\sigma=L/R=+/-$ and $r=1,\ldots,N$. The Hamiltonian density was given in \eqref{freeH}. The fermions obey the correlations \begin{align}\left\langle\psi^r_L(z)\psi^s_L(w)\right\rangle=\frac{\delta^{rs}}{z-w},\quad\left\langle\psi^r_R(\bar{z})\psi^s_R(\bar{w})\right\rangle=\frac{\delta^{rs}}{\bar{z}-\bar{w}},\label{fermioncorrelation}\end{align} where $z$ and $w$ are complex Euclidean space-time parameters $e^{2\pi(v\tau+ix)/l}$, $\tau$ is the Wick rotated Euclidean time, and $l$ is the length of the 1D wire in a closed circle. The \WZW \KM current operators are the Hermitian normal-ordered antisymmetric pairs of fermions \begin{align}J^{rs}_\sigma=-J^{sr}_\sigma=i\psi^r_\sigma\psi^s_\sigma.\end{align} They span the affine $SO(N)_1$ Lie algebra.

In this appendix, we mainly focus on the left-moving holomorphic sector $J=J_L$. The anti-holomorhpic right-moving sector follows a similar algebraic structure by conjugation $z\to\bar{z}$. The chiral current operators obey the operator product expansion ({\color{blue}\hypertarget{OPE}{OPE}}) \begin{align}\begin{split}&J^{rs}(z)J^{pq}(w)\\&=\frac{\delta^{(rs)(pq)}}{(z-w)^2}+\sum_{u<v}\frac{if_{(rs)(pq)(uv)}}{z-w}J^{uv}_L(w)+\ldots,\end{split}\label{SONOPEMaj}\end{align} where $r<s$ and $p<q$. The structure constant of the $SO(N)$ Lie algebra is $f_{(rs)(pq)(uv)}=\delta_{ur}\delta_{vq}\delta_{sp}-\delta_{ur}\delta_{vp}\delta_{sq}+\delta_{us}\delta_{rq}\delta_{vp}-\delta_{us}\delta_{vq}\delta_{rp}$. The electronic origin of the boson wire dictates that all fermion pairs, including the current operators $J^\alpha=J^{rs}$, are local and integral.

For even $N=2n$, it is sometimes convenient to express the current operators in the complexified form by pairing the Majorana fermions into Dirac ones,  such as $d^a=(\psi^a+i\psi^{n+a})/\sqrt{2}=e^{i\phi^a}$, where $a=1,\ldots,n$. The bosonized variables $\phi_a$ obey the correlations \begin{align}\left\langle\phi^a(z)\phi^b(w)\right\rangle=-\delta^{ab}\log(z-w)+\frac{i\pi}{2}\mathrm{sgn}(a-b)\label{CartanWeylcorrelation}\end{align} where the singular term implies \eqref{fermioncorrelation} after exponentiation, and the non-singular antisymmetric term ensures the Dirac fermions mutually anti-commute, $d^ad^b=-d^bd_a$ for $a\neq b$. The Cartan-Weyl current generators are the normal-ordered densities \begin{align}H^a=i\psi^{2a-1}\psi^{2a}=(d^a)^\dagger d^a=-i\partial\phi^a.\label{HCartan}\end{align} The raising and lowering roots are the normal-ordered vertex operators \begin{align}E^{\boldsymbol\alpha}=\pm i\left(d^a\right)^{\alpha_a}\left(d^b\right)^{\alpha_b}=e^{i\alpha_c\phi^c},\label{Ealpha}\end{align} where $d^+=d$ and $d^-=d^\dagger$. Here, each of the roots are labeled by a vector $\boldsymbol\alpha=(\alpha_1,\ldots,\alpha_n)$ with integral entries and of length $|\boldsymbol\alpha|=\sqrt{2}$. This means there are only two non-zero entries $\alpha_a,\alpha_b=\pm1$, for $1\leq a<b<n$. The $n(2n-1)$ roots form the simply-laced $D_n=SO(2n)$ root system of rank $n$.

For odd $N=2n+1$, the $SO(2n+1)_1$ current algebra extends from $SO(2n)_1$ by including the raising and lowering operators \begin{align}E^{\pm{\bf e}_a}=i\left(d^a\right)^\pm\psi^{2n+1}=ie^{\pm i\phi^a}\psi^{2n+1}\label{shortroots}\end{align} that associate with the short root vectors $\pm{\bf e}_a$, which is an unit integral vector of length $|{\bf e}_a|=1$ with its $a^{\mathrm{th}}$ entry being the only non-zero one. Combining the long roots from $D_n$ and these short roots, they form the non-simply-laced $B_n=SO(2n+1)$ root system of rank $n$.

Primary fields $[V]$ of the $SO(N)_1$ \WZW \KM algebra are super-selection sectors spanned by fields $V_\mu(z)$, for $\mu=1,\ldots,M$, that ``rotate'' irreducibly under the action of the $SO(N)_1$ currents \begin{align}J^\alpha(z)V_\mu(w)=\frac{1}{z-w}\left(t^\alpha\right)_\mu^\nu V_\nu(w)+\ldots,\label{SONirrep}\end{align} where $t^\alpha$ is a $M$-dimensional irreducible matrix representation of the $SO(N)$ Lie algebra generator $J^\alpha$. At level 1, these irreducible representations are the trivial, vector or spinor representations of $SO(N)$. The vector representation $[\psi]$ is fermionic and carries the conformal scaling dimension $h=1/2$. The super-selection sector is spanned by the Majorana fermions $\psi^r$, for $r=1,\ldots,N$, and obeys the \OPE \begin{align}J^{pq}(z)\psi^r(w)=\frac{i}{z-w}\left(\delta^{qr}\delta^{ps}-\delta^{pr}\delta^{qs}\right)\psi^s(w)+\ldots.\end{align} The fusion rule $[\psi]\times[\psi]=1$ indicates that fermion pairs belong to the trivial super-selection sector 1, which consists of integral combinations of local bosons. It abbreviates the \OPE \begin{align}\psi^p(z)\psi^q(w)=\frac{\delta^{pq}}{z-w}-i(1-\delta^{pq})J^{(pq)}(w)+\ldots.\end{align}

Spinor fields transform according to the spinor representations of $SO(N)$. For even $N=2n$, they decomposes into the even and odd spinors, $[s_+]$ and $[s_-]$. The super-selection sectors are spanned by products of ``half-fermions'' \begin{gather}s_{\boldsymbol\varepsilon}=e^{i\varepsilon_a\phi^a/2},\quad\boldsymbol\varepsilon=(\varepsilon_1,\ldots,\varepsilon_n)\\\begin{split}[s_+]&=\mathrm{span}\left\{e^{i\varepsilon_a\phi^a/2}:\varepsilon_a=\pm1,\prod_{a=1}^n\varepsilon_a=+1\right\},\\ [s_-]&=\mathrm{span}\left\{e^{i\varepsilon_a\phi^a/2}:\varepsilon_a=\pm1,\prod_{a=1}^n\varepsilon_a=-1\right\}.\end{split}\nonumber\end{gather} The $SO(2n)_1$ currents transform the vertex operators into each other within the same super-selection sector according to the \OPE \begin{align}H^a(z)s_{\boldsymbol\varepsilon}(w)&=\frac{-\varepsilon_a/2}{z-w}s_{\boldsymbol\varepsilon}(w)+\ldots\nonumber
\\E^{\boldsymbol\alpha}(z)s_{\boldsymbol\varepsilon}(w)&=\frac{Z^{\boldsymbol\alpha}_{\boldsymbol\varepsilon}}{z-w}s_{\boldsymbol\varepsilon+2\boldsymbol\alpha}(w)+\ldots,\quad\mbox{if $\boldsymbol\alpha\cdot\boldsymbol\varepsilon=-2$}\nonumber\\E^{\boldsymbol\alpha}(z)s_{\boldsymbol\varepsilon}(w)&=\mbox{non-singular},\quad\mbox{if $\boldsymbol\alpha\cdot\boldsymbol\varepsilon=0,2$}\label{JsOPE}\end{align} where $Z^{\boldsymbol\alpha}_{\boldsymbol\varepsilon}=\exp\left[-i\pi\alpha_a\varepsilon_b\mathrm{sgn}(a-b)/4\right]$, and $H^a$ and $E^{\boldsymbol\alpha}$ are the Cartan-Weyl generators and raising/lowering operators defined in \eqref{HCartan} and \eqref{Ealpha}. The rotated spinor $s_{\boldsymbol\varepsilon'}=s_{\boldsymbol\varepsilon+2\boldsymbol\alpha}$ has the same parity as the original, $\prod_a\varepsilon_a=\prod_a\varepsilon'_a$, and belongs to the same super-selection sector, because the raising/lowering operator flips two signs in $\boldsymbol\varepsilon=(\varepsilon_a,\ldots,\varepsilon_n)$. The above \OPE can be derived from the correlations \eqref{CartanWeylcorrelation} and the Baker-Campbell-Hausdorff formula $e^{A(z)}e^{B(w)}=e^{A(z)+B(w)+\langle A(z)B(w)\rangle}$. 

The two spinors have opposite fermion parities and can be flipped from one to the other by absorbing or emitting a fermion. This is captured by the fusion rule $[s_\pm]\times[\psi]=[s_\mp]$, which corresponds to the \OPE \begin{align}d^a(z)^{-\varepsilon_a}s_{\boldsymbol\varepsilon}(w)&=\frac{Z^a_{\boldsymbol\varepsilon}}{(z-w)^{1/2}}s_{\boldsymbol\varepsilon-2\varepsilon_a{\bf e}_a}(w)+\ldots\nonumber\\d^a(z)^{\varepsilon_a}s_{\boldsymbol\varepsilon}(w)&=\mbox{non-singular}\label{fsOPE}\end{align} where $Z^a_{\boldsymbol\varepsilon}=\exp\left[i\pi\sum_{b=1}^n\varepsilon_a\varepsilon_b\mathrm{sgn}(a-b)/4\right]$, and $d^+=d$, $d^-=d^\dagger$. The $1/(z-w)^{1/2}$ factor indicates the $\pi$ monodromy phase and the mutual semionic statistics between the fermion and the spinors. The new spinor $s_{\boldsymbol\varepsilon'}=s_{\boldsymbol\varepsilon+2\varepsilon_a{\bf e}_a}$ has opposite parity as $\prod_b\varepsilon_b'=-\prod_b\varepsilon_b$ because the fermion flip one sign in $\boldsymbol\varepsilon=(\varepsilon_a,\ldots,\varepsilon_n)$. The conjugate pair \begin{align}s_{\boldsymbol\varepsilon}(z)s_{-\boldsymbol\varepsilon}(w)=\frac{1}{(z-w)^{n/4}}+\ldots\end{align} dictates the conformal scaling dimension of each spinor to be $h_s=n/8=N/16$. A spinor $s_{\boldsymbol\varepsilon}$ has the same parity as its conjugate $s_{-\boldsymbol\varepsilon}$, and the spinor super-selection sectors are therefore self-conjugate with the fusion rule $[s_\pm]\times[s_\pm]=1$, only when $n$ is even. In this case, the anyon group $\mathcal{A}_{SO(2n)_1}=\{1,[\psi],[s_+],[s_-]\}$ has a $\mathbb{Z}_2\times\mathbb{Z}_2$ Abelian group structure under fusion product. Otherwise, when $n$ is odd, the spinor fusion rule become $[s_\pm]\times[s_\mp]=1$, and the anyon group is $\mathcal{A}_{SO(2n)_1}=\mathbb{Z}_4$ because $[s_\pm]\times[s_\pm]=[\psi]$ and $[s_\pm]^4=1$, i.e.~the spinors are of order 4. The fusion rules of $SO(2n)_1$ are summarized below (also in \eqref{SO2nfusion}). \begin{align}&[\psi]\times[\psi]=1,\quad[\psi]\times[s_\pm]=[s_\mp],\nonumber\\&[s_\pm]\times[s_\pm]=\left\{\begin{array}{*{20}l}1,&\mbox{for even $n$}\\ {} [\psi],&\mbox{for odd $n$}\end{array}\right..\label{SO2nfusionapp}\end{align}

For odd $N=2n+1$, the field content of $SO(2n+1)_1$ may be decomposed according to $SO(2n)_1\times\mathrm{Ising}$, where the Ising sector is generater by the additional Majorana fermion $\psi^{2n+1}$. The Ising \CFT has primary fields $1,\psi,\sigma$. The non-Abelian Ising twist field has conformal scaling dimension $1/16$ and follows the multi-channel fusion rule \begin{align}\sigma\times\sigma=1+\psi,\quad\psi\times\sigma=\sigma,\end{align} where $\psi=\psi^{2n+1}$ when embedded in $SO(2n+1)_1$. The Ising twist field is non-local and has semionic mutual statistics with $\psi^{2n+1}$. Since the fermion pairs $\psi^r\psi^{2n+1}$ are local, the Ising twist field must appear simultaneously with a $SO(2n)_1$ spinor $s_{\boldsymbol\varepsilon}$. The $SO(2n+1)_1$ spinor super-selection sector $[\sigma]$ is spanned by the normal-ordered products $\sigma_{\boldsymbol\varepsilon}=\sigma s_{\boldsymbol\varepsilon}$ \begin{align}[\sigma]=\mathrm{span}\left\{\sigma_{\boldsymbol\varepsilon}=\sigma e^{i\varepsilon_a\phi^a/2}:\varepsilon_a=\pm1\right\},\end{align} where the parity $\prod_{a=1}^n\varepsilon_a$ is not fixed. The conformal scaling dimension is $h_\sigma=1/16+n/8=N/16$. The actions of the Cartan-Weyl generators $H^a$ and long roots $E^{\boldsymbol\alpha}$ only affect the Abelian component $s_{\boldsymbol\varepsilon}$ and are dictated by \eqref{JsOPE}. \begin{align}H^a(z)\sigma_{\boldsymbol\varepsilon}(w)&=\frac{-\varepsilon_a/2}{z-w}\sigma_{\boldsymbol\varepsilon}(w)+\ldots\nonumber
\\E^{\boldsymbol\alpha}(z)\sigma_{\boldsymbol\varepsilon}(w)&=\frac{Z^{\boldsymbol\alpha}_{\boldsymbol\varepsilon}}{z-w}\sigma_{\boldsymbol\varepsilon+2\boldsymbol\alpha}(w)+\ldots,\quad\mbox{if $\boldsymbol\alpha\cdot\boldsymbol\varepsilon=-2$}\nonumber\\E^{\boldsymbol\alpha}(z)\sigma_{\boldsymbol\varepsilon}(w)&=\mbox{non-singular},\quad\mbox{if $\boldsymbol\alpha\cdot\boldsymbol\varepsilon=0,2$}.\label{JsigmaOPE1}\end{align} Using the fermion action \eqref{fsOPE}, the short roots \eqref{shortroots} transform the spinors according to \begin{align}E^{\pm a}(z)\sigma_{\boldsymbol\varepsilon}(w)&=\frac{Z^a_{\boldsymbol\varepsilon}}{z-w}\sigma_{\boldsymbol\varepsilon-2\varepsilon_a{\bf e}_a}(w)+\ldots,\quad\mbox{if $\varepsilon_a=\mp$},\nonumber\\E^{\pm a}(z)\sigma_{\boldsymbol\varepsilon}(w)&=\mbox{non-singular},\quad\mbox{if otherwise}.\label{JsigmaOPE2}\end{align} The fusion rules of $SO(2n)_1$ are summarized below. \begin{align}[\psi]\times[\psi]=1,\quad [\psi]\times[\sigma]=[\sigma],\quad [\sigma]\times[\sigma]=1+[\psi].\end{align}

\section{The anyon content of \texorpdfstring{$SO(n)_2$}{SO(n) level 2} and \texorpdfstring{$U(1)_l/\mathbb{Z}_2$}{U(1)/Z2}}\label{app:orbifoldanyons}

We first review the chiral $SO(n)$ \WZW \KM algebra at level 2 and its primary field content.~\cite{GinspargLectureNotes,bigyellowbook,Blumenhagenbook} The corresponding topological phase was introduced in section~\ref{sec:SUnZ2TL}. We focus on the holomorphic left-moving sector. The current algebra is generated by the diagonal sums ${\bf J}={\bf J}^A+{\bf J}^B$ in the double $SO(n)_1^A\times SO(n)_1^B$. They obey the current \OPE \eqref{SOn2OPE}. The $SO(n)_2$ currents can also be identified as self-conjugate combinations of the $SU(n)_1$ currents $J_{ab}\sim\cos(\phi_a-\phi_b)\sim d_ad_b^\dagger+d_bd_a^\dagger$ (see \eqref{SOn2currents}), where $d_a=(\psi_a^A+i\psi_a^B)/\sqrt{2}\sim e^{i\phi_a}$ are the $n$ Dirac fermions that rotate under $SU(n)$. The hermitian $SO(n)_2$ currents are symmetric under the internal $\mathbb{Z}_2$ symmetry, $\phi_a\to-\phi_a$ (see also \eqref{Z2gauge1}), and therefore $SO(n)_2=SU(n)_1/\mathbb{Z}_2$.

The primary fields of $SO(n)_2=SU(n)_1/\mathbb{Z}_2$ together with their quantum dimensions and spins are listed in table~\ref{SOn2anyonsodd} and \ref{SOn2anyonseven}. They consist of (a) Abelian sectors with quantum dimension $d=1$, (b) non-Abelian super-selection sectors $\Phi^i$ with dimension $d=2$, for $1\leq i<n/2$, and (c) $\mathbb{Z}_2$ twist fields $\sigma$ and $\tau$ when $n$ is odd, or $\sigma^\pm$ and $\tau^\pm$ when $n$ is even. Abelian primary fields includes (i) the vacuum sector containing local fields such as the $SO(n)_2$ currents, (ii) the $\mathbb{Z}_2$ charge sector $S$ containing fermion pair combinations that are odd under $\mathbb{Z}_2$ (see \eqref{SUnZ2charge}), and (iii) when $n$ is even, the $\mathbb{Z}_2$-closed $SU(n)_1$ primary field sectors $\Psi^{n/2}_\pm$ (see \eqref{SUnZ2spinors}). When $n$ is odd, none of the $SU(n)_1$ primary fields, except the vacuum, are closed under $\mathbb{Z}_2$, and $\Psi^{n/2}_\pm$ are absent. The Abelian sectors obey the fusion rules \begin{align}\begin{split}&\Psi^{n/2}_\pm\times S=\Psi^{n/2}_\mp,\quad S\times S=1,\\&\Psi^{n/2}_\pm\times\Psi^{n/2}_\pm=\left\{\begin{array}{*{20}l}1,&\mbox{if $n\equiv0$ mod 4}\\S,&\mbox{if $n\equiv2$ mod 4}\end{array}\right..\end{split}\label{SOn2Abelianfusion}\end{align} The set of Abelian anyons form the fusion group $\mathbb{Z}_2=\{1,S\}$ when $n$ is odd, $\mathbb{Z}_2\times\mathbb{Z}_2=\{1,S,\Psi^{n/2}_\pm\}$ when $n\equiv0$ modulo 4, or $\mathbb{Z}_4=\{1,S,\Psi^{n/2}_\pm\}$ when $n\equiv2$ modulo 4.

The internal $\mathbb{Z}_2$ symmetry conjugates the $SU(n)_1$ primary fields $\Psi^i\leftrightarrow\Psi^{n-i}$, and therefore each conjugate pair forms the $d=2$ super-selection sector $\Phi^i=\Psi^i+\Psi^{n-i}$, for $1\leq i<n/2$. (See \eqref{SUnanyons} for the bosonized vertex representation of $\Psi^i$.) They do not carry a $\mathbb{Z}_2$ parity because \begin{align}\Phi^i\times S=\Phi^i.\label{SOn2supersectorfusion1}\end{align} They obey the fusion rules \begin{widetext}\begin{gather}\Phi^i\times\Phi^j=\Phi^{i+j}+\Phi^{j-i},\quad\mbox{for $1\leq i<j<n/2$ and $i+j\neq n/2$}\nonumber\\\Phi^i\times\Phi^j=\Psi^{n/2}_++\Psi^{n/2}_-+\Phi^{j-i},\quad\mbox{for $1\leq i<j<n/2$ and $i+j=n/2$}\nonumber\\\Phi^i\times\Phi^i=1+S+\Phi^{2i},\quad\mbox{for $i\neq n/4$}\label{SOn2supersectorfusion2}\\\Phi^{n/4}\times\Phi^{n/4}=1+S+\Psi^{n/2}_++\Psi^{n/2}_-\nonumber\\\Psi^{n/2}_\pm\times\Phi^i=\Phi^{n/2-i}\nonumber\end{gather}\end{widetext} where we identify $\Phi^i=\Phi^{n-i}$ if $n/2<i<n$.

When $n$ is odd, there are two $\mathbb{Z}_2$ twist fields, $\sigma$ and $\tau$, that differ from each other by a $\mathbb{Z}_2$ charge, $\sigma=\tau\times S$ and $\tau=\sigma\times S$. They have quantum dimension $d=\sqrt{n}$. When $n$ is even, there are four $\mathbb{Z}_2$ twist fields $\sigma^\pm$ and $\tau^\pm$. They have quantum dimension $d=\sqrt{n/2}$. The $\pm$ sign is flipped when absorbing a $SU(n)_1$ primary field $\Psi^i$ with odd degree $i$. In addition to the fusion rules (\ref{Z2fluxfusion0}-\ref{Z2fluxfusion3}), when $n$ is even, the $\mathbb{Z}_2$ fluxes also obey the following product rules when combining with $\Psi^{n/2}_\pm$. \begin{align}\left\{\begin{array}{*{20}l}\Psi^{n/2}_\pm\times\sigma^\pm=\sigma^\pm,&\Psi^{n/2}_\pm\times\tau^\pm=\tau^\pm,\\\Psi^{n/2}_\pm\times\sigma^\mp=\tau^\mp,&\Psi^{n/2}_\pm\times\tau^\mp=\sigma^\mp,\end{array}\right.\quad\mbox{if $n\equiv0$ mod 4},\nonumber\\\left\{\begin{array}{*{20}l}\Psi^{n/2}_\pm\times\sigma^\pm=\tau^\mp,&\Psi^{n/2}_\pm\times\tau^\pm=\sigma^\mp,\\\Psi^{n/2}_\pm\times\sigma^\mp=\sigma^\pm,&\Psi^{n/2}_\pm\times\tau^\mp=\tau^\pm,\end{array}\right.\quad\mbox{if $n\equiv2$ mod 4}.\label{Z2fluxfusionapp}\end{align}

The braiding phase between anyon $x$ and $y$ with a fixed admissible fusion channel $z$ is $R^{xy}_zR^{yx}_z=e^{2\pi i(h_z-h_x-h_y)}$, where the anyon spins, $h_x$, are listed in table~\ref{SOn2anyonsodd} and \ref{SOn2anyonseven}. (For $SO(n)_2$, there is no fusion degeneracies. The fusion number $N_{xy}^z$ is either 0 or 1 depending on whether $z$ is an admissible fusion channel of $x\times y$. Hence, the exchange ``matrices'' $R^{xy}_z$ are $U(1)$ scalars.) The modular $S$-matrix is \begin{align}S_{xy}=\frac{1}{\mathcal{D}}\sum_zd_zN_{xy}^ze^{2\pi i(h_z-h_x-h_y)}\label{Smatrix}\end{align} where the total quantum dimension of $SO(n)_2$ is $\mathcal{D}=2\sqrt{n}$. The modular $T$-matrix is $T_{xy}=e^{i\pi(1-n)/12}e^{2\pi ih_x}\delta_{xy}$ where the anyon spins $h_x$ are listed in table~\ref{SOn2anyonsodd} and \ref{SOn2anyonseven}. Together the $S$ and $T$ matrices generate a representation of the modular group $SL(2,\mathbb{Z})$. They obey \begin{align}S^2=(ST^\dagger)^3,\quad S^4=\mathbb{I},\label{SL2Z}\end{align} where $C=S^2$ is the conjugation matrix so that $C_{x\bar{x}}=1$ if $x$ and $\bar{x}$ are conjugate pair, $x\times\bar{x}=1+\ldots$, and $C_{xx'}=0$ if $x'\neq\bar{x}$. The fusion rules (\ref{Z2fluxfusion0}-\ref{Z2fluxfusion3}) and (\ref{SOn2Abelianfusion}-\ref{Z2fluxfusionapp}) are related to the $S$-matrix by the Verlinde formula~\cite{Verlinde88} \begin{align}N_{x_1x_2}^y=\sum_{z}\frac{S_{x_1z}S_{x_2z}S_{yz}^\ast}{S_{1z}},\label{Verlinde}\end{align} where the fusion number $N_{x_1x_2}^y$ describes $x_1\times x_2=\sum_yN_{x_1x_2}^yy$. The modular $S$-matrices of $SO(n)_2=SU(n)_1/\mathbb{Z}_2$ are listed in table~\ref{tab:SOn2Smatrixodd} and \ref{tab:SOn2Smatrixeven}.

\begin{table}[htbp]
\centering
\begin{tabular}{c|ccccc}
&1&$S$&$\Phi^i$&$\sigma$&$\tau$\\\hline
1&1&1&2&$\sqrt{n}$&$\sqrt{n}$\\
$S$&1&1&2&$-\sqrt{n}$&$-\sqrt{n}$\\
$\Phi^{i'}$&2&2&$4c(ii')$&0&0\\
$\sigma$&$\sqrt{n}$&$-\sqrt{n}$&0&$\sqrt{n}$&$-\sqrt{n}$\\
$\tau$&$\sqrt{n}$&$-\sqrt{n}$&0&$-\sqrt{n}$&$\sqrt{n}$
\end{tabular}
\caption{The modular $S$-matrix $\mathcal{D}S_{xy}=2\sqrt{n}S_{xy}$ from \eqref{Smatrix} of $SO(n)_2=SU(n)_1/\mathbb{Z}_2$, for odd $n$. The rows and columns are arranged in the same order of the primary fields in table~\ref{SOn2anyonsodd}. $c(ii')=\cos(2\pi ii'/n)$, for $i,i'=1,\ldots,(n-1)/2$.}\label{tab:SOn2Smatrixodd}
\end{table}


\begin{table}[htbp]
\centering
\begin{tabular}{c|cccccc}
&1&$S$&$\Psi^{n/2}_s$&$\Phi^i$&$\sigma^s$&$\tau^s$\\\hline
1&1&1&1&2&$\sqrt{n/2}$&$\sqrt{n/2}$\\
$S$&1&1&1&2&$-\sqrt{n/2}$&$-\sqrt{n/2}$\\
$\Psi^{n/2}_{s'}$&1&1&$(-1)^{n/2}$&$2(-1)^i$&$\sqrt{n/2}e^{i\pi n/4}ss'$&$\sqrt{n/2}e^{i\pi n/4}ss'$\\
$\Phi^{i'}$&2&2&$2(-1)^{i'}$&$4c(ii')$&0&0\\
$\sigma^{s'}$&$\sqrt{n/2}$&$-\sqrt{n/2}$&$\sqrt{n/2}e^{i\pi n/4}ss'$&0&$\Sigma^{ss'}_{\sigma\sigma}$&$\Sigma^{ss'}_{\sigma\tau}$\\
$\tau^{s'}$&$\sqrt{n/2}$&$-\sqrt{n/2}$&$\sqrt{n/2}e^{i\pi n/4}ss'$&0&$\Sigma^{ss'}_{\tau\sigma}$&$\Sigma^{ss'}_{\tau\tau}$
\end{tabular}
\caption{The modular $S$-matrix $\mathcal{D}S_{xy}=2\sqrt{n}S_{xy}$ from \eqref{Smatrix} of $SO(n)_2=SU(n)_1/\mathbb{Z}_2$, for even $n$. The rows and columns are arranged in the same order of the primary fields in table~\ref{SOn2anyonseven}. $c(ii')=\cos(2\pi ii'/n)$, for $i,i'=1,\ldots,n/2-1$ and $s,s'=\pm$. The $\Sigma$ matrices are defined in \eqref{SmatrixSigma}. The $S$-matrix of $U(1)_n/\mathbb{Z}_2$ is obtained by replacing $e^{n/2}_\pm\leftrightarrow\Psi^{n/2}_\pm$ and $\mathcal{E}^i\leftrightarrow\Phi^i$.}\label{tab:SOn2Smatrixeven}
\end{table}

\begin{align}\begin{split}\begin{pmatrix}\Sigma^{ss'}_{\sigma\sigma}&\Sigma^{ss'}_{\sigma\tau}\\\Sigma^{ss'}_{\tau\sigma}&\Sigma^{ss'}_{\tau\tau}\end{pmatrix}=\begin{cases}
\sqrt{n}\left(\begin{smallmatrix}1&0&-1&0\\0&1&0&-1\\-1&0&1&0\\0&-1&0&1\end{smallmatrix}\right),\quad\mbox{if $n\equiv0$ mod 8},\\
\sqrt{n}\left(\begin{smallmatrix}0&1&0&-1\\1&0&-1&0\\0&-1&0&1\\-1&0&1&0\end{smallmatrix}\right),\quad\mbox{if $n\equiv4$ mod 8},\\
\sqrt{\frac{n}{2}}e^{\frac{i\pi}{4}}\left(\begin{smallmatrix}1&-i&-1&i\\-i&1&i&-1\\-1&i&1&-i\\i&-1&-i&1\end{smallmatrix}\right),\quad\mbox{if $n\equiv2$ mod 8},\\
\sqrt{\frac{n}{2}}e^{-\frac{i\pi}{4}}\left(\begin{smallmatrix}1&i&-1&-i\\i&1&-i&-1\\-1&-i&1&i\\-i&-1&i&1\end{smallmatrix}\right),\quad\mbox{if $n\equiv6$ mod 8}.\end{cases}
\end{split}\label{SmatrixSigma}\end{align}



Next, we review the modular data of the $U(1)_l/\mathbb{Z}_2$ orbifold~\cite{Ginsparg88,DijkgraafVerlindeVerlinde88,DijkgraafVafaVerlindeVerlinde89} and its primary field (anyon) content, where the level $l$ is $n$ ($4n$) when $n$ is even (odd). The corresponding topological phase was introduced in section~\ref{sec:U1Z2orbifold}. The orbifold \CFT is identical to the coset $[SO(n)_1]^2/SO(n)_2$.~\cite{GinspargLectureNotes} The topological phase can therefore be identified (as an \hyperlink{UMTC}{UMTC}) as the relative tensor product in \eqref{cosetPHequivalence}, which we repeat here \begin{align}U(1)_n/\mathbb{Z}_2=\left[SO(n)_1^A\times SO(n)_1^B\right]\boxtimes\overline{SO(n)_2},\label{U1Z2condensationpicture}\end{align} for even $n$. Here, $\overline{SO(n)_2}$ is the time-reversal of $SO(n)_2$ with the same fusion rules, $N_{\overline{x}\overline{y}}^{\overline{z}}=N_{xy}^z$, but conjugated spins, $h_{\overline{x}}=-h_x$. The following set of anyon pairs \begin{align}\mathcal{B}=\left\{1,\psi^A\psi^B\overline{S},s_\pm^As_\pm^B\overline{\Psi^{n/2}_\pm}\right\}\label{U1Z2condensedbosons}\end{align} are condensed in the relative tensor product $\boxtimes$. The Abelian bosons in $\mathcal{B}$ have trivial mutual statistics and follows $\mathbb{Z}_2\times\mathbb{Z}_2$ ($\mathbb{Z}_4$) fusion group structure when $n\equiv0$ ($n\equiv2$) modulo 4.

The $U(1)_n/\mathbb{Z}_2$ primary fields together with their quantum dimensions and spins are listed in table~\ref{Uorbifoldanyons} (by substituting $l=n$). They consists of (a) four Abelian fields $1,S,e^{n/2}_\pm$, (b) the non-Abelian super-selection sectors $\mathcal{E}^i=e^i+e^{-i}$, for $i=1,\ldots,n/2-1$, and (c) the $\mathbb{Z}_2$ twist fields $\sigma^\pm$ and $\tau^\pm$. In the anyon condensation picture \eqref{U1Z2condensationpicture}, all bosons in \eqref{U1Z2condensedbosons} belong to the vacuum sector. The non-vacuum anyons in $U(1)_n/\mathbb{Z}_2$ can be identified with anyons in $\left[SO(n)_1^A\times SO(n)_1^B\right]\boxtimes\overline{SO(n)_2}$ according to \begin{align}\begin{split}&S\equiv\psi^A\psi^B\equiv\overline{S},\quad\sigma^\pm\equiv s_\pm^A\overline{\sigma^\pm},\quad\tau^\pm\equiv s_\pm^A\overline{\tau^\pm},\\&e^{n/2}_\pm\equiv\left\{\begin{array}{*{20}l}s_\pm^As_\pm^B\equiv\overline{\Psi^{n/2}_\pm},&\mbox{if $n\equiv0$ mod 4}\\s_\pm^As_\mp^B\equiv\psi^A\overline{\Psi^{n/2}_\pm},&\mbox{if $n\equiv2$ mod 4}\end{array}\right.,\\&\mathcal{E}^i\equiv\left\{\begin{array}{*{20}l}\overline{\Phi^i},&\mbox{for even $i$}\\\psi^A\overline{\Phi^i},&\mbox{for odd $i$}\end{array}\right..\end{split}\end{align} Consequently, $U(1)_n/\mathbb{Z}_2$ have the identical fusion rules as $SO(n)_2=SU(n)_1/\mathbb{Z}_2$ (for even $n$) by replacing $e^{n/2}_\pm\leftrightarrow\Psi^{n/2}_\pm$ and $\mathcal{E}^i\leftrightarrow\Phi^i$ in (\ref{Z2fluxfusion1}-\ref{Z2fluxfusion3}) and (\ref{SOn2Abelianfusion}-\ref{Z2fluxfusionapp}). The modular $S$-matrix of $U(1)_n/\mathbb{Z}_2$ can be computed using \eqref{Smatrix}. It is identical to that of $SO(n)_2=SU(n)_1/\mathbb{Z}_2$ and is summarized in table~\ref{tab:SOn2Smatrixeven} under the same anyon label replacement.~\footnote{We observe the following differences in the $S$-matrices of $U(1)_n/\mathbb{Z}_2$ between our results and ref.~\cite{DijkgraafVafaVerlindeVerlinde89}. While they are identical when $n\equiv0,2$ mod 8, the $S$-matrices are conjugated when $n\equiv6$ and the $S_{\sigma^s\sigma^{s'}}$, $S_{\tau^s\tau^{s'}}$, $S_{\sigma^s\tau^{s'}}$ entries are unequal when $n\equiv4$ mod 8.} Together with the $T$-matrix $T_{xy}=e^{-i\pi/12}e^{2\pi ih_x}\delta_{xy}$, they represent the modular group $SL(2,\mathbb{Z})$, and obey \eqref{SL2Z} and \eqref{Verlinde}.

\section{Discrete gauge theories with cyclic, dihedral or dicyclic gauge groups}\label{app:DGT}
We review the topological orders of discrete gauge theories~\cite{BaisDrielPropitius92,Propitius-1995,PropitiusBais96,Preskilllecturenotes} relevant to the twist liquid orbifold phases of $SO(2n)_1$. The discrete gauge theories encountered in this article include \begin{enumerate}\item $D^{[u]}(\mathbb{Z}_k)$ of the cyclic gauge group $\mathbb{Z}_k$ with arbitrary order $k$ and Dijkgraaf-Witten invariant $[u]=[0]$ or $[k/2]$ in $H^3(\mathbb{Z}_k,U(1))=\mathbb{Z}_k$, \item $D^{[u,v,w]}(D_{2k})$ of the dihedral gauge group $D_{2k}$ with even degree $2k$ (order $4k$) and Dijkgraaf-Witten invariant $[u,v,w]=[k,0,0]$ or $[k,1,1]$ in $H^3(D_{2k},U(1))=\mathbb{Z}_k\times\mathbb{Z}_2\times\mathbb{Z}_2$, and \item $D^{[u]}(Q_{4k})$ of the dicyclic gauge group $Q_{4k}$ with odd degree $k$ (order $4k$) and Dijkgraaf-Witten invariant $[u]=[2k]$ in $H^3(Q_{4k},U(1))=\mathbb{Z}_{4k}$.\end{enumerate} They appear in the non-chiral components in the $SO(2n)_1/\mathbb{Z}_2$, $SO(2n)_1/\mathbb{Z}_k$ and $SO(2n)_1/D_k$ twist liquids (see \eqref{Z2orbifoldseven}, \eqref{Zkorbifolds} and \eqref{Dkorbifolds} respectively). In this appendix, while focusing on the above gauge theories, we present a more comprehensive review that (i) includes $D^{[u]}(D_k)$ of the dihedral gauge group $D_k$ with odd degree, and (ii) covers all possible Dijkgraaf-Witten deformations~\cite{DijkgraafWitten90,DijkgraafPasquierRoche91,AltschulerCoste92,BaisvanDrielPropitius93,Propitius-1995} $[u]$ and $[u,v,w]$. 

The global dihedral $D_k=\mathbb{Z}_2\ltimes\mathbb{Z}_k$ symmetry of $SO(2n)_1$ was defined in \eqref{Dksymm}, where $k=n$ (or $n/2$) when $n$ is odd (resp.~even). The dihedral group contains $\mathbb{Z}_k$ as a normal subgroup. The conjugation symmetry group is the quotient group $\mathbb{Z}_2=D_k/\mathbb{Z}_k$, which can also be (non-canonically) embedded as an abnormal subgroup in $D_k$. (See the split exact decompositions \eqref{DkDickgroupextension} for $k$ odd or \eqref{Dkevengroupextension} for $k$ even.) The gauge groups in (1), (2) and (3) are subgroups of the extension $\widehat{D_k}$ of the dihedral symmetry group $D_k$ by the anyon fusion group $\mathcal{A}=\{1,\psi,s\pm\}$ of the Abelian $SO(2n)_1$ topological order, where $\mathcal{A}=\mathbb{Z}_2^2$ when $n$ is even or $\mathcal{A}=\mathbb{Z}_4$ when $n$ is odd. We first classify the group extensions. 

In general, a global symmetry group $G$ is extended by the fusion group $\mathcal{A}$ of Abelian anyons into a quantum symmetry group~\cite{Wenspinliquid02,EtingofNikshychOstrik10,TeoHughesFradkin15,BarkeshliBondersonChengWang14} $\widehat{G}$ \begin{equation}\begin{tikzcd}1\to\mathcal{A}\arrow[r,hookrightarrow,"i"]&\widehat{G}\arrow[r,rightarrow,yshift=0.5ex,"p"]&G\to1\arrow[l,dashrightarrow,yshift=-0.5ex,"j"]\end{tikzcd},\label{appgroupextension}\end{equation} 
 where the above sequence of (forward) group homomorphisms is exact so that the inclusion map $i$ is injective, the projection map $p$ is surjective, and the image of $i$ is identical to the kernel of $p$, $\mathrm{Im}(i)=p^{-1}(1)$. $\mathcal{A}$ is a normal subgroup of $\widehat{G}$ and the global symmetry group is identical to the quotient $G=\widehat{G}/\mathcal{A}$. The (non-canonical) function $j:G\to\widehat{G}$ assigns to each $g$ in $G$, an element $j(g)=\widehat{g}$ in the quantum group $\widehat{G}$ that projects back to the same element $p(j(g))=g$. The function $j$ is not necessarily a group homomorphism and may not preserve the group product. The group extension respects the pre-assigned $G$-action on $\mathcal{A}$ so that $\widehat{g}a\widehat{g}^{-1}=g\cdot a$. Most symmetries in this article are inner automorphisms and associate to trivial $G$-actions, $g\cdot a=a$, except for the conjugation symmetry, which is an outer automorphism that switches the $SO(2n)_1$ spinors, $\mathbb{Z}_2:s_+\leftrightarrow s_-$, when $n$ is odd. 

Group extensions \eqref{appgroupextension} are classified by the group cohomology $H^2(G,\mathcal{A})$. When the $G$-action on $\mathcal{A}$ is trivial, the Abelian group $\mathcal{A}$ sits inside the center of $\widehat{G}$ and commute with all group elements. In this case, $H^2(G,\mathcal{A})$ classifies central extensions. The trivial cohomology corresponds to the direct product $\widehat{G}=G\times\mathcal{A}$. When the $G$-action is non-trivial, the extension is not central and must be non-Abelian. The trivial cohomology in $H^2(G,\mathcal{A})$ corresponds to the semi-direct product $\widehat{G}=G\ltimes\mathcal{A}$, where the group product is $(g_1,a_1)\cdot(g_2,a_2)=(g_1g_2,(g_2\cdot a_1)a_2)$. In these cases, the exact decomposition \eqref{appgroupextension} is said to split, the extension is ``symmorphic'' and $j$ can be chosen to be a group homomorphism that embeds $G\to G\times\mathcal{A}$ (or $G\to G\ltimes\mathcal{A}$). Non-trivial cohomology elements in $H^2(G,\mathcal{A})$ corresponds to ``non-symmorphic'' extensions where $\widehat{G}$ does not decompose into a (semi)direct product and the function $j$ in \eqref{appgroupextension} fails to be a group homomorphism.

The function $j$ preserves the group product when $h(g_1,g_2)=j(g_1)j(g_2)[j(g_1g_2)]^{-1}=\widehat{g_1}\widehat{g_2}(\widehat{g_1g_2})^{-1}$ is the identity. Since $p(h(g_1,g_2))=1$, the exactness of \eqref{appgroupextension} requires $h(g_1,g_2)$ to belong in $\mathcal{A}$. The function $h:G^2\to\mathcal{A}$ obey the 2-cocycle condition \begin{align}dh(g_1,g_2,g_3)=\frac{\left(g_1\cdot h(g_2,g_3)\right)h(g_1,g_2g_3)}{h(g_1g_2,g_3)h(g_1,g_2)}=1\label{2cocyclecondition}\end{align} because $\mathcal{A}$ is Abelian and the group product is associative, $\widehat{g_1}(\widehat{g_2}\widehat{g_3})=(\widehat{g_1}\widehat{g_2})\widehat{g_3}$. The function $j$ in \eqref{appgroupextension} is not unique, and there are other functions $j'=\lambda j$ that respect the projection, $p(j'(g))=g$. Since $p(\lambda(g))=p(j'(g)j(g)^{-1})=gg^{-1}=1$, exactness of \eqref{appgroupextension} requires $\lambda=j'j^{-1}$ to have image in $\mathcal{A}$. This modifies the $h$ function to $h'=hd\lambda$, where the 2-coboundary is \begin{align}d\lambda(g_1,g_2)=\frac{\left(g_1\cdot\lambda(g_2)\right)\lambda(g_1)}{\lambda(g_1g_2)}.\end{align} $h'$ obeys the cocyle condition \eqref{2cocyclecondition} because $d(d\lambda)=1$. The equivalent classes $[h]=[hd\lambda]$ of 2-cocycles (modulo 2-coboundaries) form the group cohomology~\cite{Cohomologybook} \begin{align}H^2(G,\mathcal{A})=\frac{\mathrm{ker}\left(d:C^2(G,\mathcal{A})\to C^3(G,\mathcal{A})\right)}{\mathrm{Im}\left(d:C^1(G,\mathcal{A})\to C^2(G,\mathcal{A})\right)}\end{align} where $C^r(G,\mathcal{A})$ is the Abelian group of $r$-cochains -- functions that map $G^r\to\mathcal{A}$. 

Here, we summarize the extension $\widehat{D_k}$ of the global symmetry group $D_k$ of $SO(2n)_1$, where $D_k=\left\langle\mu,m|\mu^2=m^k=(\mu m)^2=1\right\rangle$. These quantum symmetry groups can be found in section~\ref{sec:SO2nZ2TL}, \ref{sec:SO2nZkTL}, \ref{SO2nDkTL}, and the corresponding cohomology classifications can be found in appendix~\ref{app:cohomology}. (1) When $n=2k\equiv0$ modulo 4, the symmetry is ``non-symmorphically'' extended to $\widehat{D_k}=D_{2k}\times\mathbb{Z}_2$, which corresponds to the cohomology class $[x,1,1]$ in $H^2(D_k,\mathcal{A})=\mathcal{A}^3$ where $x=s_\pm$ (or $x=\psi$) when $n\equiv0$ (resp.~$n\equiv4$) modulo 8. The degree $2k$ dihedral group is \begin{align}D_{2k}=\left\langle\mu,\hat{m},x\left|\begin{array}{*{20}l}\hat{m}^k=x,\\\mu^2=\hat{m}^{2k}=(\mu\hat{m})^2=1\end{array}\right.\right\rangle.\end{align} It contains the subgroups $\mathbb{Z}_2=\langle\mu|\mu^2=1\rangle$, which is the conjugation symmetry group, and $\mathbb{Z}_{2k}=\langle\hat{m}|\hat{m}^{2k}=1\rangle$, which non-trivially extends the $\mathbb{Z}_k$ symmetry group. (2) When $n=2k\equiv2$ modulo 4, the symmetry is ``non-symmorphically'' extended to $\widehat{D_k}=Q_{4k}\times\mathbb{Z}_2$, which corresponds to the cohomology class $[\psi]$ in $H^2(D_k,\mathcal{A})=\mathcal{A}$. The dicyclic group is \begin{align}Q_{4k}&=\left\langle\hat{\mu},m,\psi\left|\begin{array}{*{20}l}\hat{\mu}^2=(\hat{\mu}m)^2=\psi,\\\psi^2=m^k=1\end{array}\right.\right\rangle\nonumber\\&=\left\langle\hat{\mu},\hat{m},\psi\left|\begin{array}{*{20}l}\hat{\mu}^2=\hat{m}^k=(\hat{\mu}\hat{m})^2=\psi,\\\psi^2=\hat{\mu}^4=\hat{m}^{2k}=1\end{array}\right.\right\rangle,\label{appDickdefinition}\end{align} where $\hat{m}=m\psi$. It contains the subgroups $\mathbb{Z}_4=\langle\hat{\mu}|\hat{\mu}^4=1\rangle$, which non-trivially extends the conjugation symmetry group $\mathbb{Z}_2$, and $\mathbb{Z}_k=\langle m|m^k=1\rangle$. (3) When $n=k$ is odd, the symmetry is ``symmorphically'' extended to $\widehat{D_k}=D_k\ltimes\mathcal{A}_{\mathrm{odd}}$, where the twofold conjugation symmetry acts non-trivially on $\mathcal{A}_{\mathrm{odd}}=\mathbb{Z}_4$ as the involution $s_+\leftrightarrow s_-$. The extension corresponds to the trivial element in $H^2(D_k,\mathcal{A}_{\mathrm{odd}})=\mathbb{Z}_2$.

Next, we describe the topological orders of the discrete gauge theory components $D^{[\omega]}(\mathbb{Z}_k)$, $D^{[\omega]}(D_k)$ and $D^{[\omega]}(Q_{4k})$ of the $SO(2n)_1/\mathbb{Z}_2$, $SO(2n)_1/\mathbb{Z}_k$ and $SO(2n)_1/D_2$ twist liquids in \eqref{Z2orbifoldseven}, \eqref{Zkorbifolds} and \eqref{Dkorbifolds}. In general, anyons in a discrete gauge theory $D(G)$ are dyons $\chi=([g],\rho)$. $[g]=\{hgh^{-1}:h\in G\}$ is the conjugacy class of the group element $g$, and $\rho:Z_g\to U(N)$ is a unitary irreducible representation of the centralizer subgroup $Z_g=\{h\in G:gh=hg\}$, which carries group elements that commute with $g$. The conjugacy class $[g]$ and the representation $\rho$ specify the gauge flux and gauge charge of the dyon. When the gauge group is the Abelian cyclic group $G=\mathbb{Z}_k=\{1,m,\ldots,m^{k-1}\}$, a conjugacy class only contains a single element $[m^j]=\{m^j\}$, and its centralizer group is the entire group $Z_{m^j}=\mathbb{Z}_k$. The irreducible representations are the homomorphism $z^b:\mathbb{Z}_k\to U(1)$ sending $m^a\to z^b(m^a)=e^{2\pi iab/k}$, where $a,b=0,1,\ldots,k-1$. Dyons in $D(\mathbb{Z}_k)$ are the flux-charge composites $\chi^{ab}=(m^a,z^b)$. Table~\ref{tab:charactersDkodd}, \ref{tab:charactersDkodd} and \ref{tab:charactersDkodd} summarize the conjugacy classes and irreducible representations of the dihdral group $D_k$ with odd and even degree $k$, and the dicyclic group $Q_{4k}$ with odd degree $k$. Possible centralizer subgroups are $\mathbb{Z}_2$, $D_2=\mathbb{Z}_2\times\mathbb{Z}_2$, $\mathbb{Z}_4$, $\mathbb{Z}_k$, $\mathbb{Z}_{2k}$, and the entire group $D_k$ or $Q_{4k}$. Irreducible representations of the Abelian centralizers $Z$ form the groups of homomorphisms $\mathrm{Hom}(Z,U(1))$, which is isomorphic to $Z$. The quantum dimension of a dyon $\chi=([g],\rho)$ is the integer $d_\chi=|[g]|\dim(\rho)$. The total quantum dimension of the discrete gauge theory is the order of the group \begin{align}|G|=\sqrt{\sum_\chi d^2_\chi}.\end{align} The number of dyons (i.e.~the ground state degeneracy on a torus) is $\mathrm{GSD}(D(\mathbb{Z}_k))=k^2$, $\mathrm{GSD}(D(D_k))=(k^2+7)/2$ when $k$ is odd, $\mathrm{GSD}(D(D_k))=(k^2+28)/2$ when $k$ is even, and $\mathrm{GSD}(D(\mathrm{Dick}_k))=2(k^2+7)$ when $k$ is odd. The dyons and their quantum dimensions of $D(D_k)$ and $D(Q_{4k})$ are listed in table~\ref{tab:dyonsDkodd}, \ref{tab:dyonsDkeven} and \ref{tab:dyonsDick}.

\begin{table}[htbp]
\centering
\begin{tabular}{c|ccc}
&$[1]$&$[\mu]$&$[m^j]$\\\hline
$A_0$&1&1&1\\
$A_1$&1&$-1$&1\\
$E_l$&2&0&$2\cos(2\pi jl/k)$
\end{tabular}
\caption{Character table of $D_k$ when $k$ is odd. $\rho=A_{0,1},E_l$ are irreducible representations. $j,l=1,\ldots,(k-1)/2$. Numerical entries are $\mathrm{Tr}(\rho(g))$.}\label{tab:charactersDkodd}
\end{table}

\begin{table}[htbp]
\centering
\begin{tabular}{c|ccccc}
&$[1]$&$[\mu]$&$[\mu m]$&$[m^j]$&$[m^{k/2}]$\\\hline
$A_0$&1&1&1&1&1\\
$A_1$&1&$-1$&$-1$&1&1\\
$B_0$&1&1&$-1$&$(-1)^j$&$(-1)^{k/2}$\\
$B_1$&1&$-1$&1&$(-1)^j$&$(-1)^{k/2}$\\
$E_l$&2&0&0&$2\cos(2\pi jl/k)$&$2(-1)^l$
\end{tabular}
\caption{Character table of $D_k$ when $k$ is even $\rho=A_{0,1},B_{0,1},E_l$ are irreducible representations. $j,l=1,\ldots,k/2-1$. Numerical entries are $\mathrm{Tr}(\rho(g))$.}\label{tab:charactersDkeven}
\end{table}

\begin{table}[htbp]
\centering
\begin{tabular}{c|ccccc}
&$[1]$&$[\hat{\mu}]$&$[\hat{\mu}^2]$&$[\hat{\mu}^3]$&$[\hat{m}^j]$\\\hline
$A_0$&1&1&1&1&1\\
$A_1$&1&$e^{i\pi/2}$&$-1$&$e^{-i\pi/2}$&$(-1)^j$\\
$A_2$&1&$-1$&1&$-1$&1\\
$A_3$&1&$e^{-i\pi/2}$&$-1$&$e^{i\pi/2}$&$(-1)^j$\\
$E_l$&2&0&$2(-1)^l$&0&$2(-1)^{jl}\cos(2\pi jl/k)$
\end{tabular}
\caption{Character table of $Q_{4k}$ when $k$ is odd. $\rho=A_{0,1,2,3},E_l$ are irreducible representations. $j,l=1,\ldots,k-1$. Numerical entries are $\mathrm{Tr}(\rho(g))$.}\label{tab:charactersDick}
\end{table}

\begin{table}[htbp]
\centering
\begin{tabular}{llll}
dyons&&$d_\chi$&$\theta_\chi=e^{2\pi ih_\chi}$\\\hline
$([1],A_\nu)$                    &$\nu=0,1$                     &1                   &1\\
$([1],E_l)$                      &$l=1,\ldots,(k-1)/2$          &2                   &1\\
\multirow{2}{*}{$([m^j],z^b)$}   &$j=1,\ldots,(k-1)/2$,         &\multirow{2}{*}{2}  &\multirow{2}{*}{$e^{2\pi ijb/k}\vartheta_{m^j}$}\\
                                 &$b=0,\ldots,k-1$              &                    &\\
$([\mu],\zeta^\lambda)$          &$\lambda=0,1$                 &$k$                 &$(-1)^\lambda\vartheta_\mu$
\end{tabular}
\caption{Dyons $\chi=([g],\rho)$ in $D^{[u]}(D_k)$, their quantum dimensions $d_\chi$ and topological spins $\theta_\chi$, when $k$ is odd. $\zeta^\lambda(\mu^A)=(-1)^{\lambda A}$ are irreducible representations of $Z_\mu=\mathbb{Z}_2={\mu^A:A=0,1}$. The deformation phases $\vartheta_{m^j}$ and $\vartheta_\mu$ depend on the Dijkgraaf-Witten invariant $[u]$ and are computed in \eqref{deformationDkodd}.}\label{tab:dyonsDkodd}
\end{table}

\begin{table}[htbp]
\centering
\begin{tabular}{llll}
dyons&&$d_\chi$&$\theta_\chi=e^{2\pi ih_\chi}$\\\hline
$([1],A_\nu)$             &\multirow{4}{*}{$\nu=0,1$}       &1                    &1\\
$([m^{k/2}],A_\nu)$       &                                 &1                    &$\vartheta_{m^{k/2}}$\\
$([1],B_\nu)$             &                                 &1                    &1\\
$([m^{k/2}],B_\nu)$       &                                 &1                    &$(-1)^{k/2}\vartheta_{m^{k/2}}$\\
$([1],E_l)$               &\multirow{2}{*}{$l=1,\ldots,k/2-1$}       &2           &1\\
$([m^{k/2}],E_l)$         &                                          &2           &$(-1)^l\vartheta_{m^{k/2}}$\\
\multirow{2}{*}{$([m^j],z^b)$}           &$j=1,\ldots,k/2-1$,        &\multirow{2}{*}{2}     &\multirow{2}{*}{$e^{2\pi ijb/k}\vartheta_{m^j}$}\\
                                         &$b=0,\ldots,k-1$           &                       &\\
$([\mu],\zeta^{\lambda_1,\lambda_2})$    &\multirow{2}{*}{$\lambda_1,\lambda_2=0,1$}  &$k/2$  &$(-1)^{\lambda_1}\vartheta_\mu$\\
$([\mu m],\zeta^{\lambda_1,\lambda_2})$  &                                            &$k/2$  &$(-1)^{\lambda_1}\vartheta_{\mu m}$
\end{tabular}
\caption{Dyons $\chi=([g],\rho)$ in $D^{[u,v,w]}(D_k)$ their quantum dimensions $d_\chi$ and topological spins $\theta_\chi$, when $k$ is even. $\zeta^{\lambda_1,\lambda_2}(\mu^{A_1}m^{kA_2/2})=(-1)^{\lambda_1A_1+\lambda_2A_2}$ are irreducible representations of $Z_{\mu}=\mathbb{Z}_2\times\mathbb{Z}_2=\{1,\mu,m^{k/2},\mu m^{k/2}\}$. The deformation phases $\vartheta_{m^j}$, $\vartheta_\mu$ and $\vartheta_{\mu m}$ depend on the Dijkgraaf-Witten invariant $[u,v,w]$ and are computed in \eqref{deformationDkeven}.}\label{tab:dyonsDkeven}
\end{table}

\begin{table}[htbp]
\centering
\begin{tabular}{llll}
dyons&&$d_\chi$&$\theta_\chi=e^{2\pi ih_\chi}$\\\hline
$([1],A_\nu)$             &\multirow{2}{*}{$\nu=0,1,2,3$}            &1                  &1\\
$([\hat{\mu}^2],A_\nu)$   &                                          &1                  &$(-1)^\nu\vartheta_{\hat{\mu}^2}$\\
$([1],E_l)$               &\multirow{2}{*}{$l=1,\ldots,k-1$}         &2                  &1\\
$([\hat{\mu}^2],E_l)$     &                                          &2                  &$(-1)^l\vartheta_{\hat{\mu}^2}$\\
\multirow{2}{*}{$([\hat{m}^j],z^b)$}  &$j=1,\ldots,k-1$,  &\multirow{2}{*}{2} &\multirow{2}{*}{$e^{i\pi jb/k}\vartheta_{\hat{m}^j}$}\\
                                      &$b=0,\ldots,2k-1$  &                   &\\
$([\hat{\mu}],\zeta^\lambda)$   &\multirow{2}{*}{$\lambda=0,1,2,3$}  &$k$                &$e^{i\pi\lambda/2}\vartheta_{\hat{\mu}}$\\
$([\hat{\mu}^{-1}],\zeta^\lambda)$ &                                    &$k$                &$e^{-i\pi\lambda/2}\vartheta_{\hat{\mu}^{-1}}$
\end{tabular}
\caption{Dyons $\chi=([g],\rho)$ in $D^{[u]}(Q_{4k})$ their quantum dimensions $d_\chi$ and topological spins $\theta_\chi$, when $k$ is odd. $z^b(\hat{m}^j)=e^{i\pi bj/k}$ and $\zeta^\lambda(\hat{\mu}^A)=e^{i\pi\lambda A/2}$ are irreducible representations of $Z_{\hat{m}^j}=\mathbb{Z}_{2k}=\{\hat{m}^j:j=0,1,\ldots,2k-1\}$ and $Z_{\hat{\mu}}=\mathbb{Z}_4=\{\hat{\mu}^A:A=-1,0,1,2\}$ respectively. The deformation phases $\vartheta_{\hat{m}^j}$, $\vartheta_{\hat\mu}$ and $\vartheta_{\hat\mu^2}$ depend on the Dijkgraaf-Witten invariant $[u]$ and are computed in \eqref{deformationDick}.}\label{tab:dyonsDick}
\end{table}

A finite group $G$ can be promoted to a fusion category $\mathcal{F}^{[\omega]}(G)$. The simple objects of the tensor category are the group elements. Fusion rules are group multiplications, $g_1\times g_2=g_1g_2$. The associativity equivalence $(g_1\times g_2)\times g_3\equiv g_1\times(g_2\times g_3)$ is described by the $U(1)$ basis transformation \begin{align}\left|\vcenter{\hbox{\includegraphics[width=0.5in]{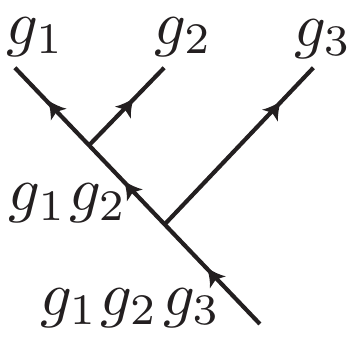}}}\right\rangle=f^{g_1g_2g_3}\left|\vcenter{\hbox{\includegraphics[width=0.5in]{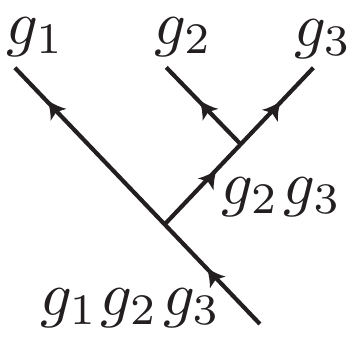}}}\right\rangle.\label{Fsymboldef}\end{align} The collection of $f$-symbols are 3-cochains in $C^3(G,U(1))$ that obey the cocycle condition (also known as the pentagon identity)~\cite{DijkgraafWitten90,DijkgraafPasquierRoche91,AltschulerCoste92,BaisvanDrielPropitius93,Propitius-1995} \begin{align}df^{g_1g_2g_3g_4}=\frac{f^{g_1g_2g_3}f^{g_1(g_2g_3)g_4}f^{g_2g_3g_4}}{f^{(g_1g_2)g_3g_4}f^{g_1g_2(g_3g_4)}}=1.\label{3cocycle}\end{align} This is because the following pentagon diagram is commutative. \begin{align}\vcenter{\hbox{\includegraphics[width=0.35\textwidth]{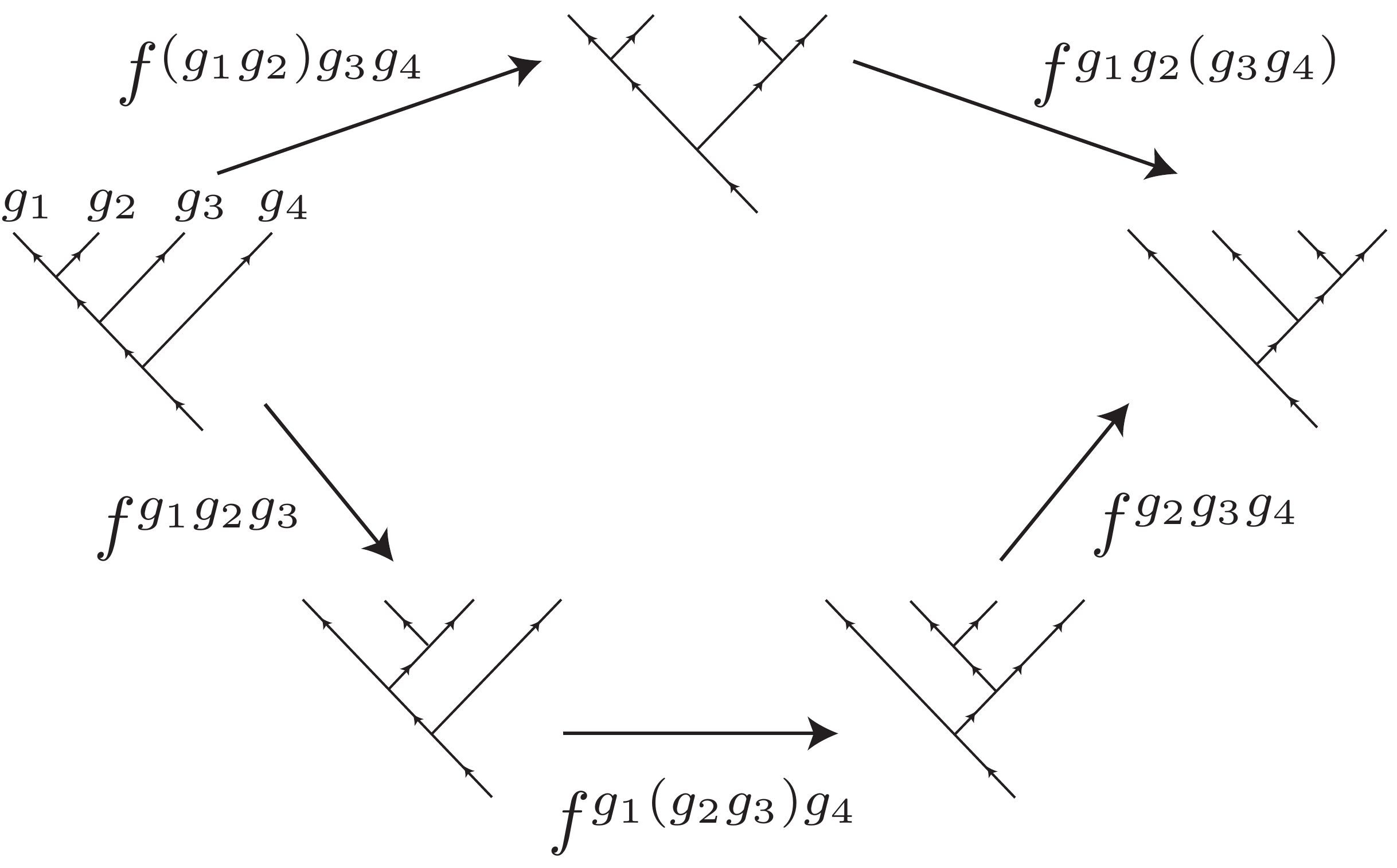}}}\end{align} The $f$-symbols are gauge dependent and can be changed under a gauge transformation of the quantum states, $f^{g_1g_2g_3}\to f^{g_1g_2g_3}d\Lambda^{g_1g_2g_3}$, where \begin{align}d\Lambda^{g_1g_2g_3}=\frac{\Lambda^{g_1g_2}\Lambda^{(g_1g_2)g_3}}{\Lambda^{g_1(g_2g_3)}\Lambda^{g_2g_3}}\label{3coboundary}\end{align} is the coboundary of $\Lambda^{g_1g_2}$, the gauge transformation of the fusion vertex $|g_1\times g_2\rangle\to\Lambda^{g_1g_2}|g_1\times g_2\rangle$. Equivalent classes of $f$-symbols $[\omega]=f\cdot dC^2(G,U(1))$ are cohomology elements in \begin{align}H^3(\mathbb{Z}_n,U(1))=\frac{\mathrm{ker}\left(d:C^3(G,U(1))\to C^4(G,U(1))\right)}{\mathrm{Im}\left(d:C^2(G,U(1))\to C^3(G,U(1))\right)}.\end{align} The cohomology class $[\omega]$ is also referred to as a Dijkgraaf-Witten invariant.~\cite{DijkgraafWitten90,DijkgraafPasquierRoche91,AltschulerCoste92,BaisvanDrielPropitius93,Propitius-1995} The group cohomology classifications and explicit cocycle representatives of cohomology elements of the relevant gauge groups $G=\mathbb{Z}_k$, $D_k$, and $Q_{4k}$ are presented in appendix~\ref{app:cohomology}. 

The discrete gauge theory $D^{[\omega]}(G)$ is the Drinfeld center~\cite{Kasselbook,BakalovKirillovlecturenotes,LevinWen05} of the fusion category $\mathcal{F}^{[\omega]}(G)$. The trivial cohomology class $[0]$ associates to the conventional (or un-deformed) gauge theory $D^{[0]}(G)$, where all the $f$-symbols can be chosen to be trivial, $f^{g_1g_2g_3}=1$. In this case, all pure gauge fluxes, whose charge components (i.e.~centralizer representations) are trivial, carry bosonic exchange statistics. When a gauge theory $D^{[\omega]}(G)$ is deformed by a non-trivial Dijkgraaf-Witten invariant $[\omega]$, the $f$-symbols are necessarily non-trivial. They modify the exchange statistics of gauge fluxes. The collection of $U(1)$ exchange phases $r^{gh}$, each relates \begin{gather}|g\times h\rangle=r^{gh}|h\times\left(h^{-1}gh\right)\rangle,\\\vcenter{\hbox{\includegraphics[width=0.4\textwidth]{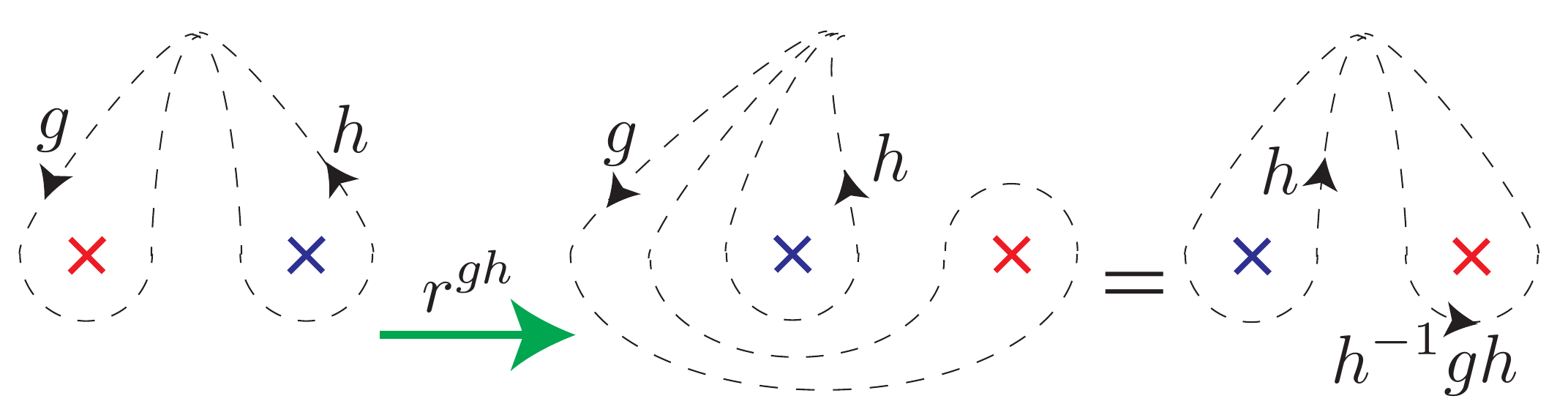}}}\nonumber\end{gather} obeys the hexagon identity~\cite{DijkgraafPasquierRoche91,AltschulerCoste92,BaisvanDrielPropitius93,Propitius-1995} \begin{align}\begin{split}&r^{g_1g_2}f^{g_2(g_2^{-1}g_1g_2)g_3}r^{(g_2^{-1}g_1g_2)g_3}\\&=f^{g_1g_2g_3}r^{g_1(g_2g_3)}f^{g_2g_3(g_3^{-1}g_2^{-1}g_1g_2g_3)}.\end{split}\label{hexagoneq}\end{align} This is because the following hexagon diagram is commutative. \begin{align}\vcenter{\hbox{\includegraphics[width=0.4\textwidth]{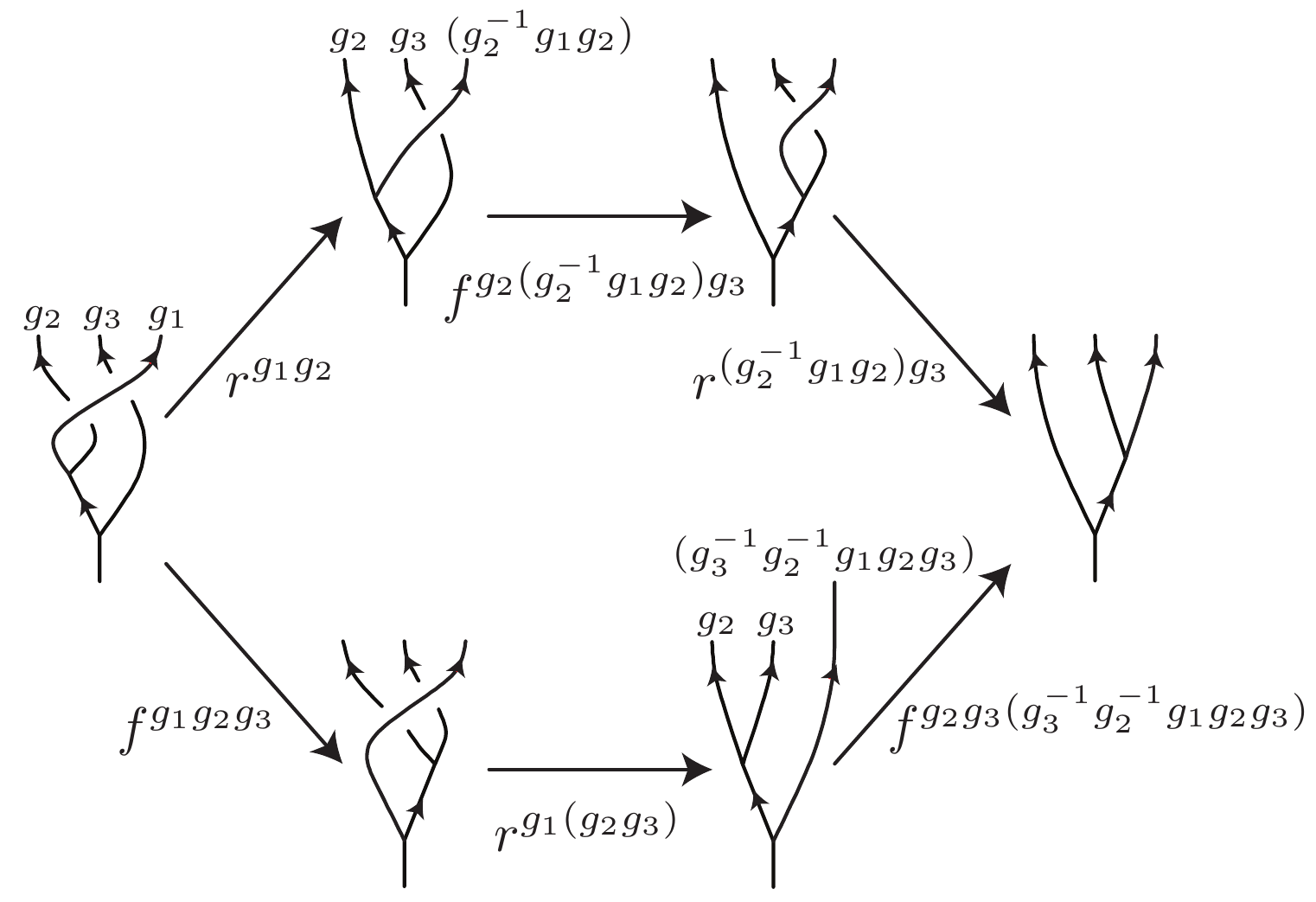}}}\end{align} The self-exchange phase of a pure gauge flux $g$ is $\vartheta_g=r^{gg}$. In general, the topological spin $\theta_\chi=e^{2\pi ih_\chi}$ of a dyon $\chi=([g],\rho)$ is the $U(1)$ phase \begin{align}\theta_\chi=\left\langle\rho(g)\right\rangle\vartheta_g,\label{dyonspin}\end{align} where $\rho:Z_g\to U(N_\rho)$ is the irreducible representation of the centralizer subgroup $Z_g$, $\rho(g)=\langle\rho(g)\rangle\mathbb{I}_{N_\rho\times N_\rho}$, and $\langle\rho(g)\rangle=\mathrm{Tr}\left(\rho(g)\right)/N_\rho$. The modular $T$-matrix is the diagonal unitary matrix $T_{\chi\chi'}=\delta_{\chi\chi'}\theta_\chi$. The modular $S$-matrix is the symmetric unitary matrix~\cite{DijkgraafPasquierRoche91,BaisvanDrielPropitius93,Propitius-1995} \begin{widetext}\begin{align}S_{\chi\chi'}=\frac{1}{|G|}\sum_{\substack{g_i\in[g],g'_j\in[g']\\g_ig'_j=g'_jg_i}}\mathrm{Tr}\left(\rho\left(x_i^{-1}g'_jx_i\right)\right)\mathrm{Tr}\left(\rho'\left({x'}_j^{-1}g_ix'_j\right)\right)r^{g_ig'_j}r^{g'_jg_i},\label{DGTSmatrix}\end{align}\end{widetext} where $\chi=([g],\rho)$ and $\chi'=([g'],\rho')$ are dyons, whose flux components are conjugacy classes $[g]=\{g_i=x_igx_i^{-1}:i=1,\ldots,|[g]|\}$ and $[g']=\{g'_j=x'_jg'{x'}_j^{-1}:j=1,\ldots,|[g']|\}$. Together the $S$ and $T$ matrices generate a representation of the modular group $SL(2,\mathbb{Z})$. They obey \begin{align}S^2=(ST^\dagger)^3,\quad S^4=\mathbb{I},\end{align} where $C=S^2$ is the conjugation matrix so that $C_{\chi\bar\chi}=1$ if $\chi$ and $\bar\chi$ are conjugate pair, $\chi\times\bar\chi=1+\ldots$, and $C_{\chi\chi'}=0$ if $\chi'\neq\bar\chi$. The fusion rule \begin{align}\chi_1\times\chi_2=\sum_{\chi}N_{\chi_1\chi_2}^\chi\chi\end{align} is determined by the non-negative integral fusion number $N_{\chi_1\chi_2}^\chi$, which is related to the $S$-matrix by the Verlinde formula~\cite{Verlinde88} \begin{align}N_{\chi_1\chi_2}^\chi=\sum_{\chi'}\frac{S_{\chi_1\chi'}S_{\chi_2\chi'}S_{\chi\chi'}^\ast}{d_{\chi'}/|G|}.\end{align} The $S$-matrix can in turn be computed by plugging in the fusion numbers and topological spins into the ribbon identity~\cite{Kitaev06} \begin{align}S_{\chi_1\chi_2}=\frac{1}{|G|}\sum_\chi d_\chi N_{\chi_1\chi_2}^\chi\frac{\theta_\chi}{\theta_{\chi_1}\theta_{\chi_2}}.\end{align}

\subsection{The \texorpdfstring{$\mathbb{Z}_k$}{Zk} gauge theory}\label{app:ZkGT}
The $f$-symbols of the quantum double $D^{[u]}(\mathbb{Z}_k)$ that represent the Dijkgraaf-Witten invariant $[u]$ in $H^3(\mathbb{Z}_k,U(1))=\mathbb{Z}_k$ can be chosen to be \begin{align}\begin{split}f^{a_1a_2a_3}&=\omega^3_u(a_1,a_2,a_3)\\&=\exp\left(\frac{2\pi iu}{k^2}a_1\left(a_2+a_3-[a_2+a_3]\right)\right)\end{split}\end{align} from the cocycle representative $\omega^3_u$ in \eqref{repHZkU1}, where $a_i\in(-k/2,k/2]\cap\mathbb{Z}$ label group elements in $\mathbb{Z}_k$ and the square bracket $[a_2+a_3]$ wraps the number back to the appropriate range $(-k/2,k/2]$ by adding or subtracting $k$ to $a_2+a_3$ if necessary. 

When the Dijkgraaf-Witten invariant is trivial, $[u]=[0]$, the solutions to the hexagon equation \eqref{hexagoneq} are homomorphisms $r^{a_1\ast}:\mathbb{Z}_k\to U(1)$ sending $a_2$ to $r^{a_1a_2}=\exp(2\pi i p_{a_1}a_2/k)$, where $p_a$ are integers. These exchange phases can be canceled by attaching the gauge charge $z^{-p_a}$ to the gauge flux $m^a$. Therefore, $p_a$ can be chosen to be 0 without loss of generality. In this case, all pure fluxes $m^a$ are bosonic because their self-exchange phases \eqref{dyonspin} are trivial, $\theta_{m^a}=\vartheta_a=r^{aa}=1$. The topological spin of a general dyon $\chi^{ab}=m^az^b$ is $\theta_{m^az^b}=e^{2\pi iab/k}$. The modular $S$-matrix defined in \eqref{DGTSmatrix} summarizes the $2\pi$ braiding $kS_{\chi^{ab}\chi^{a'b'}}=e^{2\pi i(ab'+ba')/k}$ between $\chi^{ab}$ and $\chi_{a'b'}$. The $D^{[0]}(\mathbb{Z}_k)$ gauge theory can be described by a topological Chern-Simons field theory with the action $\mathcal{S}=\int_{2+1}K_{IJ}\alpha^I\wedge d\alpha^J/(4\pi)$, where the $K$-matrix is $K=2\sigma_x$. The dyon $\chi^{ab}=m^az^b$ is associated to the integral vector ${\bf v}=(a,b)^T$, and the spin and braiding statistics are captured by $\theta_{\bf v}=e^{\pi i{\bf v}^TK^{-1}{\bf v}}$ and $nS_{{\bf u}{\bf v}}=e^{2\pi i{\bf u}^TK^{-1}{\bf v}}$.

For a general Dijkgraaf-Witten deformation $[u]$, the solution to the hexagon equation \eqref{hexagoneq} can be chosen to be~\cite{Propitius-1995} \begin{align}r_u^{a_1a_2}=\exp\left(\frac{2\pi iu}{k^2}a_1a_2\right).\end{align} The discrete gauge theory depends only on $u$ modulo $k$ because $r_{u+k}=r_ur_k$ differs from $r_u$ by $r_k^{a_1a_2}=e^{2\pi ia_1a_2/k}$, which can be canceled attaching gauge fluxes. These exchange phases $r^{a_1a_2}_u$ modify the spin and braiding statistics of the dyons. Pure fluxes $m^a$ are no longer bosonic but carry non-trivial exchange statistics $\vartheta_a=r^{aa}=e^{2\pi iua^2/k^2}$. The topological spins \eqref{dyonspin} and the modular $S$-matrix \eqref{DGTSmatrix} of dyons $\chi^{ab}=m^az^b$ are \begin{align}\begin{split}\theta_{\chi^{ab}}&=\vcenter{\hbox{\includegraphics[width=0.5in]{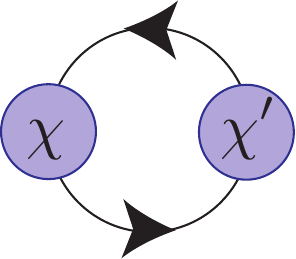}}}=e^{2\pi i\left(\frac{ab}{k}+\frac{ua^2}{k^2}\right)},\\kS_{\chi^{ab}\chi^{a'b'}}&=\vcenter{\hbox{\includegraphics[width=0.6in]{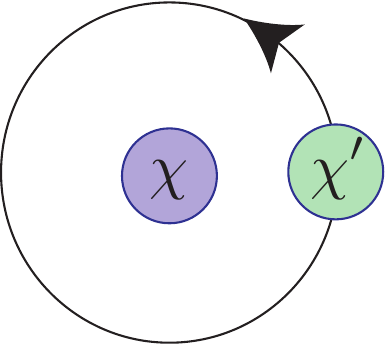}}}=e^{2\pi i\left(\frac{ab'+ba'}{k}+\frac{2uaa'}{k^2}\right)}.\end{split}\label{Zkspinbraiding}\end{align}

In section~\ref{sec:SO2nZ2TL} and \ref{sec:SO2nZkTL}, we encounter the discrete gauge theories $D^{[0]}(\mathbb{Z}_2)$, $D^{[1]}(\mathbb{Z}_2)$ and $D^{[2]}(\mathbb{Z}_4)$ in \eqref{Z2orbifoldseven} as well as $D^{[0]}(\mathbb{Z}_k)$ and $D^{[k]}(\mathbb{Z}_{2k})$ in \eqref{Zkorbifolds}. In general, for any even order $n$, by choosing $u=-n/2$ to represent the Dijkgraaf-Witten invariant $[n/2]$, the discrete gauge theory $D^{[n/2]}(\mathbb{Z}_n)$ has the non-chiral $U(1)_n\times\overline{U(1)_n}$ topological order. The theory can be described by the topological Chern-Simons field theory $\mathcal{S}=\int_{2+1}K_{IJ}\alpha^I\wedge d\alpha^J/(4\pi)$ with the $K$-matrix $K=n\sigma_z$. The dyon $\chi^{ab}=m^az^b$ is represented by the integral vector ${\bf v}=(b,a-b)$ so that the spin and braiding statistics \eqref{Zkspinbraiding} agrees with $\theta_{\bf v}=e^{\pi i{\bf v}^TK^{-1}{\bf v}}$ and $kS_{{\bf v}{\bf v}'}=e^{2\pi i{\bf v}^TK^{-1}{\bf v}'}$.

\subsection{The \texorpdfstring{$D_k$}{Dk} gauge theory}\label{app:DkGT}
The cohomology classification $H^3(D_k,U(1))$ of $f$-symbols depends on the parity of $k$ (see \eqref{listofcohomologies} in appendix~\ref{app:DGT}). We begin with dihedral groups with odd degree $k$. The $f$-symbols of $D^{[u]}(D_k)$ that represent the Dijkgraaf-Witten invariant $[u]$ in $H^3(D_k,U(1))=\mathbb{Z}_{2k}$ can be identified with the cocycle representative $\xi^3_u$ in \eqref{repH3DkoddU1}, \begin{align}f^{g_1g_2g_3}=\xi^3_u(g_1,g_2,g_3).\end{align} The solution to the hexagon equation \eqref{hexagoneq} can be chosen to be~\cite{Propitius-1995} \begin{widetext}\begin{align}r^{g_1g_2}_u=\exp\left(\frac{2\pi iu}{k^2}\left\{a_2\left[(-1)^{A_2}a_1+2A_1a_2\right]-A_1a_2^2\right\}\right)\exp\left(\frac{i\pi u}{2}A_1A_2\right),\label{rDkodd}\end{align}\end{widetext} where $g_i=(A_i,a_i)=\mu^{A_i}m^{a_i}$ are group elements in $D_k$, for $A_i=0,1$ and $a_i\in(-k/2,k/2)\cap\mathbb{Z}$. The square bracket wraps $\left[(-1)^{A_2}a_1+2A_1a_2\right]$ back to the appropriate range $(-k/2,k/2)$ by adding or subtracting $k$ to $(-1)^{A_2}a_1+2A_1a_2$ if necessary. The exchange phases $r_{u=2k}^{g_1g_2}$ and $(2k)^{\mathrm{th}}$ roots of unity and can be absorbed by attaching gauge charges. Therefore $r_{u+2k}$ and $r_u$ correspond to the equivalent gauge theory $D^{[u]}(D_k)$. The dyons and their quantum dimensions and topological spins are listed in table~\ref{tab:dyonsDkodd} under the deformation phases \begin{align}\vartheta_{m^j}=r^{m^jm^j}=e^{2\pi iuj^2/k^2},\quad\vartheta_\mu=r^{\mu\mu}=i^u.\label{deformationDkodd}\end{align} The modular $S$-matrix can be computed with the help of the character table~\ref{tab:charactersDkodd} and by substituting the exchange phases \eqref{rDkodd} in \eqref{DGTSmatrix}. It is presented in table~\ref{tab:SDkodd}. 

\begin{table}[htbp]
\centering
\begin{tabular}{c|cccc}
&$([1],A_\nu)$&$([1],E_l)$&$([m^j],z^b)$&$([\mu],\zeta^\lambda)$\\\hline
$([1],A_{\nu'})$&1&2&2&$k(-1)^{\nu'}$\\
$([1],E_{l'})$&2&4&$4c(jl')$&0\\
$([m^{j'}],z^{b'})$&2&$4c(j'l)$&$4s^u_{jbj'b'}$&0\\
$([\mu],\zeta^{\lambda'})$&$k(-1)^\nu$&0&0&$k(-1)^{\lambda+\lambda'+u}$
\end{tabular}
\caption{The $S$-matrix $2kS_{\chi\chi'}$ from \eqref{DGTSmatrix} of the $D^{[u]}(D_k)$ gauge theory, for odd $k$. We abbreviate $c(x)=\cos(2\pi x/k)$ and $s^u_{jbj'b'}=c(jb'+j'b+2ujj'/k)$. The row and column entries are arranged in the same order of the dyons in table~\ref{tab:dyonsDkodd}.}\label{tab:SDkodd}
\end{table}

When the degree $k$ is even, dihedral gauge theories $D^{[u,v,w]}(D_k)$ are classified by the Dijkgraaf-Witten invariant $[u,v,w]$ in $H^3(D_k,U(1))=\mathbb{Z}_k\times\mathbb{Z}_2\times\mathbb{Z}_2$. Their $f$-symbols can be identified with the cocycle representative $\xi^3_{u,v,w}$ in \eqref{repH3DkevenU1}, \begin{align}f^{g_1g_2g_3}=\xi^3_{u,v,w}(g_1,g_2,g_3).\end{align} The solution to the hexagon equation \eqref{hexagoneq} can be chosen to be~\cite{footnoteDkGTeven} 
\begin{align}\begin{split}r^{g_1g_2}_{u,v,w}&=\exp\left(\frac{2\pi iu}{k^2}\left\{a_2\left[(-1)^{A_2}a_1+2A_1a_2\right]_k-A_1a_2^2\right\}\right)\exp\left(\frac{i\pi}{2}\left(vA_1A_2+(v+w)[a_1]_2A_2\right)\right),\\&\quad\mbox{if $(A_1,a_1)\neq(0,k/2)$},\\r^{g_1g_2}_{u,v,w}&=\exp\left(-\frac{i\pi u}{k}a_2\right)\exp\left(\frac{i\pi}{2}(v+w)\left[\frac{k}{2}\right]_2A_2\right),\\&\quad\mbox{if $(A_1,a_1)=(0,k/2)$}.\end{split}\label{rDkeven}\end{align}
Here, $A_i=0,1$ and $a_i\in(-k/2,k/2]\cap\mathbb{Z}$ label group elements $g_i=(A_i,a_i)=\mu^{A_i}m^{a_i}$ in $D_k$. The square brackets $[\ast]_k$ and $[\ast]_2$ wrap the numbers back to the appropriate ranges, $(-k/2,k/2]$ for the former and $0,1$ for the latter, by respectively adding or subtracting $k$ or 2 if necessary. Like the previous case, the Dijkgraaf-Witten invariants $u,v,w$ have periodicities $k,2,2$ respectively because $r_{k,0,0}$, $r_{0,2,0}$ and $r_{0,0,2}$ are $(2k)^{\mathrm{th}}$ roots of unity and can be absorbed by attaching gauge charges. The dyons and their quantum dimensions and topological spins are listed in table~\ref{tab:dyonsDkeven} under the deformation phases $\vartheta_g=r^{gg}$ \begin{align}\begin{split}&\vartheta_{m^j}=e^{2\pi iuj^2/k^2},\quad\mbox{for $0<j<k/2$},\\&\vartheta_{m^{k/2}}=i^{-u},\quad\vartheta_\mu=i^v,\quad\vartheta_{\mu m}=i^{2v+w}.\end{split}\label{deformationDkeven}\end{align} The modular $S$-matrix can be computed with the help of the character table~\ref{tab:charactersDkeven} and by substituting the exchange phases \eqref{rDkeven} in \eqref{DGTSmatrix}. It is presented in table~\ref{tab:SDkeven}.

The dihedral gauge theories encountered in this article include $D^{[n/2,0,0]}(D_n)$ for $n\equiv0$ modulo 8, and $D^{[n/2,1,1]}(D_n)$ for $n\equiv4$ modulo 8 (see \eqref{Dkorbifolds}). The Dijkgraaf-Witten invariants $[u,v,w]\in\mathbb{Z}_k\times\mathbb{Z}_2\times\mathbb{Z}_2$ are fixed by the subgroup inclusion $i:\mathbb{Z}_n\hookrightarrow D_n$ and the (non-canonical) injections $j_0,j_1:\mathbb{Z}_2\to D_n$ into the two inequivalent reflections in $D_n$ (see \eqref{Dkevengroupextension}). The inclusion/injections induce the restriction homomorphisms in the group cohomology level $i^\ast:H^3(D_n,U(1))\to H^3(\mathbb{Z}_n,U(1))=\mathbb{Z}_n$ and $j_0^\ast,j_1^\ast:H^3(D_n,U(1))\to H^3(\mathbb{Z}_2,U(1))=\mathbb{Z}_2$ (see \eqref{HDkevenhomo}), all being surjective in this case and extract $i^\ast[u,v,w]=[u]$, $j_0^\ast[u,v,w]=[v]$ and $j_1^\ast[u,v,w]=[w]$ (see \eqref{Dkevenrestrictions}). The invariant $[u]=[n/2]$ is fixed by the $D^{[n/2]}(\mathbb{Z}_{n/2})$ theory in \eqref{Zkorbifolds} when $n\equiv0$ modulo 4, while $[v]=[w]=[0]$ (or $[1]$) is fixed by $D^{[0]}(\mathbb{Z}_2)$ (or $D^{[1]}(\mathbb{Z}_2)$) in \eqref{Z2orbifoldseven} when $n\equiv0$ (resp.~$n\equiv4$) modulo 8. From \eqref{deformationDkeven}, the two $\mathbb{Z}_2$ invariants dictate the topological spins of the $\mathbb{Z}_2$ fluxes $([\mu],\zeta^{\lambda_1,\lambda_2})$ and $([\mu m],\zeta^{\lambda_1,\lambda_2})$ to be bosonic/fermionic (or semionic) when $n\equiv0$ (resp.~$n\equiv4$) modulo 8. Choosing the Dijkgraaf-Witten invariants $(u,v,w)=(-n/2,0,0)$ or $(-n/2,1,1)$, the topological spins of the $\mathbb{Z}_k$ fluxes $[m^j]$ are $\vartheta_{m^j}=e^{-\pi ij^2/n}$, which match with that of the $D^{[n/2]}(\mathbb{Z}_n)$ gauge theory.

\begin{sidewaystable}
\resizebox{\columnwidth}{!}{\begin{tabular}{c|ccccccccc}
&$([1],A_\nu)$&$([m^{k/2}],A_\nu)$&$([1],B_\nu)$&$([m^{k/2}],B_\nu)$&$([1],E_l)$&$([m^{k/2}],E_l)$&$([m^j],z^b)$&$([\mu],\zeta^{\lambda_1,\lambda_2})$&$([\mu m],\zeta^{\lambda_1,\lambda_2})$\\\hline
$([1],A_{\nu'})$&1&1&1&1&2&2&2&$(-1)^{\nu'}k/2$&$(-1)^{\nu'}k/2$\\
$([m^{k/2}],A_{\nu'})$&1&$(-1)^u$&$(-1)^{k/2}$&$(-1)^{u+k/2}$&$2(-1)^l$&$2(-1)^{u+l}$&$2(-1)^b$&$(-1)^{\nu'+\lambda_2}s^{uvw}_{k/2,\mu}k/2$&$(-1)^{\nu'+\lambda_2}s^{uvw}_{k/2,\mu}k/2$\\
$([1],B_{\nu'})$&1&$(-1)^{k/2}$&1&$(-1)^{k/2}$&2&$2(-1)^{k/2}$&$2(-1)^j$&$(-1)^{\nu'}k/2$&$-(-1)^{\nu'}k/2$\\
$([m^{k/2}],B_{\nu'})$&1&$(-1)^{u+k/2}$&$(-1)^{k/2}$&$(-1)^u$&$2(-1)^l$&$2(-1)^{u+l+k/2}$&$2(-1)^{j+b}$&$(-1)^{\nu'+\lambda_2}s^{uvw}_{k/2,\mu}k/2$&$-(-1)^{\nu'+\lambda_2}s^{uvw}_{k/2,\mu}k/2$\\
$([1],E_{l'})$&2&$2(-1)^{l'}$&2&$2(-1)^{l'}$&4&$4(-1)^{l'}$&$4c(jl')$&0&0\\
$([m^{k/2}],E_{l'})$&2&$2(-1)^{u+l'}$&$2(-1)^{k/2}$&$2(-1)^{u+l'+k/2}$&$4(-1)^l$&$4(-1)^{u+l+l'}$&$4(-1)^bc(jl')$&0&0\\
$([m^{j'}],z^{b'})$&2&$2(-1)^{b'}$&$2(-1)^{j'}$&$2(-1)^{j'+b'}$&$4c(j'l)$&$4(-1)^{b'}c(j'l)$&$4s^u_{jbj'b'}$&0&0\\
$([\mu],\zeta^{\lambda'_1,\lambda'_2})$&$(-1)^\nu k/2$&$(-1)^{\nu+\lambda'_2}s^{uvw}_{k/2,\mu}k/2$&$(-1)^\nu k/2$&$(-1)^{\nu+\lambda'_2}s^{uvw}_{k/2,\mu}k/2$&0&0&0&$\Sigma^{\lambda\lambda'}_{00}$&$\Sigma^{\lambda\lambda'}_{01}$\\
$([\mu m],\zeta^{\lambda_1,\lambda_2})$&$(-1)^\nu k/2$&$(-1)^{\nu+\lambda'_2}s^{uvw}_{k/2,\mu}k/2$&$-(-1)^\nu k/2$&$-(-1)^{\nu+\lambda'_2}s^{uvw}_{k/2,\mu}k/2$&0&0&0&$\Sigma^{\lambda\lambda'}_{10}$&$\Sigma^{\lambda\lambda'}_{11}$\\
\end{tabular}}
\caption{The $S$-matrix $2kS_{\chi\chi'}$ from \eqref{DGTSmatrix} of the $D^{[u,v,w]}(D_k)$ gauge theory, for even $k$. The row and column entries are arranged in the same order of the dyons in table~\ref{tab:dyonsDkeven}. We abbreviate $c(x)=\cos(2\pi x/k)$, $s^u_{jbj'b'}=c(jb'+j'b+2ujj'/k)$ and $s^{uvw}_{k/2,\mu}=i^{-u+(v+w)[k/2]_2}$. Here, $[k/2]_2=0$ (or 1) when $k/2$ is even (odd). If $k\equiv0$ mod 4, $\Sigma^{\lambda\lambda'}_{00}=\frac{k}{2}(-1)^{\lambda_1+\lambda'_1+v}\left[1+(-1)^{\lambda_2+\lambda'_2}i^{-u}\right]$, $\Sigma^{\lambda\lambda'}_{11}=\frac{k}{2}(-1)^{\lambda_1+\lambda'_1+w}\left[1+(-1)^{\lambda_2+\lambda'_2}i^{-u}\right]$ and $\Sigma^{\lambda\lambda'}_{01}=\Sigma^{\lambda\lambda'}_{10}=0$. If $k\equiv2$ mod 4, $\Sigma^{\lambda\lambda'}_{00}=\frac{k}{2}(-1)^{\lambda_1+\lambda'_1+v}$, $\Sigma^{\lambda\lambda'}_{11}=\frac{k}{2}(-1)^{\lambda_1+\lambda'_1+w}$ and $\Sigma^{\lambda\lambda'}_{01}=\Sigma^{\lambda\lambda'}_{10}=\frac{k}{2}(-1)^{\lambda_1+\lambda_2+\lambda'_1+\lambda'_2}i^{-u+3v+w}$.}\label{tab:SDkeven}
\end{sidewaystable}

\begin{sidewaystable}
\begin{tabular}{c|ccccccc}
&$([1],A_\nu)$&$([\mu^2],A_\nu)$&$([1],E_l)$&$([\mu^2],E_l)$&$([m^j],z^b)$&$([\mu],\zeta^\lambda)$&$([\mu^3],\zeta^\lambda)$\\\hline
$([1],A_{\nu'})$&1&$(-1)^{\nu'}$&2&$2(-1)^{\nu'}$&$2(-1)^{j\nu'}$&$ki^{\nu'}$&$ki^{-\nu'}$\\
$([\mu^2],A_{\nu'})$&$(-1)^\nu$&$(-1)^{u+\nu+\nu'}$&$2(-1)^l$&$2(-1)^{u+l+\nu'}$&$2(-1)^{uj+j\nu'+b}$&$ki^{u+\nu'+2\lambda}$&$ki^{-u-\nu'+2\lambda}$\\
$([1],E_{l'})$&2&$2(-1)^{l'}$&4&$4(-1)^{l'}$&$4(-1)^{jl'}c(jl')$&0&0\\
$([\mu^2],E_{l'})$&$2(-1)^\nu$&$2(-1)^{u+l'+\nu}$&$4(-1)^l$&$4(-1)^{u+l+l'}$&$4(-1)^{uj+jl'+b}c(jl')$&0&0\\
$([m^{j'}],z^{b'})$&$2(-1)^{j'\nu}$&$2(-1)^{uj'+j'\nu+b'}$&$4(-1)^{j'l}c(j'l)$&$4(-1)^{uj'+j'l+b'}c(j'l)$&$4s^u_{jbj'b'}$&0&0\\
$([\mu],\zeta^{\lambda'})$&$ki^\nu$&$ki^{u+\nu+2\lambda'}$&0&0&0&$ki^{\lambda+\lambda'}e^{i\pi u/4}$&$ki^{\lambda-\lambda'}e^{-i\pi u/4}$\\
$([\mu^3],\zeta^{\lambda'})$&$ki^{-\nu}$&$ki^{-u-\nu+2\lambda'}$&0&0&0&$ki^{-\lambda+\lambda'}e^{-i\pi u/4}$&$ki^{-\lambda-\lambda'}e^{i\pi u/4}$\\
\end{tabular}
\caption{The $S$-matrix $4kS_{\chi\chi'}$ from \eqref{DGTSmatrix} of the $D^{[u]}(Q_{4k})$ gauge theory, for odd $k$. We abbreviate $c(x)=\cos(2\pi x/k)$ and $s^u_{jbj'b'}=c((jb'+j'b)/2+u(2/k+k/2)jj')$. The row and column entries are arranged in the same order of the dyons in table~\ref{tab:dyonsDick}.}\label{tab:SDick} 
\end{sidewaystable}

\subsection{The \texorpdfstring{$Q_{4k}$}{dicyclic} gauge theory}\label{app:DickGT}
We only consider the dicycle group $Q_{4k}$ with odd degree $k$. Its group elements are labeled by $g=(A,a)=\hat\mu^Am^a$, where $A=-1,0,1,2$ and $a\in(-k/2,k/2)$. We also abbreviate $\hat{m}=\hat\mu^2 m=(2,1)$. (See \eqref{appDickdefinition} for the group presentations.) The Dijkgraaf-Witten invariants of the $D^{[u]}(Q_{4k})$ gauge theory are cohomology elements $[u]$ in $H^3(Q_{4k},U(1))=\mathbb{Z}_{4k}$. The $f$-symbols can be identified with the cocycle representative $f^{g_1g_2g_3}=\xi^3_u(g_1,g_2,g_3)$ in \eqref{repH3DickU1}. The solution to the hexagon equation \eqref{hexagoneq} can be chosen to be \begin{widetext}\begin{align}r^{g_1g_2}_u=\exp\left(\frac{2\pi iu}{k^2}\left\{a_2\left[(-1)^{A_2}a_1+2\left[A_1\right]_2a_2\right]-\left[A_1\right]_2a_2^2\right\}\right)\exp\left(\frac{i\pi u}{8}A_1A_2\right),\label{rDick}\end{align}\end{widetext} where the square brackets, $[\ast]_k$ and $[\ast]_2$, wrap the numbers back to the appropriate ranges in $(-k/2,k/2]$ and $0,1$ respectively. Exchange phases $r^{g_1g_2}_{4k}$ can be absorbed by attaching gauge charges to fluxes. Therefore, $u$ has period $4k$. The dyons and their quantum dimensions and topological spins are listed in table~\ref{tab:dyonsDkeven} under the deformation phases $\vartheta_g=r^{gg}$ \begin{align}\vartheta_{\hat{m}^j}=\exp\left(2\pi iu\left(\frac{j^2}{k^2}+\frac{[j]_2}{4}\right)\right),\quad\vartheta_{\hat\mu^A}=\exp\left(\frac{i\pi uA^2}{8}\right),\label{deformationDick}\end{align} where $[j]_2=0,1$ is congruent to $j$ modulo 2. (Here, as a simplification, we have replaced $e^{2\pi iu[j]_k^2/k^2}$ by $e^{2\pi iuj^2/k^2}$ in $\vartheta_{\hat{m}^j}$ by attaching the $\mathbb{Z}_k$ charge $\hat{m}^j\to\hat{m}^jz^2$ when $j>k/2$ to absorb the phase difference $e^{4\pi ij/k}$.) The modular $S$-matrix can be computed with the help of the character table~\ref{tab:charactersDick} and by substituting the exchange phases \eqref{rDick} in \eqref{DGTSmatrix}. It is presented in table~\ref{tab:SDick}.

The dicyclic gauge theory component in $SO(2n)_1/D_k$ is $D^{[2k]}(Q_{4k})$, when $k=n/2$ is odd (see \eqref{Dkorbifolds}). The Dijkgraaf-Witten invariant $[u]=[2k]$ is fixed by the subgroup inclusion $i:\mathbb{Z}_k\hookrightarrow Q_{4k}$ and the (non-canonical) injection $j:\mathbb{Z}_4\to Q_{4k}$ (see \eqref{DkDickgroupextension}). They induce the restriction homomorphisms $i^\ast:H^3(Q_{4k},U(1))\to H^3(\mathbb{Z}_k,U(1))=\mathbb{Z}_k$ and $j^\ast:H^3(Q_{4k},U(1))\to H^3(\mathbb{Z}_4,U(1))=\mathbb{Z}_4$ (see \eqref{HDkDickhomo}), both being surjective in this case and relate $i^\ast[u]_{4k}=[u]_k$ and $j^\ast[u]_{4k}=[u]_4$. The Dijkgraaf-Witten invariants of $D^{[0]}(\mathbb{Z}_k)$ in \eqref{Zkorbifolds} and $D^{[2]}(\mathbb{Z}_4)$ in \eqref{Z2orbifoldseven}, when $n=2k\equiv2$ modulo 4, therefore restrict $u=2k$ modulo $4k$. The topological spins of the fluxes in \eqref{deformationDick} become $\vartheta_{\hat{m}^j}=(-1)^j$ and $\vartheta_{\hat\mu^A}=e^{i\pi kA^2/4}$. In particular, this shows $\hat{m}^k=\hat\mu^2$ is fermionic and can be identified with the fermion $\psi$ in $SO(2n)_1$.

\section{Useful group cohomology identities}\label{app:cohomology}
We summarize some relevant tools in computing group cohomologies $H^r(G,M)$.~\cite{Cohomologybook,Hatcherbook,CohomologyofFiniteGroupsbook,Adem06,ChenGuLiuWen11} We only consider the following finite groups $G$: (i) the cyclic group $\mathbb{Z}_k$ of arbitrary order $k$, (ii) the dihedral group $D_k=\mathbb{Z}_2\ltimes\mathbb{Z}_k$ of arbitrary degree $k$, and (iii) the dicyclic group $Q_{4k}=\mathbb{Z}_4\ltimes\mathbb{Z}_k$ with odd degrees $k$ (see \eqref{Dicodd}). We only examine the following Abelian coefficients group $M$: (i) the ring of integers $\mathbb{Z}$, (ii) the (multiplicative) group $U(1)$, and (iii) the anyon fusion group $\mathcal{A}=\mathbb{Z}_2^2$ of $SO(2n)_1$ when $n$ is even. We always assume $G$ acts trivially on $M$. (We omit the unnecessary odd $n$ cases when the $\mathbb{Z}_2$ conjugation acts non-trivially on $\mathcal{A}_{SO(2n)_1}=\mathbb{Z}_4$.) The group cohomology is defined to be the quotient \begin{align}H^r(G,M)=\frac{\mathrm{ker}\left(d:C^r(G,M)\to C^{r+1}(G,M)\right)}{\mathrm{Im}\left(d:C^{r-1}(G,M)\to C^r(G,M)\right)}.\label{cohomologydef}\end{align} The set of $r$-cochains, $C^r(G,M)$, consists of functions $\omega:G^r\to M$ that are not necessarily group homomorphisms. $C^r(G,M)$ is an Abelian group under addition $\omega_1+\omega_2$ if $M=\mathbb{Z}$ or multiplication $\omega_1\omega_2$ if $M=U(1)$ or $\mathcal{A}$. The differential map is defined by 
\begin{align}\begin{split}&d\omega(g_1,\ldots,g_{r+1})\\&=\omega(g_2,\ldots,g_{r+1})+\sum_{j=1}^r(-1)^j\omega(g_1,\ldots,g_{j-1},g_jg_{j+1},\ldots,g_{r+1})+(-1)^{r+1}\omega(g_1,\ldots,g_r)\\&d\omega(g_1,\ldots,g_{r+1})\\&=\omega(g_2,\ldots,g_{r+1})\left[\prod_{j=1}^r\omega(g_1,\ldots,g_{j-1},g_jg_{j+1},\ldots,g_{r+1})^{(-1)^j}\right]\omega(g_1,\ldots,g_r)^{(-1)^{r+1}}\end{split}\label{rcocyclecondition}\end{align}
depending whether $M$ is additive or multiplicative. The differential map is nilpotent and squares to zero, $d(d\omega)=0$. A $d$-exact cochain $\omega$, satisfying $d\omega=0$, is called a cocycle. The derivative of a cochain, $\omega=d\eta$, is called a coboundary. Cohomology elements in \eqref{cohomologydef} are equivalent classes (i.e.~cosets) $[\omega]=\omega+dC^{r-1}(G,M)$ for additive coefficients, or $[\omega]=\omega dC^{r-1}(G,M)$ for multiplicative coefficients.

By construction, $H^r(G,M)=0$ for any negative order $r<0$, and $H^0(G,M)=M$. The group cohomologies for $G=\mathbb{Z}_k$, $D_k$ and $Q_{4k}$ with integral coefficient are known. For the cyclic groups and positive orders $r>0$, \begin{align}H^r(\mathbb{Z}_k,\mathbb{Z})=\left\{\begin{array}{*{20}l}\mathbb{Z}_k,&\mbox{if $r$ is even}\\0,&\mbox{if $r$ is odd}\end{array}\right..\label{HZkZ}\end{align} For the dihedral groups and positive orders $r>0$, when $k$ is odd, \begin{align}H^r(D_k,\mathbb{Z})=\left\{\begin{array}{*{20}l}\mathbb{Z}_{2k},&\mbox{if $r\equiv0$ mod 4}\\\mathbb{Z}_2,&\mbox{if $r\equiv2$ mod 4}\\0,&\mbox{if $r$ is odd}\end{array}\right.,\label{HDkZodd}\end{align} and when $k$ is even,~\cite{Handel93} \begin{align}H^r(D_k,\mathbb{Z})=\left\{\begin{array}{*{20}l}\mathbb{Z}_k\times\mathbb{Z}_2^{r/2},&\mbox{if $r\equiv0$ mod 4}\\\mathbb{Z}_2^{(r-1)/2},&\mbox{if $r\equiv1,3$ mod 4}\\\mathbb{Z}_2^{(r+2)/2},&\mbox{if $r\equiv2$ mod 4}\end{array}\right..\label{HDkZeven}\end{align} For the dicyclic groups and positive orders $r>0$, we only need the cases when $k$ is odd, \begin{align}H^r(Q_{4k},\mathbb{Z})=\left\{\begin{array}{*{20}l}\mathbb{Z}_4,&\mbox{if $r\equiv1$ mod 4}\\\mathbb{Z}_{4k},&\mbox{if $r\equiv3$ mod 4}\\0,&\mbox{if $r$ is even}\end{array}\right..\label{HDickodd}\end{align} These results can be computed using technique such as the Lyndon-Hochschild-Serre spectral sequence~\cite{Lyndon48,HochschildSerre53}. The derivation is not needed and will be omitted.

The group cohomologies with the multiplicative coefficient $M=U(1)$ or $\mathbb{Z}_2^2$ can be derived from the above integral coefficient ones using the universal coefficient theorem (or the K\"unneth formula, by using $G\cong G\times1$ and $M\cong\mathbb{Z}\otimes M$) for cohomologies~\cite{Hatcherbook,ChenGuLiuWen11} \begin{align}H^r(G,M)=\left[H^r(G,\mathbb{Z})\otimes M\right]\oplus\mathrm{Tor}_1^{\mathbb{Z}}(H^{r+1}(G,\mathbb{Z}),M),\label{UCT}\end{align} where $\otimes$ is the tensor product that treats Abelian groups as modules over the ring $\mathbb{Z}$. The direct sum $\oplus$ is the Cartesian product $\times$ and the notation is used here to avoid confusion with $\otimes$. Tensor products between (additive) Abelian groups can be computed using \begin{gather}(A_1\oplus A_2)\otimes B=(A_1\otimes B)\oplus(A_2\otimes B),\nonumber\\\mathbb{Z}\otimes B=B,\quad\mathbb{Z}_k\otimes B=B/kB.\label{tensoridentities}\end{gather} The tor functor $\mathrm{Tor}_1^{\mathbb{Z}}$ between (additive) Abelian groups can be computed using the following properties \begin{align}&\mathrm{Tor}_1^{\mathbb{Z}}(A_1\oplus A_2,B)=\mathrm{Tor}_1^{\mathbb{Z}}(A_1,B)\oplus\mathrm{Tor}_1^{\mathbb{Z}}(A_2,B),\nonumber\\&\mathrm{Tor}_1^{\mathbb{Z}}(\mathbb{Z}_k,B)=B[k]=\{b\in B:kb=1\},\label{toridentities}\\&\mathrm{Tor}_1^{\mathbb{Z}}(\mathbb{Z},B)=0,\nonumber\end{align} where $B[k]$ is known as the $k$-torsion subgroup of $B$. In particular, we are interested in (i) the $H^2(G,\mathcal{A})$ classification of quantum symmetry group $\widehat{G}$ -- the $\mathcal{A}=\mathbb{Z}_2^2$ central extension \eqref{appgroupextension} of $G=\mathbb{Z}_k$ and $D_k$ -- and (ii) the Dijkgraaf-Witten invariants in $H^3(G,U(1))$, for $G=\mathbb{Z}_k$, $D_k$ and $Q_{4k}$. They are given by \begin{align}H^2(\mathbb{Z}_k,\mathcal{A})&=\left\{\begin{array}{*{20}l}0,&\mbox{for $k$ odd}\\\mathcal{A},&\mbox{for $k$ even}\end{array}\right.,\nonumber\\H^2(D_k,\mathcal{A})&=\left\{\begin{array}{*{20}l}\mathcal{A},&\mbox{for $k$ odd}\\\mathcal{A}^3,&\mbox{for $k$ even}\end{array}\right.,\nonumber\\H^3(\mathbb{Z}_k,U(1))&=\mathbb{Z}_{2k},\nonumber\\H^3(D_k,U(1))&=\left\{\begin{array}{*{20}l}\mathbb{Z}_{2k},&\mbox{for $k$ odd}\\\mathbb{Z}_k\times\mathbb{Z}_2^2,&\mbox{for $k$ even}\end{array}\right.,\nonumber\\H^3(Q_{4k},U(1))&=\mathbb{Z}_{4k},\quad\mbox{for $k$ odd}.\label{listofcohomologies}\end{align}

Next, we present explicit cocycle representatives of the elements in the above cohomologies. We begin with group cohomologies of the cyclic group $\mathbb{Z}_k$. 1-cocycles in $C^1(\mathbb{Z}_k,U(1))$ are group homomorphisms $\omega^1_u:\mathbb{Z}_k\to U(1)$ that send an (additive) group element $a$ in $\mathbb{Z}_k=\mathbb{Z}/(k\mathbb{Z})$ to \begin{align}\omega^1_u(a)=e^{2\pi iua/k},\label{omega1}\end{align} where $u=0,1,\ldots,k-1$ (mod $k$) represents a cohomology element in $H^1(\mathbb{Z}_k,U(1))=\mathrm{Hom}(\mathbb{Z}_k,U(1))=\mathbb{Z}_k$. In general, cohomologies with integral coefficients can be identified with ones with $U(1)$ coefficients. The short exact sequence of coefficient groups $0\to\mathbb{Z}\hookrightarrow\mathbb{R}\xrightarrow{\exp(2\pi i\ast)}U(1)\to1$ induces the long exact sequence of cohomologies \begin{align}\ldots\to H^r(G,\mathbb{R})&\to H^r(G,U(1))\\&\xrightarrow{\delta}H^{r+1}(G,\mathbb{Z})\to H^{r+1}(G,\mathbb{R})\to\ldots.\nonumber\end{align} The differential map (also called connecting homomorphism) is defined by $\delta\omega=d\log\omega/(2\pi i)$. The universal coefficient theorem \eqref{UCT} implies $H^r(G,F)=H^r(G,\mathbb{Z})\otimes F$ for any characteristic zero field $F$ such as $\mathbb{Q}$, $\mathbb{R}$, and $\mathbb{C}$. In particular, if $H^r(G,\mathbb{Z})$ is finite like the case when $r>0$ and $G=\mathbb{Z}_k$, $D_k$ or $Q_{4k}$, then $H^r(G,F)=0$. Therefore, \begin{equation}\begin{tikzcd}\delta:H^r(G,U(1))\arrow[r,"\cong"]&H^{r+1}(G,\mathbb{Z})\end{tikzcd},\label{HU1Ziso}\end{equation} is an isomorphism. Applying to the group homomorphisms \eqref{omega1} in $H^1(\mathbb{Z}_k,U(1))$, it generates the cocycle representatives of all cohomology elements $[\Omega^2_u]=\Omega^2_u+dC^1(\mathbb{Z}_k,U(1))$ in $H^2(\mathbb{Z}_k,\mathbb{Z})$. \begin{align}\Omega^2_u(a,b)&=\delta\omega^1_u(a,b)=\frac{d\log\omega^1_u(a,b)}{2\pi i}\nonumber\\&=\frac{u}{k}\left(a+b-[a+b]\right)\label{Omega2}\end{align} where group elements in $\mathbb{Z}_k$ are represented by $a,b=-k/2+1,\ldots,k/2$ when $k$ is even or $a,b=-(k-1)/2,\ldots,(k-1)/2$ when $k$ is odd. $[a+b]$ is the integer in the same range $\mathbb{Z}\cap(-k/2,k/2]$ that differs from $a+b$ by a multiple of $k$. The 2-cochains in \eqref{Omega2} have integral values and obey the cocycle condition \eqref{rcocyclecondition}. $\Omega^2_u\equiv\Omega^2_{u+k}$ are equivalent up to the 2-coboundary $\Omega^2_k=d\mathbb{I}$, where $\mathbb{I}(a)=a$. Hence, the index $u=0,1,\ldots,k-1$ modulo $k$ represents the cohomology class $[\Omega^2_u]$ in $H^2(\mathbb{Z}_k,\mathbb{Z})=\mathbb{Z}_k$.

In general, all group cohomology with integral coefficients $H^\ast(G,\mathbb{Z})=\bigoplus_{r\geq0}H^r(G,\mathbb{Z})$ has an (graded) ring structure, where the ring multiplication is given by a cup product $\cup:H^r(G,\mathbb{Z})\otimes H^s(G,\mathbb{Z})\to H^{r+s}(G,\mathbb{Z})$ between cochains \begin{align}\omega^r\cup\eta^s(g_1,\ldots,g_{r+s})=\omega^r(g_1,\ldots,g_r)\eta^s(g_{r+1},\ldots,g_{r+s})\label{cupproduct}\end{align} using the normal integer multiplication. Similar to differential forms, it obeys the Leibniz product rule $d(\omega^r\cup\eta^s)=(d\omega^r)\cup\eta^s+(-1)^r\omega^r\cup(d\eta^s)$. Cohomologies of the cyclic group $\mathbb{Z}_k$ (as well as the dihedral and dicyclic groups, $D_k$ and $Q_{4k}$, with odd degree $k$) are periodic, where an unit element $X$ in $H^2(\mathbb{Z}_k,\mathbb{Z})$ (resp.~$H^4(D_k,\mathbb{Z})$ and $H^4(Q_{4k},\mathbb{Z})$) generates the degree-raising isomorphisms $X\cup:H^r(\mathbb{Z}_k,\mathbb{Z})\xrightarrow{\cong}H^{r+2}(\mathbb{Z}_k,\mathbb{Z})$ (resp.~$H^r(D,\mathbb{Z})\xrightarrow{\cong}H^{r+4}(D,\mathbb{Z})$ for $D=D_k$, $Q_{4k}$). The unit element $X=[\Omega^2_{u=1}]$ in \eqref{Omega2} generates the entire cohomology (polynomial) ring $H^\ast(\mathbb{Z}_k,\mathbb{Z})=\mathbb{Z}_k[X]$. Any even degree cohomology element $[\Omega^{2s}_u]$ in $H^{2s}(\mathbb{Z}_k,\mathbb{Z})=\mathbb{Z}_k$ can be represented by the product \begin{align}\Omega^{2s}_u(a_1,\ldots,a_{2s})&=u\Omega^2_1\cup\ldots\cup\Omega^2_1(a_1,\ldots,a_{2s})\label{repHZkZ}\\&=u\prod_{j=1}^s\Omega^2_1(a_{2j-1},a_{2j})\nonumber\\&=\frac{u}{k^s}\prod_{j=1}^s\left(a_{2j-1}+a_{2j}-[a_{2j-1}+a_{2j}]\right),\nonumber\end{align} where $a_1,\ldots,a_{2s}\in\mathbb{Z}\cap(-k/2,k/2]$ represent group elements in $\mathbb{Z}_k$. The connecting isomorphism \eqref{HU1Ziso} produces explicit representatives for all odd degree cohomology elements $[\omega^{2s-1}_u]$ in $H^{2s-1}(\mathbb{Z}_k,U(1))=\mathbb{Z}_k$. \begin{align}&\omega^{2s-1}_u(a_1,\ldots,a_{2s-1})\label{repHZkU1}\\&=\exp\left(\frac{2\pi iu}{k^s}a_1\prod_{j=1}^{s-1}\left(a_{2j}+a_{2j+1}-[a_{2j}+a_{2j+1}]\right)\right)\nonumber\end{align} obeys $\delta\omega^{2s-1}_u=d\log\omega^{2s-1}_u/(2\pi i)=\Omega^{2s}_u$. In particular, the $F$-symbol of the $\mathbb{Z}_k$ fluxes of the $D^{[u]}(\mathbb{Z}_k)$ gauge theory (see appendix~\ref{app:DGT}) is chosen to be the 3-cocycle $f^{abc}=\omega^3_u(a,b,c)$ that represents the cohomology element $[u]$ in $H^3(\mathbb{Z}_k,U(1))=\mathbb{Z}_k$. 

Using the universal coefficient theorem \eqref{UCT}, when $k$ is even, the cohomology elements $[h_x]$ in $H^2(\mathbb{Z}_k,\mathcal{A})=H^2(\mathbb{Z}_k,\mathbb{Z})\otimes\mathcal{A}=\mathcal{A}$, with coefficients in the anyon fusion group $\mathcal{A}_{SO(2n)_1}=\mathbb{Z}_2^2$ for even $n$, can be represented by the cocycles \begin{align}h_x(a,b)=x^{(a+b-[a+b])/k}\label{repH2ZkA}\end{align} where $x=1,\psi,s_+,s_-$ are $SO(2n)_1$ anyons in $\mathcal{A}$ that follows the fusion rules \eqref{SO2nfusion}. When $k$ is odd, the cohomology group is trivial and $h_x\equiv h_1$ are equivalent up to a 2-coboundary for any $x$. The cohomology $[h_x]$ determines the central $\mathcal{A}$-extension of the global symmetry group $\mathbb{Z}_k=\langle m|m^k=1\rangle$ to the quantum symmetry group \begin{align}\widehat{\mathbb{Z}_k}=\left\langle\hat{m},s_\pm,\psi\left|\begin{array}{*{20}l}\hat{m}^k=x,[\hat{m},s_\pm]=1,\\s_\pm^2=\psi^2=s_+s_-\psi=1\end{array}\right.\right\rangle\label{QSGZkeven}\end{align} where $[g,h]=ghg^{-1}h^{-1}$. (See appendix~\ref{app:DGT} for a general discussion on central extensions and their relation to $H^2(G,\mathcal{A})$.) The $\mathbb{Z}_k$ fluxes now obey the extended fusion rules $\hat{m}^a\times\hat{m}^b=h_x(a,b)\times\hat{m}^{a+b}$. 

In section~\ref{sec:SO2nZkTL}, we see that the $\mathbb{Z}_k$ symmetry of $SO(2n)_1$ is (i) non-trivially extended to $\widehat{\mathbb{Z}_k}=\mathbb{Z}_{2k}\times\mathbb{Z}_2$ according to $[h_x]$, for $x=\psi$ when $k=n/2\equiv2$ modulo 4, or $x=s_\pm$ when $k=n/2\equiv0$ modulo 4, but (ii) trivially extended to $\widehat{\mathbb{Z}_k}=\mathbb{Z}_k\times\mathcal{A}$ when $n=2k\equiv2$ modulo 4. When (iii) $n=k$ is odd, the anyon fusion group becomes $\mathcal{A}_{\mathrm{odd}}=\mathbb{Z}_4$. The $\mathbb{Z}_k$ symmetry of $SO(2n)_1$ is trivially extended to $\widehat{\mathbb{Z}_k}=\mathbb{Z}_k\times\mathbb{Z}_4$ because the cohomology classification is trivial, $H^2(\mathbb{Z}_k,\mathcal{A}_{\mathrm{odd}})=H^2(\mathbb{Z}_k,\mathbb{Z})\otimes\mathcal{A}_{\mathrm{odd}}=0$ (by using the universal coefficient theorem \eqref{UCT}). For case (i), the discrete gauge theory component $D^{[k]}(\mathbb{Z}_{2k})$ in $SO(2n)_1/\mathbb{Z}_k$ (see \eqref{Zkorbifolds}) is deformed by the Dijkgraaf-Witten invariant $[k]$ in $H^3(\mathbb{Z}_{2k},U(1))=\mathbb{Z}_{2k}$ so that the primitive gauge fluxes carry non-trivial spin $h=-1/(4k)$. For case (ii) and (iii), the $D^{[0]}(\mathbb{Z}_k)$ component in $SO(2n)_1/\mathbb{Z}_k$ is un-deformed and its gauge fluxes are bosonic. 

In section~\ref{sec:SO2nZ2TL}, we see that the $\mathbb{Z}_2$ conjugation symmetry of $SO(2n)_1$ is (a) non-trivially extended to $\widehat{\mathbb{Z}_2}=\mathbb{Z}_4\times\mathbb{Z}_2$ according to $[h_{y=\psi}]$ (substitute $k=2$ in \eqref{repH2ZkA}) when $n\equiv2$ modulo 4, but (b) trivially extended to $\widehat{\mathbb{Z}_2}=\mathbb{Z}_2\times\mathcal{A}$ when $n\equiv0$ modulo 4. For case (a), the discrete gauge theory component $D^{[2]}(\mathbb{Z}_4)$ in $SO(2n)_1/\mathbb{Z}_2$ (see \eqref{Z2orbifoldseven}) is deformed by the Dijkgraaf-Witten invariant $[2]$ in $H^3(\mathbb{Z}_4,U(1))=\mathbb{Z}_4$ so that the primitive $\mathbb{Z}_4$ gauge fluxes carry non-trivial spin $h=-1/8$. For case (b), the gauge theory component $D^{[0]}(\mathbb{Z}_2)$ (or $D^{[1]}(\mathbb{Z}_2)$) in $SO(2n)_1/\mathbb{Z}_2$ is deformed by the trivial (non-trivial) invariant $[0]$ ($[1]$) in $H^3(\mathbb{Z}_2,U(1))=\mathbb{Z}_2$ when $n\equiv0$ (resp.~$n\equiv4$) modulo 8 so that the gauge flux is bosonic (semionic). Although omitted in this appendix, it can be shown that (c) when $n$ is odd, the (outer automorphic) $\mathbb{Z}_2$ conjugation symmetry of $SO(2n)_1$ is extended to $\widehat{\mathbb{Z}_2}=\mathbb{Z}_2\ltimes\mathcal{A}_{\mathrm{odd}}=D_4$. The extension split decomposes and corresponds to the trivial element in $H^2(\mathbb{Z}_2,\mathcal{A}_{\mathrm{odd}})$, where $\mathbb{Z}_2$acts as the non-trivial involution on $\mathcal{A}_{\mathrm{odd}}=\mathbb{Z}_4$. The non-chiral components $Z(\mathrm{Ising})$ or $Z(SU(2)_2)$ in $SO(2n)_1/\mathbb{Z}_2$, when $n\equiv\pm1$ or $\pm3$ modulo 8 respectively (see \eqref{Z2orbifoldsodd} and \eqref{DZ2mZ2}), are twist liquids $D^{[0]}(\mathbb{Z}_2)\sslash\mathbb{Z}_2$ that differ from each other by the Dijkgraaf-Witten invariant $[1]\in H^3(\mathbb{Z}_2,U(1))=\mathbb{Z}_2$. 

The group cohomology elements of the dihedral group $D_k$ and dicyclic group $Q_{4k}$, when $k$ is odd, can be explicitly represented by combining the previous results \eqref{repHZkZ} and \eqref{repHZkU1} of the cyclic group with the group decompositions \begin{equation}\begin{tikzcd}1\to\mathbb{Z}_k\arrow[r,hookrightarrow,"i"]&D_k\arrow[r,rightarrow,yshift=0.5ex,"p"]&\mathbb{Z}_2\to1\arrow[l,dashrightarrow,yshift=-0.5ex,"j"],\\1\to\mathbb{Z}_k\arrow[r,hookrightarrow,"i"]&Q_{4k}\arrow[r,rightarrow,yshift=0.5ex,"p"]&\mathbb{Z}_4\to1\arrow[l,dashrightarrow,yshift=-0.5ex,"j"].\end{tikzcd}\label{DkDickgroupextension}\end{equation} We label the group elements by the (additive) integral 2-tuple $g=(A,a)$, where $a=-(k-1)/2,\ldots,(k-1)/2$, and $A=0,1$ for $D_k$ or $A=-1,0,1,2$ for $Q_{4k}$. The group product is \begin{align}(A,a)\cdot(B,b)=([A+B],[(-1)^Ba+b]).\end{align} Here, $[(-1)^Ba+b]$ is the integer between $\pm(k-1)/2$ that differs from $(-1)^Ba+b$ by a multiple of $k$. $[A+B]$ is either $0,1$ for $D_k$ (or $-1,0,1,2$ for $Q_{4k}$) and is congruent to $A+B$ modulo 2 (resp.~4). The inclusion map $i$ in \eqref{DkDickgroupextension} inserts a $\mathbb{Z}_k$ group element $a$ into $(0,a)$ in $D_k$ (or $Q_{4k}$). The projection $p$ is the Abelianization map that sends $(A,a)$ to $A$ in $\mathbb{Z}_2$ (resp.~$\mathbb{Z}_4$). The short exact sequences in \eqref{DkDickgroupextension} split so that $\mathrm{ker}(p)=\mathrm{Im}(i)$ and the group homomorphism $j$ that sends $A$ back to $(A,0)$ is a one-sided inverse of $p$, i.e.~$p(j(A))=A$. A group homomorphism $f:G\to H$ induces a homomorphism between cohomologies $f^\ast:H^r(H,M)\to H^r(G,M)$ by defining the cochain $f^\ast\omega(g_1,\ldots,g_r)=\omega(f(g_1),\ldots,f(g_r))$ in $C^r(G,M)$ for any cochain $\omega(h_1,\ldots,h_r)$ in $C^r(H,M)$. The maps in \eqref{DkDickgroupextension} induces the homomorphisms between group cohomologies \begin{equation}\begin{tikzcd}H^r(\mathbb{Z}_2,M)\arrow[r,rightarrow,yshift=0.5ex,"p^\ast"]&H^r(D_k,M)\arrow[l,dashrightarrow,yshift=-0.5ex,"j^\ast"]\arrow[r,rightarrow,"i^\ast"]&H^r(\mathbb{Z}_k,M),\\H^r(\mathbb{Z}_4,M)\arrow[r,rightarrow,yshift=0.5ex,"p^\ast"]&H^r(Q_{4k},M)\arrow[l,dashrightarrow,yshift=-0.5ex,"j^\ast"]\arrow[r,rightarrow,"i^\ast"]&H^r(\mathbb{Z}_k,M).\end{tikzcd}\label{HDkDickhomo}\end{equation} Since $j^\ast\circ p^\ast=\mathbb{I}$ is the identity at the cohomology level, $p^\ast$ must be injective and $j^\ast$ must be surjective. $i^\ast$ and $j^\ast$ are restriction maps. The former restricts the group elements entries in a cochain $\omega(g_1,\ldots,g_r)$ to lie in $\mathbb{Z}_k$, i.e.~$g_i=(A_i,a_i)=(0,a_i)$. Similarly the latter restricts $g_i=(A_i,a_i)=(A_i,0)$. Exactness in general does not hold in cohomology sequences. For instance, from \eqref{HZkZ}, \eqref{HDkZodd} and \eqref{HDickodd}, we see that $i^\ast$ cannot be surjective when $r\equiv2$ modulo 4 and $M=\mathbb{Z}$. 

Similar to the cyclic group case, we begin with $H^1(D_k,U(1))=\mathrm{Hom}(D_k,U(1))=\mathbb{Z}_2$ and $H^1(Q_{4k},U(1))=\mathrm{Hom}(Q_{4k},U(1))=\mathbb{Z}_4$ when $k$ is odd. 1-cocycles are group homomorphisms $G\to U(1)$ sending $g=(A,a)$ to $\xi^1_v(g)=(-1)^{uA}$, where $v=0,1$ modulo 2 for $D_k$, or $\xi^1_v(g)=e^{i\pi uA/2}$, where $v=-1,0,1,2$ modulo 4 for $Q_{4k}$. The connecting isomorphism \eqref{HU1Ziso} produces the cocycle representatives \begin{align}\begin{split}\Xi^2_v(g_1,g_2)&=\frac{v}{2}(A_1+A_2-[A_1+A_2]),\quad\mbox{for }D_k\\\Xi^2_v(g_1,g_2)&=\frac{v}{4}(A_1+A_2-[A_1+A_2]),\quad\mbox{for }Q_{4k}\end{split}\label{repH2DkZ}\end{align} of cohomology elements in $H^2(D_k,\mathbb{Z})=\mathbb{Z}_2$ and $H^2(Q_{4k},\mathbb{Z})=\mathbb{Z}_4$ respectively. The restriction maps $j^\ast:H^2(D_k,\mathbb{Z})\to H^2(\mathbb{Z}_2,\mathbb{Z})$ and $j^\ast:H^2(Q_{4k},\mathbb{Z})\to H^2(\mathbb{Z}_4,\mathbb{Z})$ in \eqref{HDkDickhomo} are isomorphic and therefore the second order cohomology representatives matches with \eqref{Omega2}, $[\Omega^2_v]=j^\ast[\Xi^2_v]$. 

The forth order cohomology elements in $H^4(D_k,\mathbb{Z})$ can be represented by the cocycles \begin{align}&\Xi^4_u(g_1,g_2,g_3,g_4)\label{repH4DkZ}\\&=\frac{u}{k^2}(-1)^{A_3+A_4}\left((-1)^{A_2}a_1+a_2-\left[(-1)^{A_2}a_1+a_2\right]\right)\nonumber\\&\quad\quad\quad\quad\times\left((-1)^{A_4}a_3+a_4-\left[(-1)^{A_4}a_3+a_4\right]\right)\nonumber\\&\quad+\frac{u}{4}\left(A_1+A_2-[A_1+A_2]\right)\left(A_3+A_4-[A_3+A_4]\right),\nonumber\end{align} where $g_i=(A_i,a_i)$ are group elements in $D_k$. Here, the square brackets in the first (second) line have range in $-(k-1)/2,\ldots,(k-1)/2$ (resp.~$0,1$). $\Xi^4_u\equiv\Xi^4_{u+2k}$ are equivalent up to the coboundary $d\Lambda^3$ in $dC^3(D_k,\mathbb{Z})$, where $\Lambda^3(g_1,g_2,g_3)=2(-1)^{A_2+A_3}a_1((-1)^{A_3}a_2+a_3-[(-1)^{A_3}a_2+a_3])/k+kA_1(A_2+A_3-[A_2+A_3])/2$. Thus, $u=0,1,\ldots,2k-1$ modulo $2k$ represents the cohomology elements in $H^4(D_k,\mathbb{Z})=\mathbb{Z}_{2k}$. At the forth order cohomology level, both the restriction maps $i^\ast:H^4(D_k,\mathbb{Z})\to H^4(\mathbb{Z}_k,\mathbb{Z})$ and $j^\ast:H^4(D_k,\mathbb{Z})\to H^4(\mathbb{Z}_2,\mathbb{Z})$ in \eqref{HDkDickhomo} are surjective. They identify $i^\ast[\Xi^4_u]=[\Omega^4_{[u]_k}]$ in $H^4(\mathbb{Z}_k,\mathbb{Z})=\mathbb{Z}_k$ and $j^\ast[\Xi^4_u]=[\Omega^4_{[u]_2}]$ in $H^4(\mathbb{Z}_2,\mathbb{Z})=\mathbb{Z}_2$ (see \eqref{repHZkZ}). 

The forth order cohomology elements in $H^4(Q_{4k},\mathbb{Z})$ can be represented by cocycles similar to \eqref{repH4DkZ} \begin{align}&\Xi^4_u(g_1,g_2,g_3,g_4)\label{repH4DickZ}\\&=\frac{u}{k^2}(-1)^{A_3+A_4}\left((-1)^{A_2}a_1+a_2-\left[(-1)^{A_2}a_1+a_2\right]\right)\nonumber\\&\quad\quad\quad\quad\times\left((-1)^{A_4}a_3+a_4-\left[(-1)^{A_4}a_3+a_4\right]\right)\nonumber\\&\quad+\frac{u}{16}\left(A_1+A_2-[A_1+A_2]\right)\left(A_3+A_4-[A_3+A_4]\right),\nonumber\end{align} except now $A_i$ has range in $-1,0,1,2$ and $u=0,1,\ldots,4k-1$ modulo $4k$ represents the cohomology elements in $H^4(Q_{4k},\mathbb{Z})=\mathbb{Z}_{4k}$. The restriction maps $i^\ast:H^4(Q_{4k},\mathbb{Z})\to H^4(\mathbb{Z}_k,\mathbb{Z})$ and $j^\ast:H^4(Q_{4k},\mathbb{Z})\to H^4(\mathbb{Z}_4,\mathbb{Z})$ in \eqref{HDkDickhomo} are surjective. They identify $i^\ast[\Xi^4_u]=[\Omega^4_{[u]_k}]$ in $H^4(\mathbb{Z}_k,\mathbb{Z})=\mathbb{Z}_k$ and $j^\ast[\Xi^4_u]=[\Omega^4_{[u]_4}]$ in $H^4(\mathbb{Z}_4,\mathbb{Z})=\mathbb{Z}_4$ (see \eqref{repHZkZ}).

The cohomology rings $H^\ast(D_k,\mathbb{Z})$ and $H^\ast(Q_{4k},\mathbb{Z})$, when the degree $k$ is odd, are periodic (see \eqref{HDkZodd} and \eqref{HDickodd}), and an unit element $X=[\Xi^4_{u=1}]$ in $H^4$ produces the isomorphisms by the cup product \eqref{cupproduct}, $X\cup:H^r\xrightarrow{\cong}H^{r+4}$. The generator $Y=[\Xi^2_{v=1}]$ in $H^2$ is related to $X$ by the cup product $Y\cup Y=kX$ and has order 2, $2Y=0$, for $D_k$ or order 4, $4Y=0$, for $Q_{4k}$. Therefore, the cohomology rings are the polynomial quotient rings $H^\ast(D_k,\mathbb{Z})=\mathbb{Z}_{2k}[X,Y]/(2Y,kX-Y^2)$ and $H^\ast(Q_{4k},\mathbb{Z})=\mathbb{Z}_{4k}[X,Y]/(4Y,kX-Y^2)$. In other words, cohomology elements in any order can be explicitly represented as powers of $X$ and $Y$. Cohomology classes in $H^{4s}$ and $H^{4s+2}$ can be represented by cocycles \begin{align}&\Xi^{4s}_u(g_1,\ldots,g_{4s})=u\Xi^4_1\cup\ldots\cup\Xi^4_1(g_1,\ldots,g_{4s})\nonumber\\&=u\prod_{j=1}^s\Xi^4_1(g_{4j-3},g_{4j-2},g_{4j-1},g_{4j}),\\&\Xi^{4s+2}_v(g_1,\ldots,g_{4s})=v\Xi^2_1\cup\Xi^4_1\cup\ldots\cup\Xi^4_1(g_1,\ldots,g_{4s+2})\nonumber\\&=v\Xi^2_1(g_1,g_2)\prod_{j=1}^s\Xi^4_1(g_{4j-1},g_{4j},g_{4j+1},g_{4j+2}),\nonumber\end{align} where $u$ lives in $\mathbb{Z}/(k\mathbb{Z})$, and $v$ sits in $\mathbb{Z}/(2\mathbb{Z})$ for $D_k$ or $\mathbb{Z}/(4\mathbb{Z})$ for $Q_{4k}$.

Using the universal coefficient theorem \eqref{UCT}, for dihedral groups with odd degree $k$, $H^2(D_k,\mathcal{A})=H^2(D_k,\mathbb{Z})\otimes\mathcal{A}=\mathcal{A}$, where $\mathcal{A}=\mathbb{Z}_2^2$ is the anyon fusion group of $SO(2n)_1$ for $n=2k$. The cocycle representatives of cohomology elements $[h_y]$ in $H^2(D_k,\mathcal{A})$ are \begin{align}h_y(g_1,g_2)&=y^{\Xi^2_1(g_1,g_2)}=y^{(A_1+A_2-[A_1+A_2])/2}\nonumber\\&=y^{A_1A_2}\label{repH2DkA}\end{align} where $[\Xi^2_1]$ is the generator of $H^2(D_k,\mathbb{Z})$ defined in \eqref{repH2DkZ}, $g_i=(A_i,a_i)$ are group elements in $D_k$, and $y=1,\psi,s_+,s_-$ are $SO(2n)_1$ anyons in $\mathcal{A}$. The cohomology class $[h_y]$ dictates the product rules of the global quantum symmetries $\widehat{g_1}\widehat{g_2}=h_y(g_1,g_2)\widehat{g_1g_2}$. Consequently, it determines the central $\mathcal{A}$-extension of the global symmetry group $D_k=\langle\mu,m|\mu^2=m^k=(\mu m)^2=1\rangle$ to the quantum symmetry group \begin{align}\widehat{D_k}=\left\langle\begin{array}{*{20}l}\hat{\mu},\hat{m},\\s_\pm,\psi\end{array}\left|\begin{array}{*{20}l}\hat{\mu}^2=(\hat{\mu}\hat{m})^2=y,\\ {} [\hat{\mu},s_\pm]=[\hat{m},s_\pm]=1,\\\hat{m}^k=s_\pm^2=\psi^2=s_+s_-\psi=1\end{array}\right.\right\rangle\label{QSGDkodd}\end{align} where $[g,h]=ghg^{-1}h^{-1}$. Since the cohomology classification is identical to $H^2(\mathbb{Z}_2,\mathcal{A})=\mathcal{A}$, the induced homomorphism $p^\ast:H^2(\mathbb{Z}_2,\mathcal{A})\xrightarrow{\cong}H^2(D_k,\mathcal{A})$ in \eqref{HDkDickhomo} must be isomorphic (when $M=\mathcal{A}$). This correlates the the non-symmorphic quantum symmetry groups $\widehat{\mathbb{Z}_2}=\mathbb{Z}_4\times\mathbb{Z}_2$ and $\widehat{D_k}=Q_{4k}\times\mathbb{Z}_2$, both of which corresponds to the same cohomology class $[h_{y=\psi}]$, when $k=n/2$ is odd (see section~\ref{sec:SO2nZ2TL} and \ref{SO2nDkTL}).

The connecting isomorphism $\delta:H^3(D,U(1))\xrightarrow{\cong}H^4(D,\mathbb{Z})$ from \eqref{HU1Ziso}, for $D=D_k$ or $Q_{4k}$ with odd degree $k$, gives explicit cocycle representations of cohomology elements $[\xi^3_u]$ in $H^3(D,U(1))$ from \eqref{repH4DkZ} and \eqref{repH4DickZ}. The cocycle solution to $\Xi^4_u=\delta\xi^3_u=d\log\xi^3_u/(2\pi i)$ can be chosen to be~\footnote{Cocycle representatives of cohomologies in $H^3(D_k,U(1))$ for odd degree $k$, and even elements in $H^3(Q_{4k},U(1))$ can also be found in ref.~\cite{Propitius-1995}. Eq.\eqref{repH3DickU1} covers both even and odd elements in $H^3(Q_{4k},U(1))$ and is an original result in this paper.}
\begin{align}\begin{split}&\xi^3_u(g_1,g_2,g_3)\\&=\exp\left\{2\pi iu\left(\frac{1}{k^2}(-1)^{A_2+A_3}a_1\left((-1)^{A_3}a_2+a_3-\left[(-1)^{A_3}a_2+a_3\right]\right)+\frac{1}{2}A_1A_2A_3\right)\right\}\end{split}\label{repH3DkoddU1}\end{align} for $H^3(D_k,U(1))=\mathbb{Z}_{2k}$ where $u=0,1,\ldots,2k-1$ modulo $2k$, or \begin{align}\begin{split}&\xi^3_u(g_1,g_2,g_3)\\&=\exp\Bigg\{2\pi iu\left(\frac{1}{k^2}(-1)^{A_2+A_3}a_1\left((-1)^{A_3}a_2+a_3-\left[(-1)^{A_3}a_2+a_3\right]\right)\right)\\&\quad\quad+\frac{2\pi iu}{16}\left(A_1\left(A_2+A_3-[A_2+A_3]\right)\right)\Bigg\}\end{split}\label{repH3DickU1}\end{align} for $H^3(Q_{4k},U(1))=\mathbb{Z}_{4k}$ where $u=0,1,\ldots,4k-1$ modulo $4k$.
The Dijkgraaf-Witten invariant $u$ enters the discrete gauge theory $D^{[u]}(D_k)$ and $D^{[u]}(Q_{4k})$ by modifying the $F$-symbols of the gauge fluxes with the $U(1)$ phases $f^{g_1g_2g_3}=\xi^3_u(g_1,g_2,g_3)$ and subsequently affecting their topological spins (see appendix~\ref{app:DGT}). The group elements $g_i=(A_i,a_i)$ have $a_i$ ranges in $-(k-1)/2,\ldots,(k-1)/2$ and $A_i$ in $0,1$ for $D_k$ or $-1,0,1,2$ in $Q_{4k}$. Each of the square brackets wraps its entry back to the corresponding appropriate range. The restriction maps $i^\ast$ and $j^\ast$ in \eqref{HDkDickhomo} are surjectives at the $H^3(G,U(1))$ level. The $i^\ast$ map restricts the dihedral (dicyclic) group elements in $\xi(g_1,g_2,g_3)$ to have trivial twofold (fourfold) components, $g_i=(A_i,a_i)=(0,a_i)$. The $j^\ast$ map restricts the group elements to have trivial $k$-fold components, $g_i=(A_i,a_i)=(A_i,0)$. For $D_k$, the restriction maps identify $i^\ast[\xi^3_u]=[\omega^3_{[u]_k}]$ in $H^3(\mathbb{Z}_k,U(1))=\mathbb{Z}_k$ and $j^\ast[\xi^3_u]=[\omega^3_{[u]_2}]$ in $H^3(\mathbb{Z}_2,U(1))=\mathbb{Z}_2$. For $Q_{4k}$, they identify $i^\ast[\xi^3_u]=[\omega^3_{[u]_k}]$ in $H^3(\mathbb{Z}_k,U(1))=\mathbb{Z}_k$ and $j^\ast[\xi^3_u]=[\omega^3_{[u]_4}]$ in $H^3(\mathbb{Z}_4,U(1))=\mathbb{Z}_4$. (See \eqref{repHZkU1} for the cocycle representatives $\omega^3_u$ of $H^3(\mathbb{Z}_k,U(1))$.)

In section~\ref{SO2nDkTL}, we see that when $n=2k\equiv2$ modulo 4 where the $D_k$ symmetry is extended to $\widehat{D_k}=Q_{4k}\times\mathbb{Z}_2$, the discrete gauge theory component $D^{[2k]}(Q_{4k})$ of $SO(2n)_1/D_k$ (see \eqref{Dkorbifolds}) is deformed by the Dijkgraaf-Witten invariant $[2k]$ in $H^3(Q_{4k},U(1))=\mathbb{Z}_{4k}$. This is because the cohomology class relates appropriately to those of the $\mathbb{Z}_k$ and $\mathbb{Z}_4$ subgroups of $Q_{4k}$ through the restriction maps, $i^\ast[2k]=[0]\in H^3(\mathbb{Z}_k,U(1))$ and $j^\ast[2k]=[2]\in H^3(\mathbb{Z}_4,U(1))=\mathbb{Z}_4$. These match the Dijkgraaf-Witten invariant of the smaller discrete gauge theory components $D^{[0]}(\mathbb{Z}_k)$ and $D^{[2]}(\mathbb{Z}_4)$ in $SO(2n)_1/\mathbb{Z}_k$ and $SO(2n)_1/\mathbb{Z}_2$ respectively (see \eqref{Zkorbifolds} and \eqref{Z2orbifoldseven}). When $n=k$ is odd, the conjugation symmetry inside $D_k$ acts non-trivially as the involution, $\mathbb{Z}_2:s_+\leftrightarrow s_-$, on the anyon fusion group $\mathcal{A}_{\mathrm{odd}}$. The $D_k$ symmetry is symmorphically extended to $\widehat{D_k}=D_k\ltimes\mathbb{Z}_4$ and the extension corresponds to the trivial cohomology element in $H^2(D_k,\mathcal{A}_{\mathrm{odd}})$. (Using the universal coefficient theorem \eqref{UCT} and the $\otimes$ properties in \eqref{tensoridentities}, $H^2(D_k,\mathcal{A}_{\mathrm{odd}})=H^2(D_k,\mathbb{Z})\otimes\mathbb{Z}_4=\mathbb{Z}_2$. The non-trivial case is irrelevant to this article.) The non-chiral twist liquid $D^{[0]}(\mathbb{Z}_k)\sslash\mathbb{Z}_2$ in $SO(2n)_1/D_k$ (see \eqref{Dkorbifolds}) has two distinct sets of spin statistics for the $\mathbb{Z}_2$ twist fields depending on $n\equiv\pm1$ or $\pm3$ modulo 8. The difference is associated to the non-trivial cohomology element $[k]$ in $H^3(D_k,U(1))=\mathbb{Z}_{2k}$. The cohomology class is fixed by its restriction to the $\mathbb{Z}_k$ and $\mathbb{Z}_2$ subgroups, $i^\ast[k]=[0]$ in $H^3(\mathbb{Z}_k,U(1))$ and $j^\ast[k]=[1]$ in $H^2(\mathbb{Z}_2,U(1))$. 

Lastly, we present the cocycle representatives of the cohomologies in \eqref{listofcohomologies} of the dihedral symmetry group $D_k$ of $SO(2n)_1$ when the degree $k=n/2$ is even. The dihedral group element can be presented by $g=(A,a)$, where $A=0,1$, $a$ ranges in $-k/2+1,\ldots,k/2$. The group product is \begin{align}(A,a)\cdot(B,b)=([A+B],[(-1)^Ba+b])\end{align} where the first square bracket takes the value 0 or 1 that is congruent to $A+B$ modulo 2, and the second square bracket wraps $(-1)^Ba+b$ back to the appropriate range $(-k/2,k/2]$ by adding or subtracting $k$ if necessary. Unlike the odd degree case, the cohomology groups $H^\ast(D_k,\mathbb{Z})$ in \eqref{HDkZeven} are not periodic because $D_k$ contains a non-cyclic subgroup $D_2=\mathbb{Z}_2\times\mathbb{Z}_2$, whose elements are $g=(A,a)=(A,\alpha k/2)$ where $A,\alpha=0,1$. We only present the cocycle representatives in $H^2(D_k,\mathcal{A})$ and $H^3(D_k,U(1))$. Using the universal coefficient theorem \eqref{UCT}, \begin{align}H^2(D_k,\mathcal{A})&=\left[H^2(D_k,\mathbb{Z})\otimes\mathcal{A}\right]\oplus\mathrm{Tor}(H^3(D_k,\mathbb{Z}),\mathcal{A})\nonumber\\&=\mathcal{A}^2\oplus\mathcal{A}=\mathcal{A}^3,\label{H2DkevenA}\end{align} where $\mathcal{A}=\{1,\psi,s_+,s_-\}=\mathbb{Z}_2^2$ is the anyon fusion group for $SO(2n)_1$. Here we have used the cohomology group results $H^2(D_k,\mathbb{Z})=\mathbb{Z}_2^2$ and $H^3(D_k,\mathbb{Z})=\mathbb{Z}_2$ from \eqref{HDkZeven}, and the properties of the tensor product and tor functor in \eqref{tensoridentities} and \eqref{toridentities}. The cocycle representatives of cohomology elements $[h_{x,y,z}]$ in $H^2(D_k,\mathcal{A})=\mathcal{A}^3$ are \begin{align}h_{x,y,z}(g_1,g_2)=x^{(a_1+a_2-[a_1+a_2])/k}y^{(A_1+a_1)A_2}z^{a_1A_2},\label{repH2DkevenA}\end{align} where $x,y,z=1,\psi,s_\pm$ are $SO(2n)_1$ anyons, and $g_i=(A_i,a_i)$ are group elements in $D_k$. The cohomology class $[h_{x,y,z}]$ specifies the product rule of quantum symmetries $\widehat{g_1}\widehat{g_2}=h_{x,y,z}(g_1,g_2)\widehat{g_1g_2}$. It fixes the central $\mathcal{A}$-extension of the global symmetry group $D_k=\langle\mu,m|\mu^2=m^k=(\mu m)^2=1\rangle$ to the quantum symmetry group \begin{align}\widehat{D_k}=\left\langle\begin{array}{*{20}l}\hat{\mu},\hat{m},\\s_\pm,\psi\end{array}\left|\begin{array}{*{20}l}\hat{m}^k=x,\hat{\mu}^2=y,(\hat{\mu}\hat{m})^2=z,\\ {} [\hat{\mu},s_\pm]=[\hat{m},s_\pm]=1,\\s_\pm^2=\psi^2=s_+s_-\psi=1\end{array}\right.\right\rangle\label{QSGDkeven}\end{align} where $[g,h]=ghg^{-1}h^{-1}$.

The split exact decomposition \begin{equation}\begin{tikzcd}1\to\mathbb{Z}_k\arrow[r,hookrightarrow,"i"]&D_k\arrow[r,rightarrow,yshift=0.5ex,"p"]&\mathbb{Z}_2\to1\arrow[l,dashrightarrow,yshift=-0.5ex,"{j_0,j_1}"]\end{tikzcd}\label{Dkevengroupextension}\end{equation} induces the (non-exact) group homomorphisms between cohomologies \begin{equation}\begin{tikzcd}H^r(\mathbb{Z}_2,M)\arrow[r,rightarrow,yshift=0.5ex,"p^\ast"]&H^r(D_k,M)\arrow[l,dashrightarrow,yshift=-0.5ex,"{j_0^\ast,j_1^\ast}"]\arrow[r,rightarrow,"i^\ast"]&H^r(\mathbb{Z}_k,M).\end{tikzcd}\label{HDkevenhomo}\end{equation} The inclusion $i$ embeds $\mathbb{Z}_k$ in $D_k$ by restricting to group elements $g=(A,a)=(0,a)$. The projection $p$ sends $g=(A,a)$ in $D_k$ to $A=0,1$ in $\mathbb{Z}_2$. The projection has two one-sided inverses $j_0$ and $j_1$ sending $A$ in $\mathbb{Z}_2$ to $(A,0)$ and $(A,1)$ in $D_k$ respectively. Since $p(j_{0,1}(g))=g$, $j_{0,1}^\ast\circ p^\ast$ is isomorphic and therefore $p^\ast$ ($j_{0,1}^\ast$) must be injective (surjective). Unlike the odd degree case, here when $k$ is even, there are two inequivalent mirror conjugacy classes $[\mu]$ and $[\mu m]$ in $D_k$. Consequently, the restriction maps $j^\ast_0$ and $j^\ast_1$ are not equivalent in general. For cohomologies with $M=\mathcal{A}$ coefficient, the restriction maps $i^\ast:H^2(D_k,\mathcal{A})\to H^2(\mathbb{Z}_k,\mathcal{A})$, $j_{0,1}^\ast:H^2(D_k,\mathcal{A})\to H^2(\mathbb{Z}_2,\mathcal{A})$ and the induced homomorphism $p^\ast:H^2(\mathbb{Z}_2,\mathcal{A})\to H^2(D_k,\mathcal{A})$ in \eqref{HDkevenhomo} relates the cohomology classes \begin{align}\begin{split}&i^\ast[h_{x,y,z}]=[h_x],\quad p^\ast[h_y]=[h_{1,y,y}],\\&j_0^\ast[h_{x,y,z}]=[h_y],\quad j_1^\ast[h_{x,y,z}]=[h_z],\end{split}\end{align} where $[h_x]\in H^2(\mathbb{Z}_k,\mathcal{A})$ and $[h_y],[h_z]\in H^2(\mathbb{Z}_2,\mathcal{A})$. In section~\ref{sec:SO2nZkTL}, we see that the $\mathbb{Z}_k$ symmetry of $SO(2n)_1$ is non-trivially extended to $\widehat{\mathbb{Z}_k}=\mathbb{Z}_{2k}\times\mathbb{Z}_2$ according to $[h_{x=\psi}]$ when $k=n/2\equiv2$ modulo 4, or $[h_{x=s_\pm}]$ when $k=n/2\equiv0$ modulo 4. In section~\ref{sec:SO2nZ2TL}, we see that the $\mathbb{Z}_2$ conjugation symmetry of $SO(2n)_1$ is trivially extended to $\widehat{\mathbb{Z}_2}=\mathbb{Z}_2\times\mathcal{A}$ and thus associates to the trivial cohomology classes $y=z=1$. Together they fix the cohomology indices $(x,y,z)=(\psi,1,1)$ when $k=n/2\equiv2$ modulo 4, or $(x,y,z)=(s_\pm,1,1)$ when $k=n/2\equiv0$ modulo 4. In both cases, the quantum symmetry group \eqref{QSGDkeven} is the non-symmorphic extension $\widehat{D_k}=D_{2k}\times\mathbb{Z}_2$ (see section~\ref{SO2nDkTL}).

The cocycle representatives of the cohomology elements $[\xi^3_{u,v,w}]$ in $H^3(D_k,U(1))=\mathbb{Z}_k\times\mathbb{Z}_2\times\mathbb{Z}_2$, when $k=n/2$ is even (see \eqref{listofcohomologies}), can be chosen to be~\cite{footnoteDkGTeven} 
\begin{align}\begin{split}&\xi^3_{u,v,w}(g_1,g_2,g_3)\\&=\exp\Bigg\{2\pi i\left(\frac{u}{k^2}(-1)^{A_2+A_3}a_1\left((-1)^{A_3}a_2+a_3-\left[(-1)^{A_3}a_2+a_3\right]\right)\right)\\&\quad\quad\quad+2\pi i\left(\frac{v}{2}A_1A_2A_3+\frac{v+w}{2}a_1A_2A_3\right)\Bigg\},\end{split}\label{repH3DkevenU1}\end{align}
where $g_i=(A_i,a_i)$ are group elements in $D_k$, $u=0,1,\ldots,k-1$ modulo $k$ and $v,w=0,1$ modulo 2. The square bracket $\left[(-1)^{A_3}a_2+a_3\right]$ wraps $(-1)^{A_3}a_2+a_3$ back to the appropriate range $-k/2+1,\ldots,k/2$ by adding or subtracting $k$ if necessary. The cocycles $\xi^3_{u+k,v,w}$, $\xi^3_{u,v+2,w}$ and $\xi^3_{u,v,w+2}$ differ from $\xi^3_{u,v,w}$ only by 3-coboundaries. The restriction maps $i^\ast:H^3(D_k,U(1))\to H^3(\mathbb{Z}_k,U(1))$ and $j_0,j_1:H^3(D_k,U(1))\to H^3(\mathbb{Z}_2,U(1))$ in \eqref{HDkevenhomo} are surjective. $j_{0,1}^\ast$ are one-sided inverses of the injective induced homomorphism $p^\ast:H^3(\mathbb{Z}_2,U(1))\to H^3(D_k,U(1))$. They relate $[\xi^3_{u,v,w}]$ to \begin{align}\begin{split}&i^\ast[\xi^3_{u,v,w}]=[\omega^3_u],\quad p^\ast[\omega^3_v]=[\xi^3_{0,v,v}],\\&j_0^\ast[\xi^3_{u,v,w}]=[\omega^2_v],\quad j_1^\ast[\xi^3_{u,v,w}]=[\omega^2_w],\end{split}\label{Dkevenrestrictions}\end{align} where $\omega^3_u(a_1,a_2,a_3)=e^{2\pi iu\left(a_1+a_2-[a_1+a_2]\right)/k}$ represents a cohomology class in $H^3(\mathbb{Z}_k,U(1))=\mathbb{Z}_k$ (see \eqref{repHZkU1}), $\omega^3_v(A_1,A_2,A_3)=(-1)^{vA_1A_2A_3}$ and $\omega^3_w(A_1,A_2,A_3)=(-1)^{wA_1A_2A_3}$ represent cohomology classes in $H^3(\mathbb{Z}_2,U(1))=\mathbb{Z}_2$. The Dijkgraaf-Witten invariant $[u,v,w]$ in $H^3(D_k,U(1))=\mathbb{Z}_k\times\mathbb{Z}_2\times\mathbb{Z}_2$ enters the discrete gauge theory $D^{[u,v,w]}(D_k)$ by modifying the $F$-symbols of the gauge fluxes with the $U(1)$ phases $f^{g_1g_2g_3}=\xi^3_{u,v,w}(g_1,g_2,g_3)$ and subsequently affecting their topological spins (see appendix~\ref{app:DGT}).

In section~\ref{SO2nDkTL}, we see that when $n=2k\equiv0$ modulo 4 where the $D_k$ symmetry is extended to $\widehat{D_k}=D_{2k}\times\mathbb{Z}_2$, the discrete gauge theory component, $D^{[k,0,0]}(D_{2k})$ for $n\equiv0$ modulo 8 ($D^{[k,1,1]}(D_{2k})$, for $n\equiv4$ modulo 8), of $SO(2n)_1/D_k$ is deformed by the Dijkgraaf-Witten invariant $[k,0,0]$ (resp.~$[k,1,1]$) in $H^3(D_{2k},U(1))=\mathbb{Z}_{2k}\times\mathbb{Z}_2\times\mathbb{Z}_2$ (see \eqref{Dkorbifolds}). The cohomology class is determined by the restriction maps \eqref{Dkevenrestrictions}, $i^\ast[k,0,0]=i^\ast[k,1,1]=[k]$ in $H^3(\mathbb{Z}_{2k},U(1))=\mathbb{Z}_{2k}$ and $j_0^\ast[k,0,0]=j_1^\ast[k,0,0]=[0]$ (resp.~$j_0^\ast[k,1,1]=j_1^\ast[k,1,1]=[1]$) in $H^3(\mathbb{Z}_2,U(1))=\mathbb{Z}_2$. So that these classes match the Dijkgraaf-Witten invariants of the smaller discrete gauge theories $D^{[k]}(\mathbb{Z}_{2k})$ and $D^{[0]}(\mathbb{Z}_2)$ (resp.~$D^{[1]}(\mathbb{Z}_2)$) in \eqref{Zkorbifolds} and \eqref{Z2orbifoldseven}. The $[k]$ in $[k,\ast,\ast]$ determines the spin statistics of the $\mathbb{Z}_{2k}$ fluxes. The $\mathbb{Z}_2$ components $[0,0]$ (or $[1,1]$) dictates that the two inequivalent $\mathbb{Z}_2$ fluxes are all bosonic or fermionic (resp.~semionic).


\end{document}